\documentstyle[12pt]{report}
\makeindex

\font\titlesf=cmssbx10 scaled \magstep3

\def\RR{{\rm I\!\hskip-1pt R}}
\def\QQ{{\rm I\!\!\!Q}}
\def\Square{\mathchoice{\square{6pt}}{\square{5pt}}{\square{4pt}}
    {\square{3pt}}}
\def\square#1{\mathop{\mkern0.5\thinmuskip\vbox{\hrule\hbox{\vrule
    \hskip#1 \vrule height#1 width 0pt \vrule}\hrule}\mkern0.5\thinmuskip}}

\newsavebox{\fblocka}
\savebox{\fblocka}(22,1.5)[l]
{\multiput(4,2)(7,0){3}{\circle{4}}
\put(19.35,3.5){}
}

\newsavebox{\fblockb}
\savebox{\fblockb}(22,1.5)[l]
{\multiput(4,2)(7,0){3}{\circle{4}}
\put(12.4,3.5){\line(1,0){4.37}}
\put(12.4,0.5){\line(1,0){4.37}}
}

\newsavebox{\fblockc}
\savebox{\fblockc}(22,1.5)[l]
{\multiput(4,2)(7,0){3}{\circle{4}}
\put(5.3,3.5){\line(1,0){4.37}}
\put(5.3,0.5){\line(1,0){4.37}}
}

\newsavebox{\fblockd}
\savebox{\fblockd}(22,1.5)[l]
{\multiput(4,2)(7,0){3}{\circle{4}}
\put(5.3,3.5){\line(1,0){4.37}}
\put(5.3,0.5){\line(1,0){4.37}}
\put(12.9,2){\line(1,0){3.1}}
}

\newsavebox{\fblocke}
\savebox{\fblocke}(22,1.5)[l]
{\multiput(4,2)(7,0){3}{\circle{4}}
\put(5.3,3.5){\line(1,0){4.37}}
\put(5.3,0.5){\line(1,0){4.37}}
\put(12.4,3.5){\line(1,0){4.37}}
\put(12.4,0.5){\line(1,0){4.37}}
}

\newsavebox{\fblockf}
\savebox{\fblockf}(22,1.5)[l]
{\multiput(4,2)(7,0){3}{\circle{4}}
\put(5.3,3.5){\line(1,0){4.37}}
\put(5.3,0.5){\line(1,0){4.37}}
\put(12.35,3.5){\line(1,0){4.37}}
\put(12.35,0.5){\line(1,0){4.37}}
\put(12.8,2.5){\line(1,0){3.2}}
\put(12.8,1.5){\line(1,0){3.2}}
}

\newsavebox{\blocka}
\savebox{\blocka}(29,1.5)[l]
{\multiput(4,2)(7,0){4}{\circle{4}}
\put(19.35,3.5){}
}

\newsavebox{\blockb}
\savebox{\blockb}(29,1.5)[l]
{\multiput(4,2)(7,0){4}{\circle{4}}
\put(19.35,3.5){\line(1,0){4.37}}
\put(19.35,0.5){\line(1,0){4.37}}
}

\newsavebox{\blockc}
\savebox{\blockc}(29,1.5)[l]
{\multiput(4,2)(7,0){4}{\circle{4}}
\put(12.4,3.5){\line(1,0){4.37}}
\put(12.4,0.5){\line(1,0){4.37}}
}

\newsavebox{\blockd}
\savebox{\blockd}(29,1.5)[l]
{\multiput(4,2)(7,0){4}{\circle{4}}
\put(5.3,3.5){\line(1,0){4.37}}
\put(5.3,0.5){\line(1,0){4.37}}
}

\newsavebox{\blocke}
\savebox{\blocke}(29,1.5)[l]
{\multiput(4,2)(7,0){4}{\circle{4}}
\put(12.4,3.5){\line(1,0){4.37}}
\put(12.4,0.5){\line(1,0){4.37}}
\put(19.35,3.5){\line(1,0){4.37}}
\put(19.35,0.5){\line(1,0){4.37}}
\put(19.8,2.5){\line(1,0){3.2}}
\put(19.8,1.5){\line(1,0){3.2}}
}

\newsavebox{\blockf}
\savebox{\blockf}(29,1.5)[l]
{\multiput(4,2)(7,0){4}{\circle{4}}
\put(12.4,3.5){\line(1,0){4.37}}
\put(12.4,0.5){\line(1,0){4.37}}
\put(19.9,2){\line(1,0){3.1}}
}

\newsavebox{\blockg}
\savebox{\blockg}(29,1.5)[l]
{\multiput(4,2)(7,0){4}{\circle{4}}
\put(5.3,3.5){\line(1,0){4.37}}
\put(5.3,0.5){\line(1,0){4.37}}
\put(12.9,2){\line(1,0){3.1}}
}

\newsavebox{\blockh}
\savebox{\blockh}(29,1.5)[l]
{\multiput(4,2)(7,0){4}{\circle{4}}
\put(12.4,3.5){\line(1,0){4.37}}
\put(12.4,0.5){\line(1,0){4.37}}
\put(19.35,3.5){\line(1,0){4.37}}
\put(19.35,0.5){\line(1,0){4.37}}
}

\newsavebox{\blockj}
\savebox{\blockj}(29,1.5)[l]
{\multiput(4,2)(7,0){4}{\circle{4}}
\put(19.35,3.5){\line(1,0){4.37}}
\put(19.35,0.5){\line(1,0){4.37}}
\put(5.3,3.5){\line(1,0){4.37}}
\put(5.3,0.5){\line(1,0){4.37}}
}

\newsavebox{\blockk}
\savebox{\blockk}(29,1.5)[l]
{\multiput(4,2)(7,0){4}{\circle{4}}
\put(5.3,3.5){\line(1,0){4.37}}
\put(5.3,0.5){\line(1,0){4.37}}
\put(12.4,3.5){\line(1,0){4.37}}
\put(12.4,0.5){\line(1,0){4.37}}
}

\newsavebox{\blockl}
\savebox{\blockl}(29,1.5)[l]
{\multiput(4,2)(7,0){4}{\circle{4}}
\put(5.3,3.5){\line(1,0){4.37}}
\put(5.3,0.5){\line(1,0){4.37}}
\put(12.35,3.5){\line(1,0){4.37}}
\put(12.35,0.5){\line(1,0){4.37}}
\put(12.8,2.5){\line(1,0){3.2}}
\put(12.8,1.5){\line(1,0){3.2}}
}

\newsavebox{\blockm}
\savebox{\blockm}(29,1.5)[l]
{\multiput(4,2)(7,0){4}{\circle{4}}
\put(5.3,3.5){\line(1,0){4.37}}
\put(5.3,0.5){\line(1,0){4.37}}
\put(19.9,2){\line(1,0){3.1}}
\put(12.9,2){\line(1,0){3.1}}
}

\newsavebox{\blockn}
\savebox{\blockn}(29,1.5)[l]
{\multiput(4,2)(7,0){4}{\circle{4}}
\put(5.3,3.5){\line(1,0){4.37}}
\put(5.3,0.5){\line(1,0){4.37}}
\put(19.9,2){\line(1,0){3.1}}
\put(12.35,3.5){\line(1,0){4.37}}
\put(12.35,0.5){\line(1,0){4.37}}
}

\newsavebox{\blocko}
\savebox{\blocko}(29,1.5)[l]
{\multiput(4,2)(7,0){4}{\circle{4}}
\put(5.3,3.5){\line(1,0){4.37}}
\put(5.3,0.5){\line(1,0){4.37}}
\put(12.35,3.5){\line(1,0){4.37}}
\put(12.35,0.5){\line(1,0){4.37}}
\put(19.35,3.5){\line(1,0){4.37}}
\put(19.35,0.5){\line(1,0){4.37}}
}

\newsavebox{\blockp}
\savebox{\blockp}(29,1.5)[l]
{\multiput(4,2)(7,0){4}{\circle{4}}
\put(5.3,3.5){\line(1,0){4.37}}
\put(5.3,0.5){\line(1,0){4.37}}
\put(12.9,2){\line(1,0){3.1}}
\put(19.35,3.5){\line(1,0){4.37}}
\put(19.35,0.5){\line(1,0){4.37}}
\put(19.9,2){\line(1,0){3.1}}
}

\newsavebox{\blockq}
\savebox{\blockq}(29,1.5)[l]
{\multiput(4,2)(7,0){4}{\circle{4}}
\put(5.3,3.5){\line(1,0){4.37}}
\put(5.3,0.5){\line(1,0){4.37}}
\put(12.35,3.5){\line(1,0){4.37}}
\put(12.35,0.5){\line(1,0){4.37}}
\put(12.8,2.5){\line(1,0){3.2}}
\put(12.8,1.5){\line(1,0){3.2}}
\put(19.9,2){\line(1,0){3.1}}
}

\newsavebox{\blockr}
\savebox{\blockr}(29,1.5)[l]
{\multiput(4,2)(7,0){4}{\circle{4}}
\put(5.3,3.5){\line(1,0){4.37}}
\put(5.3,0.5){\line(1,0){4.37}}
\put(12.35,3.5){\line(1,0){4.37}}
\put(12.35,0.5){\line(1,0){4.37}}
\put(19.35,3.5){\line(1,0){4.37}}
\put(19.35,0.5){\line(1,0){4.37}}
\put(19.8,2.5){\line(1,0){3.2}}
\put(19.8,1.5){\line(1,0){3.2}}
}

\newsavebox{\blocks}
\savebox{\blocks}(29,1.5)[l]
{\multiput(4,2)(7,0){4}{\circle{4}}
\put(5.3,3.5){\line(1,0){4.37}}
\put(5.3,0.5){\line(1,0){4.37}}
\put(12.35,3.5){\line(1,0){4.37}}
\put(12.35,0.5){\line(1,0){4.37}}
\put(12.8,2.5){\line(1,0){3.2}}
\put(12.8,1.5){\line(1,0){3.2}}
\put(19.35,3.5){\line(1,0){4.37}}
\put(19.35,0.5){\line(1,0){4.37}}
}

\newsavebox{\blockt}
\savebox{\blockt}(29,1.5)[l]
{\multiput(4,2)(7,0){4}{\circle{4}}
\put(5.3,3.5){\line(1,0){4.37}}
\put(5.3,0.5){\line(1,0){4.37}}
\put(12.35,3.5){\line(1,0){4.37}}
\put(12.35,0.5){\line(1,0){4.37}}
\put(12.8,2.5){\line(1,0){3.2}}
\put(12.8,1.5){\line(1,0){3.2}}
\put(19.35,3.5){\line(1,0){4.37}}
\put(19.35,0.5){\line(1,0){4.37}}
\put(19.9,2){\line(1,0){3.1}}
}

\newsavebox{\blocku}
\savebox{\blocku}(29,1.5)[l]
{\multiput(4,2)(7,0){4}{\circle{4}}
\put(5.3,3.5){\line(1,0){4.37}}
\put(5.3,0.5){\line(1,0){4.37}}
\put(12.35,3.5){\line(1,0){4.37}}
\put(12.35,0.5){\line(1,0){4.37}}
\put(12.8,2.5){\line(1,0){3.2}}
\put(12.8,1.5){\line(1,0){3.2}}
\put(19.35,3.5){\line(1,0){4.37}}
\put(19.35,0.5){\line(1,0){4.37}}
\put(19.8,2.5){\line(1,0){3.2}}
\put(19.8,1.5){\line(1,0){3.2}}
}

\newsavebox{\blockv}
\savebox{\blockv}(29,1.5)[l]
{\multiput(4,2)(7,0){4}{\circle{4}}
\put(5.3,3.5){\line(1,0){4.37}}
\put(5.3,0.5){\line(1,0){4.37}}
\put(12.35,3.5){\line(1,0){4.37}}
\put(12.35,0.5){\line(1,0){4.37}}
\put(12.8,2.5){\line(1,0){3.2}}
\put(12.8,1.5){\line(1,0){3.2}}
\put(19.35,3.5){\line(1,0){4.37}}
\put(19.35,0.5){\line(1,0){4.37}}
\put(19.8,2.9){\line(1,0){3.4}}
\put(19.8,1.1){\line(1,0){3.4}}
\put(19.97,1.7){\line(1,0){3.1}}
\put(19.97,2.3){\line(1,0){3.1}}
}

\begin{document}

\begin{titlepage}
\parindent=8cm
\centerline{\sc Ministry of Higher and Special Secondary Education}
\medskip
\centerline{\sc of the USSR}
\bigskip
\centerline{\LARGE Moscow State Lomonosov University}
\bigskip
\hrule
\bigskip
\centerline{\Large Physics Faculty}
\bigskip
\centerline{\large\sf Division of Experimental and Theoretical Physics}
\bigskip


{Manuscript}
\footnote{Translated by the author on October 1995. Some misprints in the
original text are corrected. Available at @xxx.lanl.gov/hep-th/9510140}

{UDK 530.12:531.51}


\bigskip
\medskip
\centerline{\large\bf Ivan Grigoryevich AVRAMIDI}
\bigskip
\bigskip

\centerline{\titlesf Covariant Methods for the Calculation of}
\medskip
\centerline{\titlesf the Effective Action in Quantum Field Theory }
\medskip
\centerline{\titlesf and Investigation of Higher--Derivative}
\medskip
\centerline{\titlesf Quantum Gravity}
\bigskip
\centerline{\large\sf (01.04.02 --- Theoretical and Mathematical Physics)}
\bigskip
\centerline{\large\sf DISSERTATION}
\bigskip
\centerline{\large for the Degree of }
\smallskip
\centerline{\large Candidate of Sciences}
\smallskip
\centerline{\large  in Physics and Mathematics}
\bigskip
\bigskip

{\sc Scientific Supervisor:}

{\sc Professor V. R. Khalilov}

\bigskip
\bigskip
\vfill\vfill
\centerline{Moscow -- 1986}
\bigskip
\end{titlepage}
%
%
%
%
%
%

\tableofcontents
\markboth{\sc Contents}{\sc Contents}
%
%
%
%
%
%
%
%
%
%
%
%
%
%
%

\addcontentsline{toc}{chapter}{Introduction}
\chapter*{Introduction}
\markboth{\sc Introduction}{\sc Introduction}

The classical theory of macroscopic gravitational phenomena, i.e.,
Einstein's General Relativity (GR) [1, 2], cannot be treated as a complete
self--consistent theory in view of a number of serious difficulties that were
not overcome since the creation of GR [3].

This concerns, first of all, the problem of space--time singularities,
which are unavoidable in the solutions of the Einstein equations [1--4]. Close
to these singularities GR becomes incomplete as it cannot predict what is
coming from the singularity. In other words, the causal structure of
space--time breaks down at the singularities [4].

Another serious problem of GR is the problem of the energy of the gravitational
field [5], which was critically analyzed, in particular, in the papers of
Logunov and collaborators [6--8]. 
In the papers [9, 10] a new relativistic theory
of gravitation (RTG) was proposed. In RTG the gravitational field is
described by a spin--2 tensor field on a basic Minkowski space--time.
Such an approach enables one to define in the usual way the energy--momentum
tensor of the gravitational field and to obtain the usual conservation laws.
The curved space--time in this approach is only an effective one that describes
the influence of the gravitational field on all the non--gravitational matter.
Therein the ``identity'' (or geometrization) principle formulated in the
papers [10, 11] is embodied.

The difficulties of the classical theory have motivated the need to
construct a quantum theory of gravitation. Also the
recent progress towards the unification of all non--gravitational interactions
[12] shows the need to include gravitation in a general scheme of an
unified quantum field theory.

The first problem in quantizing gravity is the construction of a covariant
perturbation theory [13--21]. Einstein's theory of gravitation is a typical
non--Abelian gauge theory with the diffeomorphism group as a gauge group [14].
The quantization of gauge theories faces the known difficulty connected
with the presence of constraints [22, 23]. This problem was successfully
solved in the works of Feynman [13], De~Witt [14] and Faddeev and Popov [15].

The most fruitful approach in quantum gravity is the background field method
of De~Witt [14, 26--47]. This method is a generalization of the method of
generating functionals in quantum field theory [48--50] to the case of
non--vanishing background field. Both the gravitational and the matter fields
can have the background classical part.

The basic object in the background field method is the effective action
functional. The effective action encodes, in principle, all the information
of the standard quantum field theory. It determines the elements of the
diagrammatic technique of perturbation theory, i.e., the full one--point
propagator and the full vertex functions, with regard to all quantum
corrections, and, hence, the perturbative $S$--matrix [14, 33, 45].
On the other hand, the effective action gives at once the physical amplitudes
in real external classical fields and describes all quantum effects in
external fields [27, 28] (the vacuum polarization of quantized fields,
particle creation etc.) [53--58]. The effective action functional is the most
appropriate tool for investigating 
the structure of the physical vacuum in various
models of quantum field theory with spontaneous symmetry breaking
(Higgs vacuum, gluon condensation, superconductivity) [24, 25, 51, 52].

The effective action makes it possible  
to take into account the back--reaction of the
quantum processes on the classical background, i.e., to obtain the effective
equations for the background fields [33, 34, 45, 46, 59--61]. In this way,
however, one runs into the difficulty connected with the dependence of the
off--shell effective action on the gauge and the parametrization of the
quantum field. In the paper [34] a gauge--invariant effective action (which
still depends parametrically on the gauge fixing and the parametrization) was
constructed. The explicitly reparametrization--invariant functional that does
not depend on the gauge--fixing (``unique'' effective action) was constructed
in the papers [45, 46]. The ``unique'' effective action was studied in the
paper [47] in different models of quantum 
field theory (including Einstein
gravity) and in the paper of the author and Barvinsky [173] in case of
higher--derivative quantum gravity.

Thus, the calculation of the effective action is of high interest
from the point of view of the general formalism 
as well as for concrete applications.
The only practical method for the calculation of the effective action is the
perturbative expansion 
in the number of loops [48--50]. All the fields are split
in a background classical part and quantum perturbations propagating on this
background. The part of the classical action, which is quadratic in quantum
fields, determines the propagators of the quantum fields in external fields,
and higher--order terms reproduce the vertices of perturbation theory
[33].

At one--loop level, the contribution of the gravitational loop is of the
same order as the contributions of matter fields [62, 63]. At usual energies
much lower than the Planck energy ($E\ll \hbar c^5/G\approx 10^{19} GeV$) the
contributions of additional gravitational loops are highly suppressed.
Therefore, a semi--classical concept applies when the quantum matter fields
together with the linearized perturbations of the gravitational field interact
with the external background gravitational field (and, probably, with the
background matter fields) [53--56, 62, 63]. This approximation is known as 
one--loop quantum gravity [33, 64--66].

To evaluate the effective action it is necessary to find, first of all,
the Green functions of the quantum fields in the external classical
fields of different nature. The Green functions in external fields were
investigated by a number of authors. Fock [67] proposed a method for solving
the wave equations in external electromagnetic fields by an integral transform
in the proper time parameter (``fifth'' parameter). Schwinger [68, 69]
generalized the proper--time method and applied it to the calculation of the
one--loop effective action. De~Witt [26, 28] reformulated the proper--time
method in geometrical language and applied it to the case of external
gravitational field. Analogous questions for the elliptic operators were
investigated by mathematicians [70--76]. In the papers [77, 78] the
standard Schwinger--\-De~Witt technique was generalized to the case of
arbitrary differential operators satisfying the condition of causality.

The proper--time method gives at once the Green functions 
in the neighbourhood of the light cone. 
Therefore, it is the most suitable tool for investigation of the ultraviolet
divergences (calculation of counter--terms, $\beta$--functions and
anomalies). The most essential advantage of the proper--time method is that
it is explicitly covariant and enables one to introduce various covariant
regularizations of divergent integrals. The most popular are the analytic
regularizations: dimensional regularization, 
$\zeta$--function regularization 
etc. [53, 54]. There are a lot of works
along this line of investigation 
over the last years [79--106]. Although most of the
papers restrict themselves to the one--loop approximation, the proper--time
method is applicable at higher loops too. In the papers [37, 39, 40, 43,
44] it was applied to analyze two--loop divergences in various models
of quantum field theory including Einstein's quantum gravity.

Another important area, where the Schwinger--De~Witt pro\-per\---time me\-thod is
successfully applied, is the vacuum polarization of massive
quantum fields by external background fields. When the
Compton wave length $\lambda=\hbar/mc$, corresponding to the field mass
$m$, is much smaller than the characteristic scale $L$ of the background
field change, the proper--time method gives immediately the expansion of
the effective action in a series in the small parameter $(\lambda/L)^2$
[57, 58, 107]. The coefficients of this expansion are proportional to
the so--called De~Witt coefficients and are local invariants, constructed
from the external fields and their covariant derivatives. In the papers
[98, 45] the general structure of the Schwinger--De~Witt expansion of the
effective action was discussed. It was pointed out that there is a need
to go beyond the limits of the local expansion by the summation of the
leading derivatives of the external fields in this expansion. In the paper
[45], based on some additional assumptions concerning the convergence of
the corresponding series and integrals, the leading derivatives of the
external fields were summed up and a non--local expression for the one--loop
effective action in case of massless field was obtained.

Thus, so far, effective and explicitly covariant methods for calculation of
the effective action in arbitrary background fields are absent. All the
calculations performed so far 
concern either the local structures of the
effective action or some model specific background fields (constant fields,
homogenous spaces etc.) [53, 54].

That is why the development of general methods for covariant
calculations of the effective action, 
which is especially needed in the quantum
theory of gauge fields and gravity, is an actual and new area of research.
The papers of the author [171, 172] are devoted to the development of this
line of investigation. Therein an explicitly covariant and effective
technique for the calculation of De~Witt coefficients is elaborated. This
technique is applicable in the most general case of arbitrary external
fields and spaces and can be easily adopted to analytic calculations on
computers. In these papers [171, 172] the renormalized one--loop effective
action for massive scalar, spinor and vector fields in external
gravitational field up to terms of order $O(1/m^4)$ is calculated.

In spite of impressive progress in one--loop quantum gravity, a complete
self--consistent quantum theory of gravitation does not exist at present 
[108].
The difficulties of quantum gravity are connected, in the first line, with
the fact that there is not any consistent way to eliminate the ultraviolet
divergences arising in perturbation theory [109, 110]. It was found
[111--117] that in the one--loop approximation the pure Einstein gravity is
finite on mass--shell (or renormalizable in case of non--vanishing cosmological
constant). However, two--loop Einstein gravity is no longer
renormalizable on shell [118].
On the other hand, the interaction with the matter fields also leads to
non--renormalizability on mass--shell even in one--loop approximation
[119--128].

Among various approaches to the problem of ultraviolet divergences in
quantum gravity (supergravity [129--131], 
resummation [132--134] etc. [109, 110]) 
an important place is occupied by the modification of the
gravitational Lagrangian by adding quadratic terms in the curvature of
general type (higher--derivative theory of gravitation). This theory was
investigated by various authors both at the classical and at the quantum
levels [135--161].

The main argument against higher--derivative quantum gravity is the
presence of ghosts in the linearized perturbation theory on flat background,
that breaks down the unitarity of the theory [140--151]. There were
different attempts to solve this problem by the summation of radiative
corrections in the propagator in the momentum representation [146--149,
153--155]. However, at present they cannot be regarded as convincing
in view of causality violation, which results from
the unusual analytic properties of the $S$--matrix.
It seems likely that the problem of unitarity can be solved only beyond
the limits of perturbation theory [158].

Ultraviolet behavior of higher--derivative quantum gravity was studied in
the papers [140--157]. However, the one--loop counterterms were first
obtained in the paper of Julve and Tonin [145]. The most detailed
investigation of the ultraviolet behaviour of higher--derivative quantum
gravity was carried out in the papers of Fradkin and Tseytlin [153--156].
In these papers, an inconsistency was found in the calculations of Julve
and Tonin. The one--loop counterterms were recalculated 
in higher--derivative quantum gravity of general type as well as in
conformally invariant  
models and in conformal supergravity [155, 156].
The main conclusion of the papers [153--156] is that higher--derivative
quantum gravity is asymptotically free in the ``physical'' region of
coupling constants, which is characterized by the absence of tachyons on
the flat background. The presence of reasonably arbitrary matter does not
affect this conclusion.

Thus, the investigation of the ultraviolet behaviour of 
higher--derivative quantum gravity is an important and actual problem in
the general program of constructing a consistent quantum gravity.
This problem is just the one that the author was concern with in the papers
[173, 174]. Therein the off--shell one--loop divergences of higher--derivative
quantum gravity in arbitrary covariant gauge of the quantum field were
calculated. It was shown that the results of previous authors contain a
numerical error in the coefficient 
of the $R^2$--divergent term. The correction of
this mistake radically changes the asymptotic properties of the theory in the
conformal sector. Although the conclusion of [153--156] about the asymptotic
freedom in the tensor sector of the theory remains true, the conformal sector
exhibits just opposite ``zero--charge'' behavior in the ``physical'' region of
coupling constants considered in all previous papers [140--155]. In the
``unphysical'' region of coupling constants, which corresponds to the positive
definiteness of the part of the Euclidean action quadratic in curvature, the
``zero--charge'' singularities at finite energies are absent.

This dissertation is devoted to further development of the covariant methods
for calculation of the effective action in quantum field theory and quantum
gravity, and to the investigation of the ultraviolet behaviour of
higher--\-de\-ri\-va\-ti\-ve quantum gravity.

In Chap. 1. the background field method is presented. Sect. 1.1
contains a short functional formulation of quantum field theory in
the form that is convenient for subsequent discussion. In Sect. 1.2 the
standard proper--time method with some extensions is presented in detail.
Sect. 1.3 is concerned with the questions connected with the problem of
ultraviolet divergences, regularization, renormalization and the
renormalization group.

In Chap. 2 the explicitly covariant technique for the calculation of the
De~Witt coefficients is elaborated. In Sect. 2.1 the methods of covariant
expansions of arbitrary fields in curved space with arbitrary linear connection
in the generalized covariant Taylor series and the Fourier integral are
formulated in the most general form. In Sect. 2.2 all the quantities that will
be needed later are calculated in form of covariant Taylor series. In Sect.
2.3, based on the method of covariant expansions, the covariant technique for
the calculation of the De~Witt coefficients in matrix terms is developed. The
corresponding diagrammatic formulation of this technique is given. The
technique developed leads to the explicit
calculation of De~Witt coefficients 
as well as to the analysis of their general
structure. The possibility to use the corresponding analytic manipulations
on computers is pointed out.
In Sect. 2.4 the calculation of the De~Witt coefficients $a_3$ and
$a_4$ at coinciding points is presented. In Sect. 2.5. the one--loop effective
action for massive scalar, spinor and vector fields in an external
gravitational field is calculated up to terms of order $1/m^4$. 

In Chap. 3 the general structure of the Schwinger--De~Witt asymptotic expansion
is analyzed and the partial summation of various terms is carried out. In Sect.
3.1 the method for summation of the asymptotic series due to Borel (see, e.g., [167], 
sect. 11.4) is presented and its application to quantum field theory is
discussed. In Sect. 3.2. the covariant methods for investigations of the
non--local structure of the effective action are developed. In Sect. 3.3 the
terms of first order in the external 
field in De~Witt coefficients are calculated
and their summation is carried out. The non--local expression for the Green
function at coinciding points, up to terms 
of second order in external fields, is
obtained. The massless case is considered too. It is shown that in the
conformally invariant case the Green 
function at coinciding points is finite at
first order in external fields. In Sect. 3.4. the De~Witt coefficients at
second order in external fields are calculated. The summation of the
terms quadratic in external fields 
is carried out, and the explicitly covariant non--local expression for the
one--loop effective action up to terms of third order in external fields is
obtained. All the formfactors, their ultraviolet asymptotics and imaginary
parts in the pseudo--Euclidean region above the threshold are obtained
explicitly. The massless case in four-- and two--dimensional spaces is considered
too. In Sect. 3.5 all terms without covariant derivatives of the
external fields in De~Witt coefficients, 
in the case of scalar field, are picked 
out. It is shown that in this case the asymptotic series of the covariantly
constant terms diverges. By making use of the Borel summation procedure of the
asymptotic series, the Borel sum of the 
corresponding semi--classical series is
calculated. An explicit expression for the
one--loop effective action, non--analytic in the background fields,
is obtained up to the terms with covariant derivatives of the
external fields. 

Chap. 4 is devoted to the investigation of higher--derivative quantum
gravity. In Sect. 4.1  the standard procedure of quantizing the gauge theories
as well as the formulation of the unique effective action is presented. In
Sect. 4.2 the one--loop divergences of higher--derivative quantum gravity with
the help of the methods of the generalized Schwinger--De~Witt technique are
calculated. The error in the coefficient of the $R^2$--divergent term, due to
previous authors, is pointed out. In Sect. 4.3 the dependence of the
divergences of the effective action on the gauge of the quantum field is
analyzed. The off--shell divergences of the standard effective action in
arbitrary covariant gauge, 
and the divergences of the unique effective action,
are calculated. In Sect. 4.4 the corresponding renormalization--group equations
are solved and the ultraviolet asymptotics of the coupling constants are
obtained. It is shown that in the conformal sector of the theory there is no
asymptotic freedom in the ``physical'' region
of the coupling constants. The presence of the low--spin matter fields does not
change this general conclusion: higher--derivative quantum gravity
necessarily goes beyond the limits of the weak conformal coupling at high
energies. The physical interpretation of such ultraviolet behaviour is
discussed. It is shown that the asymptotic freedom both in tensor and conformal
sectors  is realized in the ``unphysical'' region of coupling constants, which
corresponds to the positive--definite Euclidean action. In Sect. 4.5 the
effective potential (i.e., the effective action on the de Sitter background) in
higher--derivative quantum gravity is calculated. The determinants of the
second-- and fourth--order operators are calculated with the help of the
technique of the generalized $\zeta$--function. The correctness of the result
for the $R^2$--divergence obtained in Sect. 4.2, 
as well as the correctness of the
results for the arbitrary gauge and for the unique effective action obtained in
Sect. 4.3, are maintained. Both the effective
potential in arbitrary gauge and the unique effective potential are
calculated. The unique effective equations 
for the background field, i.e., for the
curvature of de Sitter space, that do not depend on the gauge and the
parametrization of the quantum field, are obtained. The first quantum correction
to the background curvature caused by the quantum effects is found.

In the Conclusion the main results obtained in the dissertation are 
summarized. 

In the dissertation we use the unit system $\hbar=c=G=1$ (except for some expressions
where these quantities are contained explicitly) and the notation and the sign
conventions of the book [1].

%
%
%
%
%
%
%
%
%
%
%
%
%
%
%
%

\chapter{Background field method in quantum field theory}
\markboth{\sc Chapter 1. Background field method}{\sc Chapter 1. Background
field method}
%
%
%
%
%
%
%
%
%
\section{Generating functional, Green functions
and the effective action}

Let us consider an arbitrary field $\varphi(x)$ on a $n$--dimensional space--time 
given by its contravariant components $\varphi^A(x)$ that transform
with respect to some (in general, reducible) representation of the 
group of general transformations of the coordinates. 
The field components $\varphi^A(x)$ can be
of both bosonic and fermionic nature. The fermion components are treated as
anticommuting Grassmanian variables [162], i.e., 
$$
\varphi^A\varphi^B=(-1)^{AB}\varphi^B\varphi^A,
\eqno(1.1)
$$
where the indices in the exponent of the $(-1)$ are equal to $0$ for bosonic
indices and to $1$ for the fermionic ones.

For the construction of a local action functional $S(\varphi)$ one also needs a
metric of the configuration space $E_{AB}$, i.e., a scalar product
$$
(\varphi_1,\varphi_2)=\varphi_1^AE_{AB}\varphi_2^B,
\eqno(1.2)
$$
that enables one to define the covariant fields components
$$
\varphi_A=\varphi^BE_{BA}, \qquad \varphi^B=\varphi_AE^{-1\ AB},
\eqno(1.3)
$$
where $E^{-1\ AB}$ is the inverse matrix
$$
E^{-1\ AB}E_{BC}=\delta^A_{\ C}, \qquad E_{AC}E^{-1\ CB}=\delta^{\ B}_{A}.
\eqno(1.4)
$$
The metric $E_{AB}$ must be nondegenerate both in bose--bose and fermi--fermi
sectors and satisfy the symmetry conditions
$$
E_{AB}=(-1)^{A+B+AB}E_{BA}, \qquad E^{-1\ AB}=(-1)^{AB}E^{-1\ BA}.
\eqno(1.5)
$$
 In the case of gauge--invariant field theories we assume that the corresponding
ghost fields are included in the set of the fields $\varphi^A$ and the action
$S(\varphi)$ is modified by inclusion of the gauge fixing and the ghost terms.
To reduce the writing we will follow, hereafter, the condensed notation of
De~Witt [26, 33] and substitute the mixed set of indices $(A,x)$, where $x$
labels the space--time point, by one small Latin index $i\equiv(A,x)$:
$\varphi^i\equiv\varphi^A(x)$. The combined summation--\-integration should be
done over the repeated upper and lower small Latin indices
$$
\varphi_{1\,i}\varphi^i_2\equiv\int d^n x \varphi_{1\,A}(x)\varphi_2^A(x).
\eqno(1.6)
$$

Now let us single out in the space--time two causally connected in-- and out--
regions, that lie in the past and in the future respectively relative to the
region, which is of interest from the dynamical standpoint. Let us define the
vacuum states $|$in,vac$>$ and $|$out,vac$>$ in these regions and consider the
vacuum--vacuum transition amplitude
$$
\left<{\rm out, vac|in, vac}\right>\equiv \exp\left\{{i\over \hbar}W(J)\right\}
\eqno(1.7)
$$
in presence of some external classical sources $J_i$ vanishing in in-- and out--
regions.

The amplitude (1.7) can be expressed in form of a formal functional integral
[48--50]
$$
\exp\left\{{i\over \hbar}W(J)\right\}=\int d \varphi {\cal
M}(\varphi)\exp\left\{{i\over \hbar}\left[S(\varphi)+J_i\varphi^i
\right]\right\},
\eqno(1.8)
$$
where ${\cal M}(\varphi)$ is a measure functional, which should be determined
by the canonical quantization of the theory [20, 45]. The integration in
(1.8) should be taken over all fields sa\-tis\-fying the boundary conditions
determined by the vacuum states $|{\rm in,vac}>$ and $|{\rm out,vac}>$. The
functional $W(J)$ is of central interest in quantum field theory. It is the
generating functional for the Schwinger averages
$$
\left<\varphi^{i_1}\cdots\varphi^{i_k}\right>
=\exp\left\{-{i\over \hbar}W(J)\right\}
\left({\hbar\over i}\right)^k
{\delta_L^k\over \delta J_{i_1}\cdots \delta J_{i_k}}\exp\left\{{i\over
\hbar}W(J)\right\},
\eqno(1.9)
$$
where
$$
\left<F(\varphi)\right>\equiv
{\left<{\rm out, vac|T }\left(F(\varphi)\right)|{\rm in, vac}\right>
\over \left<{\rm out, vac|in, vac}\right>},
\eqno(1.10)
$$
$\delta_L$ is the left functional derivative and `T' is the operator of
chronological ordering.

The first derivative of the functional $W(J)$ gives the mean field (according
to the tradition we will call it the background field)
$$
\left<\varphi^i\right>\equiv\Phi^i(J)={\delta_L \over \delta J_i} W(J),
\eqno(1.11)
$$
the second derivative determines the one--point propagator
$$
\left<\varphi^i\varphi^k\right>=\Phi^i\Phi^k + {\hbar\over i}{\cal G}^{ik},
\eqno(1.12)
$$
$$
{\cal G}^{ik}(J)={\delta_L^2 \over \delta J_i \delta J_k}W(J),
$$
and the higher derivatives give the many--point Green functions
$$
{\cal G}^{i_1\cdots i_k}(J)={\delta_L^k \over \delta J_{i_1}\cdots \delta
J_{i_k}}W(J).
\eqno(1.13)
$$

The generating functional for the vertex functions, the effective action
$\Gamma(\Phi)$, is defined by the functional Legendre transform:
$$
\Gamma(\Phi)=W(J)-J_i\Phi^i,
\eqno(1.14)
$$
where the sources are expressed in terms of the background fields, $J=J(\Phi)$,
by inversion of the functional equation $\Phi=\Phi(J)$, (1.11).

The first derivative of the effective action gives the sources
$$
{\delta_R \over \delta \Phi^i}\Gamma(\Phi)\equiv \Gamma_{,i}(\Phi)=-J_i(\Phi),
\eqno(1.15)
$$
the second derivative determines the one--point propagator
$$
{\delta_L\delta_R \over \delta \Phi^i \delta \Phi^k}\Gamma(\Phi)\equiv {\cal
D}_{ik}(\Phi), \qquad {\cal D}_{ik}=(-1)^{i+k+ik}{\cal D}_{ki},
\eqno(1.16)
$$
$$
{\cal D}_{ik}{\cal G}^{kn}=-\delta_i^{\ n},
$$
where $\delta_R$ is the right functional derivative, $\delta_i^{\
n}=\delta_A^{\ B}\delta(x,x')$, and $\delta(x,x')$ is the delta--function. The
higher derivatives determine the vertex functions
$$
\Gamma_{i_1\cdots i_k}(\Phi)
={\delta_R^k \over \delta \Phi^{i_1}\cdots \delta \Phi^{i_k}}\Gamma(\Phi).
\eqno(1.17)
$$

{}From the definition (1.14) and the equation (1.8) it is easy to obtain the
functional equation for the effective action
$$
\exp\left\{{i\over \hbar}\Gamma(\Phi)\right\}=
\int d\varphi {\cal M}(\varphi)\exp\left\{{i\over \hbar}
\left[S(\varphi)-\Gamma_{,i}(\Phi)(\varphi^i-\Phi^i)\right]\right\}.
\eqno(1.18)
$$
Differentiating the equation (1.15) with respect to the sources one can express
all the many--point Green functions (1.13) in terms of the vertex functions
(1.17) and the one--point propagator (1.12). If one uses the diagrammatic
technique, where the propagator is represented by a line and the vertex
functions by vertexes, then each differentiation with respect to the sources
adds a new line in previous diagrams by all possible ways. Therefore, a
many--point Green function is represented by all kinds of tree diagrams with a
given number of external lines.

Thus when using the effective action functional for the construction of the
$S$--matrix (when it exists) one needs only the tree diagrams, since all quantum
corrections determined by the loops are already included in the full one--point
propagator and the full vertex functions. Therefore, the effective equations (1.15),
$$
\Gamma_{,i}(\Phi)=0,
\eqno(1.19)
$$
(in absence of classical sources, $J=0$) describe the dynamics of the background fields
with regard to all quantum corrections.

The possibility to work directly with the effective action is an obvious
advantage. First, the effective action contains all the information needed to 
construct the standard $S$--matrix [14, 29, 45]. Second, it gives the
effective equations (1.19) that enable one to take into account the influence of
the quantum effects on the classical configurations of the background fields                                                                                                                                                                                                                                                                                                                 [59, 60].

In practice, the following difficulty appear on this way. The background
fields, as well as all other Green functions, are not uniquely defined objects.
They depend on the parametrization of the quantum field [45, 46]. Accordingly,
the effective action is not unique too. It depends essentially on the
parametrization of the quantum field off mass--shell, i.e., for background fields
that do not satisfy the classical equations of motion
$$
S_{,i}(\Phi)=0.
\eqno(1.20)
$$
On mass--shell, (1.20), the effective action is a well defined quantity and leads to
the correct $S$--matrix [14, 29, 30, 49].

A possible way to solve this difficulty was recently proposed in the papers
[45, 46], where an effective action functional was constructed, that is
explicitly invariant with respect to local reparametrizations of quantum fields
(``unique'' effective action). This was done by introducing a metric and a
connection in the configuration space. Therein, [45, 46], the unique effective
action for the gauge field theories was constructed too. We will consider the
consequences of such a definition of the effective action in Chap. 5 when
investigating the higher--derivative quantum gravity.

The formal scheme of quantum field theory, described above, begins to take on a
concrete meaning in the framework of perturbation theory in the number of loops
[48--50] (i.e., in the Planck constant $\hbar$):
$$
\Gamma(\Phi)=S(\Phi)+\sum\limits_{k\ge 1} \hbar^{k-1}\Gamma_{(k)}(\Phi).
\eqno(1.21)
$$
Substituting the expansion (1.21) in (1.18), shifting the integration variable
in the functional integral $\varphi^i=\Phi^i+\sqrt\hbar h^i$, expanding the
action $S(\varphi)$ and the measure ${\cal M}(\varphi)$ in quantum fields $h^i$
and equating the coefficients at equal powers of $\hbar$, we get the recurrence
relations that uniquely define all the coefficients $\Gamma_{(k)}$. All the
functional  integrals are Gaussian and can be calculated in the standard way
[24]. As the result the diagrammatic technique for the effective action is
reproduced. The elements of this technique are the bare one--point propagator,
i.e., the Green function of the differential operator
$$
\Delta_{ik}(\varphi)={\delta_L\delta_R\over\delta\varphi^i\delta\varphi^k}
S(\varphi),
\eqno(1.22)
$$
and the local vertexes, determined by the classical action $S(\varphi)$ and the
measure ${\cal M}(\varphi)$.

In particular, the one--loop effective action has the form
$$
\Gamma_{(1)}(\Phi)=-{1\over 2i}\log{{\rm sdet}\,\Delta(\Phi)\over
{\cal M}^2(\Phi)},
\eqno(1.23)
$$
where
$$
{\rm sdet}\,\Delta=\exp\left({\rm str}\,\log \Delta\right)
\eqno(1.24)
$$
is the functional Berezin superdeterminant [162], and
$$
{\rm str}\,F=(-1^i)F^i_{\ i}=\int d^n x(-1)^AF^A_{\ A}(x)
\eqno(1.25)
$$
is the functional supertrace.

The local functional measure ${\cal M}(\varphi)$ can be taken in the form of the
superdeterminant of the metric of the configuration space
$$
{\cal M}=\left({\rm sdet}\,E_{ik}(\varphi)\right)^{1/2},
\eqno(1.26)
$$
where
$$
E_{ik}(\varphi)=E_{AB}(\varphi(x))\delta(x,x').
\eqno(1.27)
$$

In this case $d \varphi {\cal M}(\varphi)$ is the volume element of the
configuration space that is invariant under the point transformations of the
fields: $\varphi(x)\to F(\varphi(x))$. Using the multiplicativity of the
superdeterminant [162], the one--loop effective action with the measure (1.26)
can be rewritten in the form
$$
\Gamma_{(1)}(\Phi)=-{1\over 2i}\log{\rm sdet}\,\hat\Delta,
\eqno(1.28)
$$
with
$$
{\rm sdet}\,\hat\Delta^i_{\ k}=E^{-1\ in}\Delta_{nk}.
\eqno(1.29)
$$
The local measure ${\cal M}(\varphi)$ can be also chosen in such a way, that the leading
ultraviolet divergences in the theory, proportional to the delta--function in
coinciding points $\delta(0)$, vanish [18, 20].

\section{Green functions of minimal differential 
operators}

The construction of Green functions of arbitrary differential operators (1.22),
(1.29) can be reduced finally to the construction of the Green functions of the
``minimal'' differential operators of second order [78] that have the form
$$
\Delta^i_{\ k}=\left\{\delta^A_{\ B}(\Square-m^2)+Q^A_{\
B}(x)\right\}g^{1/2}(x)\delta(x,x'),
\eqno(1.30)
$$
where $\Square=g^{\mu\nu}\nabla_\mu\nabla_\nu$ is the covariant d'Alambert
operator, $\nabla_\mu$ is the covariant derivative, defined by means of some
background connection ${\cal A}_\mu(x)$,
$$
\nabla_\mu\varphi^A=\partial_\mu\varphi^A+{\cal A}^A_{\ B\mu}\varphi^B,
\eqno(1.31)
$$
$g^{\mu\nu}(x)$ is the metric of the background space--time, $g(x)=-\det
g_{\mu\nu}(x)$, $m$ is the mass parameter of the quantum field and  $Q^A_{\
B}(x)$ is an arbitrary matrix.

The Green functions $G^A_{\ B'}(x,x')$ of the differential operator (1.30) are
two--point objects, which transform as the field $\varphi^A(x)$ under the
transformations of coordinates at the point $x$, and as the current
$J_{B'}(x')$ under the coordinate transformations at the point $x'$. The
indexes, belonging to the tangent space at the point $x'$, are labelled with a
prime.

We will construct solutions of the equation for the Green functions
$$
\left\{\delta^A_{\ C}(\Square-m^2)+Q^A_{\ C}\right\}G^C_{\ B'}(x,x')
=-\delta^A_{\ B}g^{-1/2}(x)\delta(x,x'),
\eqno(1.32)
$$
with appropriate boundary conditions, by means of the Fock--Schwinger--De~Witt
proper time method [67--69, 26, 28] in form of a contour integral over an
auxiliary variable $s$,
$$
G=\int\limits_C i\, d s\, \exp(-ism^2) U(s),
\eqno(1.33)
$$
where the ``evolution function'' $U(s)\equiv U^A_{\ B'}(s|x,x')$ satisfies the
equation
$$
{\partial\over \partial i s} U(s)=\left(\hat 1\Box+Q\right)U(s),\qquad
\hat 1\equiv \delta^A_{\ B},
\eqno(1.34)
$$
with the boundary condition
$$
 U^A_{\ B'}(s|x,x')\Big\vert_{\partial C}=-\delta^A_{\
B}g^{-1/2}(x)\delta(x,x'),
 \eqno(1.35)
 $$
where $\partial C$ is the boundary of the contour $C$.

The evolution equation (1.34) is as difficult to solve exactly as the initial
equation (1.32). However, the representation of the Green functions in form of
the contour integrals over the proper time, (1.33), is more convenient to use for the
construction of the
asymptotic expansion of the Green functions in inverse powers of the mass and
for the study of the behavior of the Green functions and their derivatives on
the light cone, $x\to x'$, as well as for the regularization and
renormalization of the divergent vacuum expectation values of local variables
(such as the energy--momentum tensor, one--loop effective action etc.).

Deforming the contour of integration, $C$, over $s$ in (1.33) we can get
different Green functions for the same evolution function. To obtain
the causal Green function (Feynman propagator) one has to integrate over $s$
from $0$ to $\infty$ and add an infinitesimal negative imaginary part to the $m^2$ [26,
28]. It is this contour that we mean hereafter.

Let us single out in the evolution function a rapidly oscillating factor that
reproduces the initial condition (1.35) at $s\to 0$:
$$
U(s)=i(4\pi s)^{-n/2}\Delta^{1/2}
\exp\left(-{\sigma\over 2is}\right){\cal P}\Omega (s),
\eqno(1.36)
$$
where $\sigma(x,x')$ is half the square of the geodesic distance between the
points $x$ and $x'$,
$$
\Delta(x,x')=-g^{-1/2}(x)
\det\left(-\nabla_{\mu'}\nabla_\nu\sigma(x,x')\right)g^{-1/2}(x')
\eqno(1.37)
$$
is the Van~Fleck--Morette determinant, ${\cal P}\equiv {\cal P}^A_{\ B'}(x,x')$
is the parallel displacement operator of the field along the geodesic from the
point $x'$ to the point $x$. We assume that there exists 
the only geodesic connecting the
points $x$ and $x'$, the points $x$ and $x'$ being not conjugate, and
suppose the two--point functions $\sigma(x,x')$, $\Delta(x,x')$ and $ {\cal
P}^A_{\ B'}(x,x')$ to be single--valued differentiable functions of the
coordinates of the points $x$ and $x'$. When the points $x$ and $x'$ are close
enough to each other this will be always the case [2, 26, 28].

The introduced ``transfer function'' $\Omega(s)\equiv\Omega^{A'}_{\
B'}(s|x,x')$ transforms as a scalar at the point $x$ and as a matrix at the
point $x'$ (both its indexes are primed). This function is regular in $s$ at
the point $s=0$, i.e.,
$$
\Omega^{A'}_{\ B'}(0|x,x')\Big\vert_{x\to x'}=\delta^{A'}_{\ B'}
\eqno(1.38)
$$
independently on the way how $x\to x'$.

If one assumes that there are no boundary surfaces in space--time (that we
will do hereafter), then the transfer function is analytic also in $x$ close to
the point $x=x'$ for any $s$, i.e., there exist finite coincidence limits of the
transfer function and its derivatives at $x=x'$ that do not depend on the way
how $x$ approaches $x'$.

Using the equations for the introduced functions [26, 28],
$$
\sigma=\frac{1}{2}\sigma_\mu\sigma^\mu, \qquad
\sigma_\mu\equiv\nabla_\mu\sigma,
\eqno(1.39)
$$
$$
\sigma^\mu\nabla_\mu{\cal P}=0, \qquad
{\cal P}^A_{\ B'}(x',x')=\delta^{A'}_{\ B'},
\eqno(1.40)
$$
$$
\sigma^\mu\nabla_\mu\log\,\Delta^{1/2}=\frac{1}{2}(n-\Square\sigma),
\eqno(1.41)
$$
we obtain from (1.34) and (1.36) the transfer equation for the function
$\Omega(s)$:
$$
\left({\partial\over\partial is }
+{1\over is}\sigma^\mu\nabla_\mu\right)
\Omega(s)={\cal P}^{-1}
\left(\hat 1\Delta^{-1/2}\Square\Delta^{1/2}+Q\right){\cal P}\Omega(s).
\eqno(1.42)
$$

If one solves the transfer equation (1.42) in form of a power series in the
variable $s$
$$
\Omega(s)=\sum\limits_{k\ge 0}{(is)^k\over k!}a_k,
\eqno(1.43)
$$
then from (1.38) and (1.42) one gets the recurrence relations for the De~Witt
coefficients $a_k$
$$
\sigma^\mu\nabla_\mu a_0=0, \qquad {a_0}^{A'}_{\ B'}(x',x')=\delta^{A'}_{\ B'},
\eqno(1.44)
$$
$$
\left(1+{1\over k}\sigma^\mu\nabla_\mu\right)a_k={\cal P}^{-1}
\left(\hat 1\Delta^{-1/2}\Square\Delta^{1/2}+Q\right){\cal P}a_{k-1}.
\eqno(1.45)
$$

\noindent
{}From the equations (1.44) it is easy to find the zeroth coefficient
$$
{a_0}^{A'}_{\ B'}(x,x')=\delta^{A'}_{\ B'}.
\eqno(1.46)
$$
The other coefficients are calculated usually by differentiating the relations
(1.45) and taking the coincidence limits [26, 28]. However such method of
calculations is very cumbersome and non--effective. In this way only the
coefficients $a_1$ and $a_2$ at
coinciding points were calculated [26, 83, 84].
The same coefficients as well as the coefficient $a_3$ at coinciding points
were calculated
in the paper [76] by means of a completely different noncovariant method. Thus,
up to now a manifestly covariant method for calculation of the De~Witt
coefficients is absent.

In the Chap. 2 we develop a manifestly covariant and very  convenient and
effective calculational technique that enables  to calculate explicitly
arbitrary De~Witt coefficients $a_k$ as well as to analyze their general
structure. Moreover, the elaborated technique is very algorithmic and can be
easily realized on computers.

Let us stress that the expansion (1.43) is asymptotic and does not reflect
possible nontrivial analytical properties of the transfer function, which are
very important when doing the contour integration in (1.33). The expansion in
the power series in proper time, (1.43), corresponds physically to the expansion
in the dimensionless parameter that is equal to the ratio of the Compton wave
length, $\lambda=\hbar/mc$, to the characteristic scale of variation of the
external fields, $L$. That means that it corresponds to the expansion in the
Planck constant $\hbar$ in usual units [57, 58]. This is the usual
semi--classical approximation of quantum mechanics.
This approximation is good enough for the study of the light cone singularities
of the Green functions, for the regularization and renormalization of the
divergent coincidence limits of the Green functions and their derivatives in a
space--time point, as well as for the calculation of the vacuum polarization of
the massive fields in the case when the Compton wave length $\lambda$ is much
smaller than the characteristic length scale $L$, \  $\lambda/L=\hbar/(mcL)\ll 1$.

At the same time the expansion in powers of the proper time (1.43) does not
contain any information about all effects that depend non--analytically on the
Planck constant $\hbar$ (such as the particle creation and the vacuum
polarization of massless fields) [28, 58]. Such effects can be described only
by summation of the asymptotic expansion (1.43). The exact summation in general
case is, obviously, impossible. One can, however, pick up the leading terms in
some approximation and sum them up in the first line. Such partial summation of
the asymptotic (in general, divergent) series is possible only by employing
additional physical assumptions about the analytical structure of the exact
expression and corresponding analytical continuation.

In the Chap. 3 we will carry out the partial summation of the terms that are
linear and quadratic in external fields as well as the terms without the
covariant derivatives of the external fields.

\section{Divergences, regularization, re\-nor\-ma\-li\-za\-tion
 and the re\-nor\-ma\-li\-za\-tion group}

A well known problem of quantum field theory is the presence of the ultraviolet
divergences that appear by practical calculations in perturbation theory. They
are exhibited by the divergence of many integrals (over coordinates and
momentums) because of the singular behavior of the Green functions at small
distances. The Green functions are, generally speaking, distributions, i.e.,
linear functionals defined on smooth finite functions [48, 49]. Therefore,
numerous products of Green functions appeared in perturbation theory cannot be
defined correctly.

A consistent scheme for eliminating the ultraviolet divergences and obtaining
finite results is the theory of renormalizations [48, 49], that can be carried
out consequently in renormalizable field theories. First of all, one has to
introduce an intermediate regularization to give, in some way, the finite
values to the formal divergent expressions. Then one should  single out the
divergent part and include the counterterms in the classical action that
compensate the corresponding divergences. In renormalizable field theories one
introduces the counterterms that have the structure of the individual terms of
the classical action. They are interpreted in terms of renormalizations of the
fields, the masses and the coupling constants.

By the regularization some new parameters are introduced: a dimensionless
regularizing parameter $r$ and a dimensionful renormalization parameter $\mu$.
After subtracting the divergences and going to the limit $r\to 0$ the
regularizing parameter disappears, but the renormalization parameter $\mu$
remains and enters the finite renormalized expressions. In renormalized quantum
field theories the change of this parameter is compensated by the change of the
coupling constants of the renormalized action, $g_i(\mu)$, that are defined at the
renormalization point characterized by the energy scale $\mu$.
The physical quantities do not depend on the choice of the renormalization
point $\mu$ where the couplings are defined, i.e., they are renormalization
invariant. The transformation of the renormalization parameter $\mu$ and the
compensating transformations of the parameters of the renormalized action
$g_i(\mu)$ form the group of renormalization transformations [48, 163, 109].
The infinitesimal form of these transformations determines the differential
equations of the renormalization group that are used for investigating the
scaling properties (i.e., the behavior under the homogeneous scale
transformation) of the renormalized coupling parameters $g_i(\mu)$, the
many--point Green functions and other quantities. In particular, the equations
for renormalized coupling constants have the form [109]
$$
\mu{d\over d\mu}\bar g_i(\mu)=\beta_i(\bar g(\mu)),
\eqno(1.47)
$$
where $\bar g_i(\mu)=\mu^{-d_i} g_i(\mu)$ are the dimensionless coupling
parameters ($d_i$ is the dimension of the coupling $g_i$), and $\beta_i(\bar
g)$ are the Gell--Mann--Low $\beta$--functions.

Let us note, that  among the parameters $g_i(\mu)$ there are also
non--essential couplings [109] (like the renormalization constants of the fields
$Z_r(\mu)$) that are not invariant under the redefinition of the fields. The
equations for the renormalization constants $Z_r(\mu)$ have more simple form
[109]
$$
\mu{d\over d\mu}Z_r(\mu)=\gamma_r(\bar g(\mu))Z_r(\mu),
\eqno(1.48)
$$
where $\gamma_r(\bar g)$ are the anomalous dimensions.

The physical quantities (such as the matrix elements of the $S$--matrix on the
mass--shell) do not depend on the details of the field definition and,
therefore, on the non--essential couplings. On the other hand, the off
mass--shell
Green functions depend on all coupling constants including the non--essential
ones. We will apply the renormalization group equa\-tions for the investigation
of the ultraviolet behavior of the higher--derivative quantum gravity in Chap.
4.

Let us illustrate the procedure of eliminating the ultraviolet divergences
by the example of the Green function of the minimal differential operator
(1.30) at coinciding points, $G(x,x)$, and the corresponding one--loop effective
action $\Gamma_{(1)}$, (1.28). Making use of the Schwinger--De~Witt
representation for the Green function (1.33) we have
$$
G(x,x)=\int\limits_0^\infty i\,ds\, i (4\pi i
s)^{-n/2}\exp(-ism^2)\Omega(s|x,x),
\eqno(1.49)
$$
$$
\Gamma_{(1)}={1\over 2}\int\limits_0^\infty {ds\over s}(4\pi i
s)^{-n/2}\exp(-ism^2)\int d^n x g^{1/2}{\rm str}\,\Omega(s|x,x).
\eqno(1.50)
$$

It is clear that in four--dimensional space--time ($n=4$) the integrals over the
proper time in (1.49) and (1.50) diverge at the lower limit. Therefore, they
should be regularized. To do this one can introduce in the proper time integral
a regularizing function $\rho(is\mu^2; r)$ that depends on the regularizing
parameter $r$ and the renormalization parameter $\mu$. In the limit $r\to 0$
the regularizing function must tend to unity, and for $r\ne 0$ it must ensure
the convergence of the proper time integrals (i.e., it must approach zero
sufficiently rapidly at $s\to 0$ and be bounded at $s\to \infty$ by a
polynomial). The concrete form of the function $\rho$ does not matter. In
practice, one uses the cut--off regularization, the Pauli--Villars one, the
analytical one, the dimensional one, the $\zeta$--function regularization and
others [54, 48, 49].

The dimensional regularization is one of the most convenient for the practical
calculations (especially in massless and gauge theories) as well as for general
investigations [22, 50, 54, 111--114]. The theory is formulated in the space of
arbitrary dimension $n$ while the topology and the metric of the additional
$n-4$ dimensions can be arbitrary. To preserve the physical dimension of all
quantities in the $n$--dimensional space--time it is necessary to introduce the
dimensionful parameter $\mu$. All integrals are calculated in that region of
the complex plane of $n$ where they converge. It is obvious that for ${\rm
Re}\, n <C$, with some constant $C$, the integrals (1.49) and (1.50) converge
and define analytic functions of the dimension $n$. The analytic continuation
of these functions to the neighborhood of the physical dimension leads to
singularities in the point $n=4$. After subtracting these singularities we
obtain analytical functions in the vicinity of the physical dimension, the
value of this function
in the point $n=4$ defines the finite value of the initial expression.

Let us make some remarks on the dimensional regularization. The
analytical continuation of all the relations of the theory to the complex plane
of the dimension $n$ is not single--valued, since the values of a function of
complex variable in discrete integer values of the argument do not define
the unique analytical function [165]. There is also an arbitrariness connected
with the subtraction of the divergences. Together with the poles in $(n-4)$ one
can also subtract some finite terms (non--minimal renormalization). It is also
not necessary to take into account the dependence on the dimension of some
quantities (such as the volume element $d^n\,x\,g^{1/2}(x)$, background fields,
curvatures etc.). On the other hand, one can specify the dependence of all
quantities on the additional $n-4$ coordinates in some special explicit way and
then calculate the integrals over the $n-4$ dimensions. This would lead to an
additional factor that will give, when expanding in $n-4$, additional finite
terms. This u
uncertainty affects only the finite renormalization terms that should be
determined from the experiment.

Using the asymptotic expansion (1.43) we obtain in this way from (1.49) and
(1.50) the Green function at coinciding points and the one--loop effective
action in  dimensional regularization
\begin{eqnarray}\setcounter{equation}{51}
G(x,x)&=&{i\over(4\pi )^2}\left\{\left({2\over n-4}+{\bf C}+\log {m^2\over
4\pi\mu^2}\right)(m^2-a_1(x,x))-m^2\right\}\nonumber\\[10pt]
& &+G_{\rm ren}(x,x),
\end{eqnarray}
$$
G_{\rm ren}(x,x)={i\over(4\pi )^2}\sum\limits_{k\ge 2}{a_k(x,x)\over
k(k-1)m^{2(k-1)}},
\eqno(1.52)
$$
\begin{eqnarray}\setcounter{equation}{53}
\Gamma_{(1)}&=&{1\over 2(4\pi )^2}\Biggl\{
-\frac{1}{2}\left({2\over n-4}+{\bf C}
+\log {m^2\over 4\pi\mu^2}\right)
(m^4-2m^2A_1+A_2)\nonumber\\[10pt]
& & \qquad\qquad
+\frac{3}{4}m^4A_0-m^2A_1\Biggr\}+\Gamma_{(1){\rm ren}},
\end{eqnarray}
$$
\Gamma_{(1){\rm ren}}={1\over 2(4\pi )^2}\sum\limits_{k\ge 3}{A_k\over
k(k-1)(k-2)m^{2(k-2)}},
\eqno(1.54)
$$
where ${\bf C}\approx 0.577$ is the Euler constant and
$$
A_k\equiv\int\,d^n x\, g^{1/2}{\rm str}\,a_k(x,x).
\eqno(1.55)
$$
Here all the coefficients $a_k$ and $A_k$ are $n$--dimensional. However in that
part, which is analytical in $n-4$, one can treat them as $4$--dimensional.

%
%
%
%
%
%
%
%
%
%
%
%
%
%
%
%

\chapter{Technique for the calculation of 
De~Witt coefficients and its
applications}
\markboth{\sc Chapter 2. Calculation of De~Witt coefficients}{\sc Chapter 2.
Calculation of De~Witt coefficients}
%
%
%
%
%
%
%
%
%
\section{Covariant expansions of field variables in 
cur\-ved space}

Let us single out a small regular region in the space, fix a point $x'$ in it,
and connect any  other point $x$ with the point $x'$ by a geodesic $x=x(\tau)$,
$x(0)=x'$, where $\tau$ is an affine parameter.

The world function $\sigma(x,x')$ (in terminology of [2]), introduced in the
Chap. 1, has the form
$$
\sigma(x,x')=\frac{1}{2}\tau^2\dot x^2(\tau),
\eqno(2.1)
$$
where
$$
\dot x^\mu(\tau)={ d \over d \tau}x^\mu(\tau).
\eqno(2.2)
$$
The first derivatives of the function $\sigma(x,x')$ with respect to
coordinates are proportional to the tangent vectors to the geodesic at the
points $x$ and $x'$ [2],
$$
\sigma^\mu=\tau\dot x^\mu(\tau), \qquad
\sigma^{\mu'}=-\tau\dot x^\mu(0),
\eqno(2.3)
$$
where
$$
\sigma_\mu=\nabla_\mu\sigma, \qquad
\sigma_{\mu'}=\nabla_{\mu'}\sigma.
$$
Herefrom, it follows, in particular, the basic identity (1.39), that the
function $\sigma(x,x')$ satisfies,
$$
(D-2)\sigma=0, \qquad D\equiv\sigma^\mu\nabla_\mu,
\eqno(2.4)
$$
the coincidence limits
$$
[\sigma]=[\sigma^\mu]=[\sigma^{\mu'}]=0,
\eqno(2.5)
$$
$$
[f(x,x')]\equiv\lim_{x\to x'} f(x,x'),
\eqno(2.6)
$$
and the relation between the tangent vectors
$$
\sigma^\mu=-g^\mu_{\ \nu'}\sigma^{\nu'},
\eqno(2.7)
$$
where $g^\mu_{\ \nu'}(x,x')$ is the parallel displacement operator of vectors
along the geodesic from the point $x'$ to the point $x$. The non--primed
(primed) indices are lowered and risen by the metric tensor in the point $x$
$(x')$.

By differentiating the basic identity (2.4) we obtain the relations
$$
(D-1)\sigma^\mu=0,\qquad
\sigma^\mu=\xi^\mu_{\ \nu}\sigma^\nu,
\eqno(2.8)
$$
$$
(D-1)\sigma^{\mu'}=0,\qquad
\sigma^{\mu'}=\eta^{\mu'}_{\ \nu}\sigma^\nu,
\eqno(2.9)
$$
where
$$
\xi^\mu_{\ \nu}=\nabla_\nu\sigma^\mu,\qquad
\eta^{\mu'}_{\ \nu}=\nabla_\nu\sigma^{\mu'}.
\eqno(2.10)
$$
Therefrom the coincidence limits follow
$$
[\xi^\mu_{\ \nu}]=-[\eta^{\mu'}_{\ \nu}]=\delta^\mu_{\ \nu},
\eqno(2.11)
$$
$$
[\nabla_{(\mu_1}...\nabla_{\mu_k)}\sigma^\nu]
=[\nabla_{(\mu_1}...\nabla_{\mu_k)}\sigma^{\nu'}]=0,\qquad
(k\ge 2).
\eqno(2.12)
$$

Let us consider a field $\varphi=\varphi^A(x)$ and an affine connection 
${\cal A}_\mu={\cal A}^A_{\ B\mu}(x)$, that defines the covariant derivative (1.31) and
the commutator of covariant derivatives,
$$
[\nabla_\mu, \nabla_\nu ]\varphi ={\cal R}_{\mu\nu}\varphi,
\eqno(2.13)
$$
$$
{\cal R}_{\mu\nu}=\partial_\mu{\cal A}_\nu
-\partial_\nu{\cal A}_\mu+[{\cal A}_\mu, {\cal A}_\nu].
\eqno(2.14)
$$

Let us define the parallel displacement operator of the field $\varphi$ along
the geodesic from the point $x'$ to the point $x$, ${\cal P}={\cal P}^A_{\
B'}(x,x')$, to be the solution of the equation of parallel transport,
$$
D{\cal P}=0,
\eqno(2.15)
$$
with the initial condition
$$
[{\cal P}]={\cal P}(x,x)=\hat 1.
\eqno(2.16)
$$
Herefrom one can obtain the coincidence limits
$$
\left[\nabla_{(\mu_1}...\nabla_{\mu_k)}{\cal P}\right]=0,\qquad
(k\ge 1).
\eqno (2.17)
$$
In particular, when $\varphi=\varphi^\mu$ is a vector field, and the connection
${\cal A}_\mu=\Gamma^\alpha_{\ \mu\beta}$ is the Christoffel connection, the
equations (2.15) and (2.16) define the parallel displacement operator of
the vectors: ${\cal P}=g^\mu_{\ \nu'}(x,x')$.

Let us transport the field $\varphi$ parallel along the geodesic to the point
$x'$
$$
\bar\varphi=\bar\varphi^{C'}(x)={\cal P}^{C'}_{\ A}(x',x)\varphi^A(x)
={\cal P}^{-1}\varphi,
\eqno(2.18)
$$
where ${\cal P}^{-1}={\cal P}^{C'}_{\ A}(x',x)$ is the parallel displacement
operator along the opposite path (from the point $x$ to the point $x'$ along
the geodesic):
$$
{\cal P}{\cal P}^{-1}=\hat 1.
\eqno(2.19)
$$
The obtained object $\bar\varphi$, (2.18), is a scalar under the coordinate
transformations at the point $x$, since it does not have any non--prime indices.
By considering $\bar\varphi$ as a function of the affine parameter $\tau$, let
us expand it in the Taylor series
$$
\bar\varphi=\sum\limits_{k\ge 0}{1\over k!}\tau^k\left[{d^k\over d
\tau^k}\bar\varphi\right]_{\tau=0}.
\eqno(2.20)
$$
Noting that $d/d\tau=\dot x^\mu\partial_\mu$,
$\partial_\mu\bar\varphi=\nabla_\mu\bar\varphi$ and using the equation of
the geodesic, $\dot x^\mu\nabla_\mu \dot x^\nu=0$, and the equations (2.3), (2.9) and
(2.18), we obtain
$$
\varphi={\cal P}\sum_{k\ge 0}{(-1)^k\over k!}\sigma^{\mu'_1}\cdots
\sigma^{\mu'_k}\varphi_{\mu'_1\cdots\mu'_k},
\eqno(2.21)
$$
where
$$
\varphi_{\mu'_1\cdots\mu'_k}
=\left[\nabla_{(\mu_1}\cdots\nabla_{\mu_k)}\varphi\right].
\eqno(2.22)
$$
The equation (2.21) is the generalized covariant Taylor series for arbitrary
field with arbitrary affine connection in a curved space.

Let us show that the series (2.21) is the expansion in a complete set of
eigenfunctions of the operator $D$, (2.4). The vectors $\sigma^\mu$ and
$\sigma^{\mu'}$ are the eigenfunctions of the operator $D$ with the eigenvalues
equal to 1 (see eqs.  (2.8) and (2.9)). 
Therefore, one can construct the eigenfunctions with
arbitrary positive integer eigenvalues:
$$
\vert 0>\equiv 1,
$$
$$
\vert n>\equiv\vert \nu'_1...\nu'_n>
={(-1)^n\over n!}\sigma^{\nu'_1}\cdots
\sigma^{\nu'_n}, \qquad (n\ge 1),
\eqno(2.23)
$$
$$
D\vert n>=n\vert n>.
\eqno(2.24)
$$
We have
$$
|n_1>\otimes\cdots\otimes|n_k>={n \choose n_1, \dots, n_k}|n>,
\eqno(2.25)
$$
where
$$
{n \choose n_1, \dots, n_k}={n!\over n_1!\cdots n_k!},
\qquad n=n_1+\cdots+n_k .
\eqno(2.26)
$$

Let us note that there exist more general eigenfunctions of the form
$(\sigma)^z|n>$ with arbitrary eigenvalues $(n+2z)$
$$
D\,(\sigma)^z\vert n>=(n+2z)(\sigma )^z\vert n>.
\eqno(2.27)
$$
However, for non--integer or negative $z$ these functions are not analytic in
coordinates of the point $x$ in the vicinity of the point $x'$. For positive
integer $z$ they reduce to the linear combinations of the functions (2.23).
Therefore, we restrict ourselves to the functions (2.23) having in mind to
study only regular fields near the point $x'$.

Let us introduce the conjugate functions
$$
<m\vert\equiv<\mu'_1\cdots\mu'_m\vert=(-1)^m g^{\mu_1}_{\mu'_1}\cdots
g^{\mu_m}_{\mu'_m}\nabla_{(\mu_1}\cdots\nabla_{\mu_m)}\delta (x,x')
\eqno(2.28)
$$
and the scalar product
$$
<m|n>=\int\,d^n x\,<\mu'_1\cdots\mu'_m\vert\nu'_1...\nu'_n>.
\eqno(2.29)
$$
Using the coincidence limits (2.11) and (2.12) it is easy to prove that the set
of the eigenfunctions (2.23) and (2.28) is orthonormal
$$
<m\vert n>=\delta_{mn}1_{(n)}, \qquad
1_{(n)}\equiv\delta^{\nu_1\cdots\nu_n}_{\mu_1\cdots\mu_n}
=\delta^{\nu_1}_{(\mu_1}\cdots\delta^{\nu_n}_{\mu_n)}.
\eqno(2.30)
$$
The introduced scalar product (2.29) reduces to the coincidence limit of the
symmetrized covariant derivatives,
$$
<m\vert\varphi>
=\left[\nabla_{(\mu_1}\cdots\nabla_{\mu_m)}\varphi\right].
\eqno(2.31)
$$
Therefore, the covariant Taylor series (2.21) can be rewritten in a compact
form
$$
\vert\varphi>={\cal P}\sum_{n\ge 0}\vert n><n\vert\varphi>.
\eqno(2.32)
$$
Herefrom it follows the condition of completeness of the set of the eigenfunctions (2.23)
in the space of scalar functions regular in the vicinity of the point $x'$
$$
1\!\!{\rm I}=\sum_{n\ge 0}\vert n><n\vert,
\eqno(2.33)
$$
or, more precisely,
\begin{eqnarray}\setcounter{equation}{34}
\delta(x,y)&=&\sum_{n\ge 0}{1\over n!}
\sigma^{\mu'_1}(x,x')\cdots\sigma^{\mu'_n}(x,x')
g^{\mu_1}_{\mu'_1}(y,x')\cdots g^{\mu_n}_{\mu'_n}(y,x')\nonumber\\[10pt]
& &\times\nabla^y_{(\mu_1}\cdots\nabla^y_{\mu_n)}\delta(y,x').
\end{eqnarray}

Let us note that, since the parallel displacement operator ${\cal P}$ is an
eigenfunction of the operator $D$ with zero eigenvalue, (2.15), one can
also introduce a complete orthonormal set of `isotopic'  eigenfunctions 
${\cal P}|n>$ and $<n|{\cal P}^{-1}$.

The complete set of eigenfunctions (2.23) can be employed to present an
arbitrary linear differential operator $F$ defined on the fields $\varphi$ in
the form
$$
F=\sum\limits_{m,n\ge 0}{\cal P}|m><m|{\cal P}^{-1}F{\cal P}|n><n|{\cal
P}^{-1},
\eqno(2.35)
$$
where
$$
<m|{\cal P}^{-1}F{\cal P}|n>
=\left[\nabla_{(\mu_1}\cdots \nabla_{\mu_m)}
{\cal P}^{-1}F{\cal P}{{(-1)^n}\over {n!}}
\sigma^{\nu'_1}\cdots \sigma^{\nu'_n}\right]
\eqno(2.36)
$$
are the ``matrix elements'' of the operator $F$ (2.35). The matrix elements
(2.36) are expressed finally in terms of the coincidence limits of the
derivatives of the coefficient functions of the operator $F$, the parallel
displacement operator ${\cal P}$ and the world function $\sigma$.

For calculation of the matrix elements of differential operators (2.36) as well
as for constructing the covariant Fourier integral it is convenient to make a
change of the variables
$$
x^\mu=x^\mu(\sigma^{\nu'}, x^{\lambda'}),
\eqno(2.37)
$$
i.e., to consider a function of the coordinates $x^\mu$ as the function of the
vectors $\sigma^{\nu'}(x,x')$ and the coordinates $x^{\lambda'}$.

The derivatives and the differentials in old and new variables are connected by
the relations
$$
\partial_\mu=\eta^{\nu'}_{\ \mu}\bar\partial_{\nu'}, \qquad
\bar\partial_{\nu'}=\gamma^{\mu}_{\ \nu'}\partial_{\mu},
$$
$$
dx^\mu=\gamma^{\mu}_{\ \nu'}d\sigma^{\nu'},\qquad
d\sigma^{\nu'}=\eta^{\nu'}_{\ \mu}dx^\mu,
\eqno(2.38)
$$
where $\partial_\mu=\partial/\partial x^\mu$,
$\bar\partial_{\mu'}=\partial/\partial\sigma^{\mu'}$, $\eta^{\nu'}_{\ \mu}$ is
defined in (2.10), $\gamma^{\mu}_{\ \nu'}$ are the elements of the inverse
matrix,
$$
\gamma=\eta^{-1},
\eqno(2.39)
$$
and $\eta$ is a matrix with elements $\eta^{\nu'}_{\ \mu}$.

{}From the coincidence limits (2.11) it follows that for close points $x$ and
$x'$
$$
\det \eta\ne 0,\qquad \det \gamma\ne 0,
\eqno(2.40)
$$
and, therefore, the change of variables (2.37) is admissible.

The corresponding covariant derivatives are connected by analogous relations
$$
\nabla_\mu=\eta^{\nu'}_{\ \mu}\bar\nabla_{\nu'}, \qquad
\bar\nabla_{\nu'}=\gamma^{\mu}_{\ \nu'}\nabla_{\mu}.
\eqno(2.41)
$$

\noindent
{}From the definition of the matrices $\eta$, (2.10), and $\gamma$, (2.39), one can
get the relations
$$
\nabla_{_{[\lambda}}\eta^{\mu'}_{\ \nu]}=0,\qquad
\bar\nabla_{_{[\lambda'}}\gamma^\mu_{\ \nu']}=0,
\eqno(2.42)
$$
$$
\bar\nabla_{\mu'}\left(\Delta^{-1}\eta^{\mu'}_{\ \nu}\right)=0,
\eqno(2.43)
$$
where $\Delta$ is the Van~Fleck--Morette determinant (1.37)
\begin{eqnarray}\setcounter{equation}{44}
\Delta(x,x')&=&g^{1/2}(x')g^{-1/2}(x)\det(-\eta)
=g^{1/2}(x')g^{-1/2}(x)\det(-\gamma)^{-1}\nonumber\\[10pt]
&=&\det(-\bar\eta)=\left(\det(-\bar\gamma)\right)^{-1}=\left(\det X\right)^{1/2},
\end{eqnarray}
and $\bar\eta$, $\bar\gamma$ and $X$ are matrices with elements
$$
\bar\eta^{\mu'}_{\ \nu'}=g^\nu_{\ \nu'}\eta^{\mu'}_{\ \nu}, \qquad
\bar\gamma^{\mu'}_{\ \nu'}=g^{\mu'}_{\ \mu}\gamma^{\mu}_{\ \nu'},
\eqno(2.45)
$$
$$
X^{\mu'\nu'}=\eta^{\mu'}_{\ \mu}g^{\mu\nu}\eta^{\nu'}_{\ \nu}.
\eqno(2.46)
$$

Let us note that the conjugate eigenfunctions (2.28) of the operator $D$ can be
expressed in terms of the operators $\bar\nabla$, (2.41),
$$
<m\vert\equiv<\mu'_1\cdots\mu'_m\vert
=\bar\nabla_{(\mu'_1}\cdots\bar\nabla_{\mu'_m)}\delta (x,x').
\eqno(2.47)
$$
Therefore, the coefficients of the covariant Taylor series (2.21), (2.22),
(2.31) and (2.32) can be also written in terms of the operators $\bar\nabla$:
$$
<m\vert\varphi>
=(-1)^m\left[\bar\nabla_{(\mu'_1}\cdots\bar\nabla_{\mu'_m)}\varphi\right].
\eqno(2.48)
$$

The commutator of the operators $\bar\nabla$, when acting on the parallel
displacement operator ${\cal P}$, has the form
$$
[\bar\nabla_{\mu'}, \bar\nabla_{\nu'}]{\cal P}
=\bar{\cal R}_{\mu'\nu'}{\cal P},
\eqno(2.49)
$$
where
\begin{eqnarray}\setcounter{equation}{50}
\bar{\cal R}_{\mu'\nu'}
&=&\gamma^\mu_{\ \mu'}\gamma^\nu_{\ \nu'}{\cal P}^{-1}{\cal R}_{\mu\nu}{\cal P}
\nonumber\\[10pt]
&=&\bar\nabla_{\mu'}\bar{\cal A}_{\nu'}-\bar\nabla_{\nu'}\bar{\cal A}_{\mu'}
+[\bar{\cal A}_{\mu'}, \bar{\cal A}_{\nu'}],
\end{eqnarray}
$$
\bar{\cal A}_{\mu'}={\cal P}^{-1}\bar\nabla_{\mu'}{\cal P}.
\eqno(2.51)
$$

The quantity (2.51) introduced here satisfies the equation (2.15),
$$
\sigma^{\mu'}\bar{\cal A}_{\mu'}=0.
\eqno(2.52)
$$

On the other hand, when acting on the objects $\bar\varphi$, (2.18), that do
not have non--primed indices, the operators $\bar\nabla$ commute with each other
$$
[\bar\nabla_{\mu'}, \bar\nabla_{\nu'}]\bar\varphi=0.
\eqno(2.53)
$$

Thus the vectors $\sigma^{\nu'}$ and the operators $\bar\nabla_{\mu'}$, (2.41), play
the role analogous to that of usual coordinates and the operator of
differentiation in the tangent space at the point $x'$. In particular,
$$
[\bar\nabla_{\mu'}, \sigma^{\nu'}]=\delta^{\nu'}_{\ \mu'}.
\eqno(2.54)
$$

Therefore, one can construct the covariant Fourier integral  in the tangent
space at the point $x'$ in the usual way using the variables $\sigma^{\mu'}$. So that
for the fields $\bar\varphi$, (2.18), we have
$$
\bar\varphi(k)=\int\,d^n x\,g^{1/2}(x)\Delta(x,x')
\exp(ik_{\mu'}\sigma^{\mu'})\bar\varphi(x),
$$
$$
\bar\varphi(x)=\int\,{d^n k^{\mu'}\over (2\pi)^n}\,g^{1/2}(x')
\exp(-ik_{\mu'}\sigma^{\mu'})\bar\varphi(k).
\eqno(2.55)
$$
Note, that the standard rule,
$$
\bar\nabla_{\mu'}\bar\varphi(x)=\int\,{d^n k^{\mu'}\over (2\pi)^n}
\,g^{1/2}(x')
\exp(-ik_{\mu'}\sigma^{\mu'})(-ik_{\mu'})\bar\varphi(k),
\eqno(2.56)
$$
takes place and the covariant momentum representation of the $\delta$--function
has the form
$$
\delta(x,y)=\int\,{d^n k^{\mu'}\over (2\pi)^n}\,g^{1/2}(x')g^{1/2}(x)
\Delta(x,x')
\exp\left\{ik_{\mu'}\left(\sigma^{\mu'}(y,x')
-\sigma^{\mu'}(x,x')\right)\right\}.
\eqno(2.55)
$$

\section{Structure elements of covariant expansions}

Let us calculate the quantities $\eta$, $\gamma$ and $X$ introduced in previous
section. By differentiating the equations (2.8) and (2.9) and commuting the
covariant derivatives we get
$$
D\xi+\xi(\xi-\hat 1)+S=0,
\eqno(2.58)
$$
$$
D\eta+\eta(\xi-\hat 1)=0,
\eqno(2.59)
$$
where
$$
\xi=\xi^\mu_{\ \nu}, \qquad S=S^\mu_{\ \nu}, \qquad \hat 1=\delta^\mu_{\ \nu},
$$
$$
S^\mu_{\ \nu}=R^\mu_{\ \alpha\nu\beta}\sigma^\alpha\sigma^\beta.
\eqno (2.60)
$$

By solving the eq. (2.59) with respect to the matrix $\xi$, substituting the
solution in (2.58) and taking into account (2.39) we obtain the linear equation
for the matrix $\bar\gamma$ (2.45)
$$
\left\{\hat 1\left(D^2+D\right)+\bar S\right\}\bar\gamma=0,
\eqno (2.61)
$$
where $\bar S=\bar S^{\mu'}_{\ \nu'}
=g^{\mu'}_{\ \mu}g^{\nu}_{\ \nu'}S^\mu_{\ \nu}$, with the boundary condition,
(2.39), (2.11),
$$
[\bar\gamma]=-1.
\eqno (2.62)
$$

One can solve the equation (2.61) perturbatively, treating the matrix $\bar S$
as a perturbation. Supposing
$$
\bar\gamma=-\hat 1 +\hat{\bar\gamma},
\eqno(2.63)
$$
we obtain
$$
\left\{\hat 1(D^2+D)+\bar S\right\}\hat{\bar\gamma}=\bar S, \qquad
[\hat{\bar\gamma}]=0.
\eqno (2.64)
$$
Herefrom we have formally
$$
\hat{\bar\gamma}=\left\{\hat 1(D^2+D)+\bar S\right\}^{-1}\bar S=
\sum_{k\ge 1}(-1)^{k+1}\left\{(D^2+D)^{-1}\bar S\right\}^k\cdot\hat 1.
\eqno(2.65)
$$
The formal expression (2.65) becomes meaningful in terms of the expansion in
the eigenfunctions of the operator $D$, (2.23). The inverse operator
$$
(D^2+D)^{-1}=\sum_{n\ge 0}{1\over n(n+1)}\vert n><n\vert,
\eqno(2.66)
$$
when acting on the matrix $\bar S$, is well defined, since $<0|\bar S>=0$.
Expanding the matrix $\bar S$ in the covariant Taylor matrix according to (2.21)
and  (2.32),
$$
\bar S=\sum_{n\ge 2}{(-1)^n\over (n-2)!}K_{(n)},
\eqno(2.67)
$$
where $K_{(n)}$ is the matrix with entries $K^{\mu'}_{\ \nu'(n)}$
$$
K^{\mu'}_{\ \nu'(n)}=K^{\mu'}_{\ \nu'\mu'_1\cdots\mu'_n}
\sigma^{\mu'_1}\cdots\sigma^{\mu'_n},
$$
$$
K^\mu_{\ \nu \mu_1\cdots\mu_n}=\nabla_{(\mu_1}\cdots\nabla_{\mu_{n-2}}
R^\mu_{ \ \mu_{n-1}\vert\nu\vert\mu_n)},
\eqno(2.68)
$$
we obtain
$$
\bar\gamma=-\hat 1+\sum_{n\ge 2}{(-1)^n\over n!}\gamma_{(n)},
\eqno(2.69)
$$
\begin{eqnarray}\setcounter{equation}{70}
\gamma_{(n)}&=&\gamma_{\mu_1'\cdots\mu_n'}\sigma^{\mu'_1}\cdots\sigma^{\mu'_n}
\nonumber\\[10pt]
&=&\sum_{1\le k\le[{n\over 2}]}(-1)^{k+1}(2k)!{n \choose 2k}
\sum_{{n_1,\cdots,n_k\ge 2\atop n_1+\cdots+n_k=n}}
{n-2k \choose n_1-2, \cdots, n_k-2}\nonumber\\[10pt]
& &\times
{K_{(n_k)}\over n(n+1)}\cdot
{K_{(n_k-1)}\over (n_1+\cdots +n_{k-1})(n_1+\cdots +n_{k-1}+1)}
\\[10pt]
& &\times\cdots {K_{(n_2)}\over (n_1+n_{2})(n_1+n_{2}+1)}\cdot
{K_{(n_1)}\over n_1(n_1+1)},
\end{eqnarray}
where
$$
{n \choose m}={n!\over m!(n-m)!}.
\eqno(2.70a)
$$

Let us write down some first coefficients (2.70):
$$
\gamma^\alpha_{\ \beta\mu_1\mu_2}
=\frac{1}{3}R^\alpha_{\ (\mu_1\vert\beta\vert\mu_2)},
$$
$$
\gamma^\alpha_{\ \beta\mu_1\mu_2\mu_3}
=\frac{1}{2}\nabla_{(\mu_1}R^\alpha_{\ \mu_2\vert\beta\vert\mu_3)},
\eqno(2.71)
$$
$$
\gamma^\alpha_{\ \beta\mu_1\mu_2\mu_3\mu_4}
=\frac{3}{5}\nabla_{(\mu_1}\nabla_{\mu_2}
R^\alpha_{\ \mu_3\vert\beta\vert\mu_4)}
-\frac{1}{5}R^\alpha_{\ (\mu_1\vert\gamma\vert\mu_2}
R^\gamma_{\ \mu_3\vert\beta\vert\mu_4)}.
$$
Using the solution (2.69) one can find all other quantities. The inverse matrix
$\bar\eta=\bar\gamma^{-1}$ has the form
$$
\bar\eta=-\hat 1+\sum_{n\ge 2}{(-1)^n\over n!}{\eta}_{(n)},
\eqno(2.72)
$$
\begin{eqnarray}\setcounter{equation}{73}
{\eta}_{(n)}&=&\eta_{\mu_1'\cdots\mu_n'}\sigma^{\mu'_1}\cdots\sigma^{\mu'_n}
\nonumber\\[10pt]
&=&-\sum_{1\le k\le[{n\over 2}]}
\sum_{{n_1,\cdots, n_k\ge 2\atop n_1+\cdots+n_k=n}}
{n \choose n_1,\cdots n_k}
\gamma_{(n_k)}\cdots\gamma_{(n_1)}.
\end{eqnarray}
Some first coefficients (2.73) equal
$$
\eta^\alpha_{\ \beta\mu_1\mu_2}
=-\frac{1}{3}R^\alpha_{\ (\mu_1\vert\beta\vert\mu_2)},
$$
$$
\eta^\alpha_{\ \beta\mu_1\mu_2\mu_3}
=-\frac{1}{2}\nabla_{(\mu_1}R^\alpha_{\ \mu_2\vert\beta\vert\mu_3)},
\eqno(2.74)
$$
$$
\eta^\alpha_{\ \beta\mu_1\mu_2\mu_3\mu_4}
=-\frac{3}{5}\nabla_{(\mu_1}\nabla_{\mu_2}
R^\alpha_{\ \mu_3\vert\beta\vert\mu_4)}
-\frac{7}{15}R^\alpha_{\ (\mu_1\vert\gamma\vert\mu_2}
R^\gamma_{\ \mu_3\vert\beta\vert\mu_4)}.
$$
Using (2.64) and (2.72) we find the matrix $X$:
$$
X^{\mu'\nu'}=g^{\mu'\nu'}
+\sum_{n\ge 2}{(-1)^n\over n!}X^{\mu'\nu'}_{\ \ (n)},
\eqno(2.75)
$$
\begin{eqnarray}\setcounter{equation}{76}
X^{\mu'\nu'}_{\ \ (n)}
&=&X^{\mu'\nu'}_{\ \ \ \mu_1'\cdots\mu_n'}\sigma^{\mu'_1}\cdots\sigma^{\mu'_n}
\nonumber\\[10pt]
&=&-2\eta^{(\mu'\nu')}_{\ \ (n)}
+\sum\limits_{2\le k \le n-2} {n \choose k}\eta^{(\mu'}_{\ \ \alpha(n-k)}
\eta^{\nu')\alpha}_{\ \ \ (k)}.
\end{eqnarray}
The lowest order coefficients (2.76) read
$$
X^{\mu\nu}_{\ \ \mu_1\mu_2}
=\frac{2}{3}R^{\mu\ \ \nu}_{\ (\mu_1\ \mu_2)},
$$
$$
X^{\mu\nu}_{\ \ \mu_1\mu_2\mu_3}
=\nabla_{(\mu_1}R^{\mu\ \ \nu}_{\ \mu_2\ \mu_3)},
\eqno(2.77)
$$
$$
X^{\mu\nu}_{\ \ \mu_1\mu_2\mu_3\mu_4}
=\frac{6}{5}\nabla_{(\mu_1}\nabla_{\mu_2}
R^{\mu\ \ \nu}_{\ \mu_3\ \mu_4)}
+\frac{8}{5}R^\mu_{\ (\mu_1\vert\alpha\vert\mu_2}
R^{\alpha\ \ \nu}_{\ \mu_3\ \mu_4)}.
$$

Finally, we find the Van~Fleck--Morette determinant (2.44):
$$
\Delta=\exp(2\zeta),
$$
$$
\zeta=\sum_{n\ge 2}{{(-1)^n}\over {n!}}{\zeta}_{(n)},
\eqno(2.78)
$$
\begin{eqnarray}\setcounter{equation}{79}
{\zeta}_{(n)}&=&\zeta_{\mu'_1\cdots\mu'_n}\sigma^{\mu'_1}\cdots\sigma^{\mu'_n}
\nonumber\\[10pt]
&=&\sum_{1\le k\le [{n\over 2}]}{1\over 2k}
\sum_{n_1,\cdots,n_k\ge 2\atop n_1+\cdots +n_k=n}
{n \choose n_1,\cdots, n_k}
{\rm tr}\left(\gamma_{(n_1)}\cdots\gamma_{(n_k)}\right).
\end{eqnarray}
The first coefficients (2.79) equal
$$
\zeta_{\mu_1\mu_2}
={{1}\over {6}}R_{\mu_1\mu_2},
$$
$$
\zeta_{\mu_1\mu_2\mu_3}
={1\over 4}\nabla_{(\mu_1}R_{\mu_2\mu_3)},
\eqno(2.80)
$$
$$
\zeta_{\mu_1\mu_2\mu_3\mu_4}
={{3}\over {10}}\nabla_{(\mu_1}\nabla_{\mu_2}
R_{\mu_3\mu_4)}
+{{1}\over {15}}R_{\alpha (\mu_1\vert\gamma\vert\mu_2}
R^{\gamma\ \ \alpha}_{\ \mu_3\ \mu_4)}.
$$

Let us calculate the quantity $\bar{\cal A}_{\mu'}$, (2.51). By differentiating
the equation (2.52) and using (2.50) we obtain
$$
(D+1)\bar{\cal A}_{\mu'}=-\bar\gamma^{\nu'}_{\ \mu'}\bar{\cal L}_{\nu'},
\eqno(2.81)
$$
where
$$
\bar{\cal L}_{\nu'}=
g^\mu_{\ \nu'}{\cal P}^{-1}{\cal R}_{\mu\alpha}{\cal P}\sigma^\alpha.
\eqno(2.82)
$$
Herefrom we have
$$
\bar{\cal A}_{\mu'}
=-(D+1)^{-1}\bar\gamma^{\nu'}_{\ \mu'}\bar{\cal L}_{\nu'},
\eqno(2.83)
$$
where the inverse operator $(D+1)^{-1}$ is defined by
$$
(D+1)^{-1}=\sum_{n\ge 0}{{1}\over {n+1}}\vert n><n\vert.
$$
Expanding  the vector $\bar{\cal L}_{\mu'}$, (2.82), in the covariant Taylor
series (2.21), (2.32)
$$
\bar{\cal L}_{\mu'}
=\sum_{n\ge 1}{{(-1)^n}\over {(n-1)!}}{\cal R}_{\mu'(n)},
\eqno(2.85)
$$
where
$$
{\cal R}_{\mu'(n)}={\cal R}_{\mu'\mu'_1\cdots\mu'_n}
\sigma^{\mu'_1}\cdots\sigma^{\mu'_n},
$$
$$
{\cal R}^\mu_{\
\mu_1\cdots\mu_n}
=\nabla_{(\mu_1}\cdots\nabla_{\mu_{n-1}}{\cal R}^\mu_{\ \mu_n)},
\eqno(2.86)
$$
we obtain
$$
\bar{\cal A}_{\mu'}
=\sum_{n\ge 1}{{(-1)^n}\over {n!}}{\cal A}_{\mu'(n)},
\eqno(2.87)
$$
\begin{eqnarray}\setcounter{equation}{88}
{\cal A}_{\mu'(n)}
&=&{\cal A}_{\mu'\mu'_1\cdots\mu'_n}\sigma^{\mu'_1}\cdots\sigma^{\mu'_n}
\nonumber\\[10pt]
&=&{{n}\over {n+1}}\left\{{\cal R}_{\mu'(n)}
-\sum_{2\le k\le n-1}{n-1\choose k}\gamma^{\alpha'}_{\ \mu'(k)}
{\cal R}_{\alpha'(n-k)}\right\}.
\end{eqnarray}
In particular, the first coefficients (2.88) have the form
$$
{\cal A}_{\mu\mu_1}={1\over 2}{\cal R}_{\mu\mu_1},
$$
$$
{\cal A}_{\mu\mu_1\mu_2}={2\over 3}\nabla_{(\mu_1}{\cal R}_{|\mu|\mu_2)},
\eqno(2.89)
$$
$$
{\cal A}_{\mu\mu_1\mu_2\mu_3}
={3\over 4}\nabla_{(\mu_1}\nabla_{\mu_2}{\cal R}_{|\mu|\mu_3)}
-{1\over 4}R_{\alpha(\mu_1|\mu|\mu_2}{\cal R}^\alpha_{\ \mu_3)}.
$$
Thus we have obtained all the needed quantities in form of covariant Taylor
series: (2.69), (2.72), (2.75), (2.78) and (2.87).

In particular case, when all the derivatives of the curvature tensors can be
neglected,
$$
\nabla_\mu R^\alpha_{\ \beta\gamma\delta}=0,\qquad
\nabla_\mu {\cal R} _{\alpha\beta} =0,
\eqno(2.90)
$$
one can solve exactly the equations (2.61), (2.81) or, equivalently, sum up all
terms in the Taylor series, which do not contain the derivatives of the
curvature tensors. In this case the Taylor series (2.69) is a power series in
the matrix $\bar S$. It is an eigenmatrix of the operator $D$,
$$
D\bar S=2\bar S,
\eqno(2.91)
$$
and, 
therefore, when acting on the series (2.69), one can present the operator $D$
in the form
$$
D=2\bar S{d\over d \bar S},
\eqno(2.92)
$$
and treat the matrix $\bar S$ as an usual scalar variable. Substituting (2.92)
in (2.61), we obtain an ordinary second order differential equation
$$
\left({d^2\over d t^2}+{2\over t}{d\over d t} +\hat 1\right)\bar\gamma=0,
\qquad t\equiv\sqrt{\bar S},
\eqno(2.93)
$$
that has the general solution
$$
\bar\gamma=(\bar S)^{-1/2}\left(C_1\sin\sqrt{\bar S}
+C_2\cos\sqrt{\bar S}\right),
\eqno(2.94)
$$
where $C_1$ and $C_2$ are the integration constants. Using the initial
condition (2.62), we have finally $C_1=-1$, $C_2=0$, i.e.,
$$
\bar\gamma =-{{\sin \sqrt{\bar S}}\over {\sqrt{\bar S}}}.
\eqno(2.95)
$$
In any case this formal expression makes sense as the series (2.69). It is
certain to converge for close points $x$ and $x'$.

Using (2.95), (2.39) and (2.46) we find easily the matrices $\bar\eta$ and $X$
$$
\bar\eta=-{{\sqrt{\bar S}}\over {\sin \sqrt{\bar S}}},
$$
$$
X={{\bar S}\over {\sin^2 \sqrt{\bar S}}},
\eqno(2.96)
$$ and the Van~Fleck--Morette determinant $\Delta$, (2.44),
$$
\Delta=\det\left({{\sqrt{\bar S}}\over {\sin \sqrt{\bar S}}}\right).
\eqno(2.97)
$$
The vector $\bar{\cal A}_{\mu'}$, (2.87), in the case (2.90) is presented as the
product of some matrix $H^{\mu'}_{\ \nu'}$, that does not depend on ${\cal
R}_{\mu\nu}$, and the vector $\bar{\cal L}_{\mu'}$, (2.82),
$$
\bar{\cal A}_{\mu'}=H_{\mu' \nu'}\bar{\cal L}^{\nu'}.
\eqno(2.98)
$$
Substituting (2.98) in (2.81) and using the equation
$$
D\bar{\cal L}_{\mu'}=\bar{\cal L}_{\mu'},
\eqno(2.99)
$$
we get the equation for the matrix $H$
$$
(D+2)H^{\mu'}_{\ \nu'}=-\bar\gamma^{\mu'}_{\ \nu'}.
\eqno(2.100)
$$
Substituting (2.92) and (2.95) in (2.100), we obtain an ordinary first order
differential equation
$$
\left(t{d\over d t}+\hat 1\cdot 2\right)H={\sin t\over t}, \qquad
t\equiv\sqrt{\bar S}.
\eqno(2.101)
$$
The solution of this equation has the form
$$
H=\bar S^{-1}\left(\hat 1-\cos\sqrt{\bar S}+C_3\right),
\eqno(2.102)
$$
where $C_3$ is the integration constant. Using the initial condition
$[\bar{\cal A}_{\mu'}]=0$, we find $C_3=0$, i.e., finally
$$
H=\bar S^{-1}\left(\hat 1-\cos\sqrt{\bar S}\right),
$$
$$
\bar{\cal A}_{\mu'}
=\left\{\bar S^{-1}\left(\hat 1-\cos\sqrt{\bar S}\right)\right\}_{\mu'\nu'}
\bar{\cal L}^{\nu'}.
\eqno(2.103)
$$

Let us apply the formulas (2.95)--(2.97) and (2.103) to the case of the De Sitter
space  with the curvature
$$
R^\mu_{\ \alpha\nu\beta}
=\Lambda(\delta^\mu_\nu g_{\alpha\beta}-\delta^\mu_\beta g_{\alpha\nu}), \qquad
\Lambda={\rm const}.
\eqno(2.104)
$$
The matrix $\bar S$, (2.60), (2.61), and the vector $\bar{\cal L}_{\mu'}$, (2.82),
have in this case the form
$$
\bar S^{\mu'}_{\ \nu'}=2\Lambda\sigma{\Pi_{\bot}}^{\mu'}_{\ \nu'}, \qquad
\bar{\cal L}_{\mu'}=-{\cal R}_{\mu'\nu'}\sigma^{\nu'},
\eqno(2.105)
$$
where
$$
{\Pi_{\bot}}^{\mu'}_{\ \nu'}=\delta^{\mu'}_{\ \nu'}
-{\sigma^{\mu'}\sigma_{\nu'}\over 2\sigma}, \qquad
{\Pi_{\bot}}^2={\Pi_{\bot}},
\eqno(2.106)
$$
$$
{\Pi_{\bot}}^{\nu'}_{\ \mu'}\sigma^{\mu'}=
\sigma_{\nu'}{\Pi_{\bot}}^{\nu'}_{\ \mu'}=0,
$$
$\Pi_{\bot}$ being the projector on the hypersurface that is orthogonal to the
vector $\sigma^{\mu'}$.
Using (2.106) we obtain for an analytic function of the matrix $\bar S$
$$
f(\bar S)=\sum\limits_{k\ge 0}c_k\bar S^k
=f(0)(\hat 1-\Pi_\bot)+f(2\Lambda\sigma)\Pi_\bot.
\eqno(2.107)
$$
Thus the formal expressions (2.95)--(2.97) and (2.103) take the concrete form
$$
\bar\gamma^{\mu'}_{\ \nu'}=-\delta^{\mu'}_{\ \nu'}\Phi
+{\sigma^{\mu'}\sigma_{\nu'}\over 2\sigma}(\Phi-1),
$$
$$
\bar\eta^{\mu'}_{\ \nu'}=-\delta^{\mu'}_{\ \nu'}\Phi^{-1}
+{\sigma^{\mu'}\sigma_{\nu'}\over 2\sigma}(\Phi^{-1}-1),
\eqno(2.108)
$$
$$
X^{\mu'\nu'}=g^{\mu'\nu'}\Phi^{-2}
+{\sigma^{\mu'}\sigma^{\nu'}\over 2\sigma}(1-\Phi^{-2}),
$$
$$
H^{\mu'}_{\ \nu'}=\delta^{\mu'}_{\ \nu'}\Psi
+{\sigma^{\mu'}\sigma_{\nu'}\over 2\sigma}\left({1\over 2}-\Psi\right),
$$
where
$$
\Phi={\sin \sqrt{2\Lambda\sigma}\over\sqrt{2\Lambda\sigma}}, \qquad
\Psi={1-\cos \sqrt{2\Lambda\sigma}\over 2\Lambda\sigma}.
$$
Using eq. (2.105) one can neglect the longitudinal terms in the matrix $H$ (2.108)
and obtain the quantity $\bar{\cal A}_{\mu'}$:
$$
\bar{\cal A}_{\mu'}=-\Psi {\cal R}_{\mu'\nu'}\sigma^{\nu'}.
\eqno(2.109)
$$
In the case (2.90) it is transverse,
$$
\bar\nabla^{\mu'}\bar{\cal A}_{\mu'}=0.
\eqno(2.110)
$$

Let us stress that all formulas obtained in present section are valid for
spaces of any dimension and signature.

\section{Technique for the calculation of De~Witt coefficients}

Let us apply the method of covariant expansions developed above to the
calculation of the De~Witt coefficients $a_k$, i.e., the coefficients of the
asymptotic expansion of the transfer function (1.43). Below we follow our paper
[172].

{}From the recurrence relations (1.45) we obtain the formal solution
$$
a_k=\left(1+{1\over k}D\right)^{-1}F
\left(1+{1\over k-1}D\right)^{-1}F\cdots(1+D)^{-1}F,
\eqno(2.111)
$$
where the operator $D$ is defined by (2.4) and the operator $F$ has the form
$$
F={\cal P}^{-1}(\hat 1\Delta^{-1/2}\Square\Delta^{1/2}+Q){\cal P}.
\eqno(2.112)
$$

In the present section we develop a convenient covariant and effective method
that gives a practical meaning to the formal expression (2.111). It will
suffice, in particular, to calculate the coincidence limits of the De~Witt
coefficients $a_k$ and their derivatives.

First of all, we suppose that there exist finite coincidence limits of the
De~Witt coefficients
$$
[a_k]\equiv\lim_{x\to x'} a_k(x,x'),
\eqno(2.113)
$$
that do not depend on the way how the points $x$ and $x'$ approach each other,
i.e., the De~Witt coefficients $a_k(x,x')$ are analytical functions of the
coordinates of the point $x$ near the point $x'$. Otherwise, i.e., if the limit
(2.113) is singular, the operator $(1+{1\over k}D)$ does not have a
single--valued inverse operator, since there exist the eigenfunctions of this
operator with eigenvalues equal to zero
$$
D\sigma^{-n/2} \vert n>=0,\qquad
\left(1+{{1}\over {k}}D\right)\sigma^{-(k+n)/2}\vert n>=0.
\eqno(2.114)
$$

Therefore, the zeroth coefficient $a_0(x,x')$ is defined in general up to an
arbitrary function $f(\sigma^{\mu'}/\sqrt\sigma;x')$, and the inverse operator
$(1+{1\over k}D)^{-1}$ is defined up to an arbitrary function
$\sigma^{-k/2}$$f_k(\sigma^{\mu'}/\sqrt\sigma;x')$.

Using the covariant Taylor series (2.21) and (2.32) for the De~Witt coefficients
$$
a_k=\sum_{n\ge 0}\vert n><n\vert a_k>
\eqno(2.115)
$$
and defining the inverse operator $(1+{1\over k}D)^{-1}$ in form of the
eigenfunctions expansion (2.35),
$$
\left(1+{{1}\over {k}}D\right)^{-1}
=\sum_{n\ge 0}{k\over {k+n}}\vert n><n\vert,
\eqno(2.116)
$$
we obtain from (2.111)
\begin{eqnarray}\setcounter{equation}{117}
\lefteqn{<n\vert a_k>
=\sum_{n_1,\cdots,n_{k-1}\ge 0}{k\over k+n}\cdot
{k-1\over k-1+n_{k-1}}\cdots{1\over 1+n_1}}
\qquad\ \,\nonumber\\[10pt]
& &\times <n\vert F\vert n_{k-1}><n_{k-1}\vert F\vert n_{k-2}>\cdots
<n_1\vert F\vert 0>,
\end{eqnarray}
where
$$
<m\vert F\vert n>
=\left[\nabla_{(\mu_1}\cdots \nabla_{\mu_m)}F{{(-1)^n}\over {n!}}
\sigma^{\nu'_1}\cdots \sigma^{\nu'_n}\right]
\eqno(2.118)
$$
are the matrix elements (2.36) of the operator $F$, (2.112).

Since the operator $F$, (2.112), is a differential operator of second order, its
matrix elements $<m|F|n>$, (2.118), do not vanish only for $n\le m+2$. Therefore,
the sum (2.117) always contains a finite number of terms, i.e., the summation
over $n_i$ is limited from above
$$
n_1\ge 0, \qquad n_i\le n_{i+1}+2, \qquad (i=1, \dots, k-1; \ n_k\equiv n).
\eqno(2.119)
$$

Thus we reduced the problem of calculation of the De~Witt coefficients to the
calculation of the matrix elements (2.118) of the operator $F$, (2.112). For the
calculation of the matrix elements (2.118) it is convenient to write the
operator $F$, (2.112), in terms of the operators $\bar\nabla_{\mu'}$, (2.41).
Using (2.112), (2.41)--(2.46) and (2.51) we obtain
\begin{eqnarray}\setcounter{equation}{120}
F&=&{\cal P}^{-1}\Delta^{1/2}\bar\nabla_{\mu'}
\Delta^{-1}X^{\mu'\nu'}\bar\nabla_{\nu'}
\Delta^{1/2}{\cal P} + \bar Q
\nonumber\\[10pt]
&=&\left\{\hat 1(\bar\nabla_{\mu'}-\zeta_{\mu'})
+\bar{\cal A}_{\mu'}\right\}X^{\mu'\nu'}\left\{\hat
1(\bar\nabla_{\nu'}+\zeta_{\nu'})
+\bar{\cal A}_{\nu'}\right\}+\bar Q
\nonumber\\[10pt]
&=&\hat 1 X^{\mu'\nu'}\bar\nabla_{\mu'}\bar\nabla_{\nu'}
+Y^{\mu'}\bar\nabla_{\mu'}+Z,
\end{eqnarray}
where
$$
\bar Q={\cal P}^{-1}Q{\cal P}, \qquad
$$
$$
\zeta_{\mu'}=\bar\nabla_{\mu'}\zeta=\bar\nabla_{\mu'}\log\Delta^{1/2},
\eqno(2.121)
$$
$$
Y^{\mu'}=\hat 1\bar\nabla_{\nu'}X^{\nu'\mu'}+2X^{\mu'\nu'}
\bar{\cal A}_{\nu'},
\eqno(2.122)
$$
$$
Z=X^{\mu'\nu'}\left(\bar{\cal A}_{\mu'}\bar{\cal A}_{\nu'}
-\hat 1\zeta_{\mu'}\zeta_{\nu'}\right)
+\bar\nabla_{\nu'}\left\{X^{\mu'\nu'}\left(\hat 1\zeta_{\mu'}
+\bar{\cal A}_{\mu'}\right)\right\}
+\bar Q.
\eqno(2.123)
$$

Now one can easily calculate the matrix elements (2.118). Using (2.54), (2.5)
and (2.48) we obtain from (2.118)
$$
<m\vert F\vert m+2>
=\hat 1\delta^{(\nu_1\cdots\nu_m}_{\mu_1\cdots\mu_m}g^{\nu_{m+1}\nu_{m+2})},
$$
$$
<m|F|m+1>=0,
$$
\begin{eqnarray}\setcounter{equation}{124}
<m|F|n>&=&{m\choose n}
\delta^{\nu_1\cdots\nu_n}_{(\mu_1\cdots\mu_n}Z_{\mu_{n+1}\cdots\mu_m)}
-{m\choose n-1}\delta^{(\nu_1\cdots\nu_{n-1}}_{(\mu_1\cdots\mu_{n-1}}
Y^{\nu_n)}_{\ \ \ \mu_n\cdots \mu_m)}
\nonumber\\[10pt]
& &
+{m\choose n-2}\hat 1\delta^{(\nu_1\cdots\nu_{n-2}}_{(\mu_1\cdots\mu_{n-2}}
X^{\nu_{n-1}\nu_n)}_{\ \ \ \ \ \ \ \ \mu_{n-1}\cdots\mu_m)},
\end{eqnarray}
where
$$
Z_{\mu_1\cdots\mu_n}
=(-1)^n\left[\bar\nabla_{(\mu'_1}\cdots\bar\nabla_{\mu'_n)}Z\right],
$$
$$
Y^{\nu}_{\ \ \mu_1\cdots\mu_n}
=(-1)^n\left[\bar\nabla_{(\mu'_1}\cdots\bar\nabla_{\mu'_n)}Y^{\nu'}\right],
$$
$$
X^{\nu_1\nu_2}_{\ \ \ \ \mu_1\cdots\mu_n}
=(-1)^n\left[\bar\nabla_{(\mu'_1}\cdots\bar\nabla_{\mu'_n)}X^{\nu'_1\nu'_2}\right].
\eqno(2.125)
$$
In (2.124) it is meant that the binomial coefficient ${n\choose k}$ is equal to
zero if $k<0$ or $n<k$.

Thus, to calculate the matrix elements (2.118) it is sufficient to have the
coincidence limits of the symmetrized covariant derivatives (2.125) of the
coefficient functions $X^{\mu'\nu'}$, $Y^{\mu'}$ and $Z$, (2.122), (2.123), i.e.,
the coefficients of their Taylor expansions (2.48), that are expressed in terms
of the Taylor coefficients of the quantities $X^{\mu'\nu'}$, (2.76), $\bar{\cal
A}_{\mu'}$, (2.88), and $\zeta$, (2.78), (2.79), found in the Sect. 2.2. From the
dimensional arguments it is obvious that for $m=n$ the matrix elements
$<m|F|n>$, (2.124), (2.125), are expressed in terms of the curvature tensors
$R^\alpha_{\ \beta\gamma\delta}$, ${\cal R}_{\mu\nu}$ and the matrix $Q$; for
$m=n+1$ --- in terms of the quantities $\nabla R$, $\nabla{\cal R}$ and $\nabla
Q$; for $m=n+2$ --- in terms of the quantities of the form $R^2$, $\nabla\nabla
R$ etc.

In the calculation of the De~Witt coefficients by means of the matrix algorithm
(2.117) a ``diagrammatic'' technique, i.e., a graphic method for
enumerating the different terms of the sum (2.117), turns out to be very
convenient and pictorial.
  The matrix elements $<m|F|n>$, (2.118), are presented by some blocks with $m$
lines coming in from the left and $n$ lines going out to the right (Fig. 1),

\begin{center}
\unitlength1mm
\begin{picture}(41,10)
\put(-2.0,-1.0){$m\,\Biggl\{$}
\put(5.0,4.0){\line(1,0){12}}
\put(7.0,0.0){$\vdots$}
\put(5.0,-1.5){\line(1,0){10}}
\put(5.0,-4.0){\line(1,0){12}}
\put(20.0,0){\circle{10}}
\put(23.0,4.0){\line(1,0){12}}
\put(32.0,0.0){$\vdots$}
\put(25.0,-1.5){\line(1,0){10}}
\put(23.0,-4.0){\line(1,0){12}}
\put(36.0,-1.0){$\Biggl\}\,n$}
\end{picture}
\end{center}
\vglue4mm
\begin{center}
Fig. 1
\end{center}
and the product of the matrix elements $<m|F|k><k|F|n>$ --- by two blocks
connected by $k$ intermediate lines (Fig. 2),

\begin{center}
\unitlength1mm
\begin{picture}(61,10)
\put(-2.0,-1.0){$m\,\Biggl\{$}
\put(5.0,4.0){\line(1,0){12}}
\put(7.0,0.0){$\vdots$}
\put(5.0,-1.5){\line(1,0){10}}
\put(5.0,-4.0){\line(1,0){12}}
\put(20.0,0){\circle{10}}
\put(23.0,4.0){\line(1,0){14}}
\put(25.0,-1.5){\line(1,0){10}}
\put(23.0,-4.0){\line(1,0){14}}
\put(26.0,-1.0){$k\,\Biggl\{$}
\put(32.0,0.0){$\vdots$}
\put(40.0,0){\circle{10}}
\put(43.0,4.0){\line(1,0){12}}
\put(52.0,0.0){$\vdots$}
\put(45.0,-1.5){\line(1,0){10}}
\put(43.0,-4.0){\line(1,0){12}}
\put(56.0,-1.0){$\Biggl\}\,n$}
\end{picture}
\end{center}
\vglue4mm
\begin{center}
Fig. 2
\end{center}
that represents the contractions of corresponding tensor indices.

To obtain the coefficient $<n|a_k>$, (2.117), one should draw all possible
diagrams with $k$ blocks connected in all possible ways by any number of
intermediate lines. When doing this, one should keep in mind that the number of
the lines, going out of any block, cannot be greater than the number of the lines,
coming in, by more than two and by exactly one (see (2.124)). Then one should
sum up all diagrams with the weight determined for each diagram by the number
of intermediate lines from (2.117). Drawing of such diagrams is of no
difficulties. Therefore, the main problem is reduced to the calculation of some
standard blocks.

Let us note that the elaborated technique does not depend at all on the
dimension and on the signature of the manifold and allows to obtain results in
most general case. It is also very algorithmic and can be easily adopted for
analytic computer calculations.

\section{De~Witt coefficients $a_3$ and $a_4$ at coinciding points}

Using the developed technique one can calculate the coincidence limits (2.113)
of the De~Witt coefficients $[a_3]$ and $[a_4]$. The coefficients $[a_1]$ and
$[a_2]$, that determine the one--loop divergences (1.51), (1.53), have long been
calculated by De~Witt [26], Christensen [83, 84] and Gilkey [76]. The
coefficient $[a_3]$, that describes the vacuum polarization of massive quantum
fields in the lowest nonvanishing approximation $\sim 1/m^2$, (1.54), was
calculated in general form first by Gilkey [76]. The coefficient $[a_4]$ in
general form has not been calculated until now.

The diagrams for the De~Witt coefficients $[a_3]$ and $[a_4]$ have the form,
(2.117),

\begin{eqnarray}\setcounter{equation}{126}
[a_3]&=&\usebox{\fblocka}
+{1\over 3}\usebox{\fblockb}
+{2\over 4}\usebox{\fblockc}
\nonumber\\[10pt]
& &
+{2\over 4}\cdot{1\over 2}\usebox{\fblockd}
+{2\over 4}\cdot{1\over 3}\usebox{\fblocke}
+{2\over 4}\cdot{1\over 5}\usebox{\fblockf},
\nonumber\\[10pt]
& &
\end{eqnarray}

\begin{eqnarray}\setcounter{equation}{127}
[a_4]&=&\usebox{\blocka}
+{ \over }{1\over 3}\usebox{\blockb}
+{ \over }{2\over 4}\usebox{\blockc}
\nonumber\\[10pt]
& &+{ \over }{3\over 5}\usebox{\blockd}
+{1\over 5}\cdot{2 \over 4}\usebox{\blocke}
+{2\over 4}\cdot{1\over 2}\usebox{\blockf}
\nonumber\\[10pt]
& &+{3\over 5}\cdot{2\over 3}\usebox{\blockg}
+{2\over 4}\cdot{1\over 3}\usebox{\blockh}
\nonumber\\[10pt]
& &
+{3\over 5}\cdot{1\over 3}\usebox{\blockj}
+{3\over 5}\cdot{2\over 4}\usebox{\blockk}
\nonumber\\[10pt]
& &
+{3\over 5}\cdot{2\over 6}\usebox{\blockl}
+{3\over 5}\cdot{2\over 3}\cdot{1\over 2} \usebox{\blockm}
\nonumber\\[10pt]
& &
+{3\over 5}\cdot{2\over 4}\cdot{1\over 2} \usebox{\blockn}
+{3\over 5}\cdot{2\over 4}\cdot{1\over 3} \usebox{\blocko}
\nonumber\\[10pt]
& &
+{3\over 5}\cdot{2\over 3}\cdot{1\over 4} \usebox{\blockp}
+{3\over 5}\cdot{2\over 6}\cdot{1\over 2} \usebox{\blockq}
\nonumber\\[10pt]
& &
+{3\over 5}\cdot{2\over 4}\cdot{1\over 5} \usebox{\blockr}
+{3\over 5}\cdot{2\over 6}\cdot{1\over 3} \usebox{\blocks}
\nonumber\\[10pt]
& &
+{3\over 5}\cdot{2\over 6}\cdot{1\over 4} \usebox{\blockt}
+{3\over 5}\cdot{2\over 6}\cdot{1\over 5} \usebox{\blocku}
\nonumber\\[10pt]
& &
+{3\over 5}\cdot{2\over 6}\cdot{1\over 7} \usebox{\blockv}.
\end{eqnarray}

\par
\bigskip

Substituting the matrix elements (2.124) in (2.126) and (2.127)  we express the
coefficients $[a_3]$ and $[a_4]$ in terms of the quantities $X$, $Y$ and $Z$,
(2.125),
$$
[a_3]=P^3+{{1}\over {2}}[P,Z_{(2)}]_+ +{{1}\over {2}}B^\mu Z_\mu
+{{1}\over {10}}Z_{(4)}+\delta_3,
\eqno(2.128)
$$
\begin{eqnarray}\setcounter{equation}{129}
[a_4]&=&P^4+{{3}\over {5}}\left[P^2,Z_{(2)}\right]_+
+{{4}\over {5}}PZ_{(2)}P+{4\over 5}[P, B^\mu Z_\mu]_+
\nonumber\\[10pt]
& &
+{2\over 5}B^\mu PZ_\mu-{2\over 5}B_\mu Y^{\nu\mu}Z_\nu
+{{1}\over {3}}Z_{(2)}Z_{(2)}
+{2\over 5}B^\mu Z_{\mu(2)}
\nonumber\\[10pt]
& &
+{2\over 5}C^\mu Z_\mu
+{{1}\over {5}}\left[P,Z_{(4)}\right]_+
+{4\over 15}{\cal D}^{\mu\nu}Z_{\mu\nu}
+{{1}\over {35}}Z_{(6)}+\delta_4,
\end{eqnarray}
where
$$
\delta_3={1\over 6}U_1^{\mu\nu}Z_{\mu\nu},
\eqno(2.130)
$$
\begin{eqnarray}\setcounter{equation}{131}
\delta_4&=&{4\over 15}PU_1^{\mu\nu}Z_{\mu\nu}
+U_1^{\mu\nu}\Biggl\{{3\over 10}Z_{\mu\nu}P+{1\over 10}PZ_{\mu\nu}
+{3\over 10}Z_\mu Z_\nu
\nonumber\\[10pt]
& &
-{3\over 10}Y^\alpha_{\ \ \mu\nu}Z_\alpha
-{1\over 5}Y^\alpha_{\ \ \mu}Z_{\alpha\nu}
+{1\over 10}X^{\alpha\beta}_{\ \ \ \mu\nu}Z_{\alpha\beta}
+{7\over 50}Z_{\mu\nu(2)}\Biggr\}
\nonumber\\[10pt]
& &
+{1\over 5}U_2^{\mu\nu\lambda}Z_{\mu\nu\lambda}
+{4\over 25}U_3^{\mu\nu\alpha\beta}Z_{\mu\nu\alpha\beta}.
\end{eqnarray}
Here the following notation is introduced:
$$
P=[Z],
$$
$$
Z_{(2)}=g^{\mu_1\mu_2}Z_{\mu_1\mu_2},
\eqno(2.132)
$$
$$
Z_{(4)}=g^{\mu_1\mu_2}g^{\mu_3\mu_4}Z_{\mu_1\cdots\mu_4},
$$
$$
Z_{(6)}=g^{\mu_1\mu_2}g^{\mu_3\mu_4}g^{\mu_5\mu_6}Z_{\mu_1\cdots\mu_6},
\eqno(2.133)
$$
$$
Z_{\mu(2)}=g^{\mu_1\mu_2}Z_{\mu\mu_1\mu_2},
$$
$$
Z_{\mu\nu(2)}=g^{\mu_1\mu_2}Z_{\mu\nu\mu_1\mu_2},
\eqno(2.134)
$$
$$
B^\mu=Z^\mu-{1\over 2}g^{\mu_1\mu_2}Y^\mu_{\ \ \mu_1\mu_2},
\eqno(2.135)
$$
$$
C^\mu=Z^\mu_{\ (2)}-{1\over 4}g^{\mu_1\mu_2}g^{\mu_3\mu_4}Y^\mu_{\ \
\mu_1\cdots\mu_4},
\eqno(2.136)
$$
$$
{\cal D}^{\mu\nu}=Z^{\mu\nu}-g^{\mu_1\mu_2}Y^{(\mu\nu)}_{\ \ \ \ \mu_1\mu_2}
+{1\over 4}\hat 1g^{\mu_1\mu_2}g^{\mu_3\mu_4}X^{\mu\nu}_{\ \ \
\mu_1\cdots\mu_4},
\eqno(2.137)
$$
$$
U_1^{\mu\nu}=-2Y^{(\mu\nu)}
+\hat 1g^{\mu_1\mu_2}X^{\mu\nu}_{\ \ \ \mu_1\mu_2},
\eqno(2.138)
$$
$$
U_2^{\mu\nu\lambda}=-Y^{(\mu\nu\lambda)}
+\hat 1g^{\mu_1\mu_2}X^{(\mu\nu\lambda)}_{\ \ \ \ \ \mu_1\mu_2},
\eqno(2.139)
$$
$$
U_3^{\mu\nu\alpha\beta}=X^{(\mu\nu\alpha\beta)},
\eqno(2.140)
$$
and ${[A,B]}_+$ denotes the anti--commutator of the matrices $A$ and $B$.

Using the formulas (2.132)--(2.140), (2.125), (2.122) and (2.123) and the
quantities $X^{\mu'\nu'}$, $\bar{\cal A}_{\mu'}$ and $\zeta$, (2.70)--(2.89), and
omitting cumbersome computations we obtain
$$
P=Q+{{1}\over {6}}\hat 1 R,
\eqno(2.141)
$$
$$
Y_{\mu\nu}={\cal R}_{\mu\nu}+{1\over 3} \hat 1 R_{\mu\nu},
\eqno(2.142)
$$
$$
U_1^{\mu\nu}=U_3^{\mu\nu\alpha\beta}=0,
\eqno(2.143)
$$
$$
Z_\mu=\nabla_\mu P-{1\over 3}J_\mu,
\eqno(2.144)
$$
$$
B_\mu=\nabla_\mu P+{1\over 3}J_\mu,
\eqno(2.145)
$$
$$
U_2^{\mu\nu\lambda}=0,
\eqno(2.146)
$$
$$
Z_{(2)}=\Square\left(Q+{1\over 5}\hat 1 R\right)
+{{1}\over {30}}\hat 1\left(R_{\mu\nu\alpha\beta}R^{\mu\nu\alpha\beta}
-R_{\mu\nu}R^{\mu\nu}\right)
+{1\over 2}{\cal R}_{\mu\nu}{\cal R}^{\mu\nu },
\eqno(2.147)
$$
$$
Z_{\mu\nu}=W_{\mu\nu}-{1\over 2}\nabla_{(\mu}J_{\nu)},
\eqno(2.148)
$$
$$
{\cal D}_{\mu\nu}=W_{\mu\nu}+{1\over 2}\nabla_{(\mu}J_{\nu)},
\eqno(2.149)
$$
$$
Z_{\mu(2)}=V_\mu+G_\mu,
\eqno(2.150)
$$
$$
C_{\mu}=V_\mu-G_\mu,
\eqno(2.151)
$$
where
$$
J_\mu=\nabla_\alpha{\cal R}^\alpha_{\ \mu},
\eqno(2.152)
$$
\begin{eqnarray}\setcounter{equation}{153}
W_{\mu\nu}&=&\nabla_{(\mu}\nabla_{\nu)}\left(Q+{3\over 20}\hat 1R\right)
+\hat 1\Biggl\{{1\over 20}\Square R_{\mu\nu}
-{{1}\over {15}}R_{\mu\alpha}R^\alpha_{\ \nu}
\nonumber\\[10pt]
& &
+{{1}\over {30}}R_{\mu\alpha\beta\gamma}R_\nu^{\ \ \alpha\beta\gamma}
+{1\over 30}R_{\alpha\beta}R^{\alpha\ \beta}_{\ \mu\ \nu}\Biggr\}
+{{1}\over {2}}{\cal R}_{\alpha(\mu}{\cal R}^\alpha_{\ \nu)},
\end{eqnarray}
\begin{eqnarray}\setcounter{equation}{154}
V_\mu&=&Q_{\mu(2)}
+{{1}\over {2}}[{\cal R}_{\alpha\beta},\nabla_\mu{\cal R}^{\alpha\beta}]_+
-{{1}\over {3}}[J^\nu,{\cal R}_{\nu\mu}]_+
\nonumber\\[10pt]
& &
+\hat 1\Biggl\{{{1}\over {5}}\nabla_\mu\Square R
+{{1}\over {9}}R^\nu_\mu\nabla_\nu R
+{{1}\over {15}}R_{\alpha\beta\gamma\delta}\nabla_\mu
R^{\alpha\beta\gamma\delta}
\nonumber\\[10pt]
& &
-{{1}\over {15}}R_{\alpha\beta}\nabla_\mu R^{\alpha\beta}\Biggr\},
\end{eqnarray}
\begin{eqnarray}\setcounter{equation}{155}
Q_{\mu(2)}&=&g^{\mu_1\mu_2}\nabla_{(\mu}\nabla_{\mu_1}\nabla_{\mu_2)}Q
\nonumber\\[10pt]
& &
=\nabla_\mu\Square Q
+[{\cal R}_{\nu\mu},\nabla^\nu Q]
+{{1}\over {3}}[J_\mu,Q]
+{{2}\over {3}}R^\nu_\mu\nabla_\nu Q,
\end{eqnarray}
\begin{eqnarray}\setcounter{equation}{156}
G_\mu&=&-{1\over 5}\Square J_\mu
-{{2}\over {15}}[{\cal R}_{\alpha\mu},J^\alpha]
-{{1}\over {10}}[{\cal R}_{\alpha\beta},\nabla_\mu{\cal R}^{\alpha\beta}]
\nonumber\\[10pt]
& &
-{{2}\over {15}}R^{\alpha\beta}\nabla_\alpha{\cal R}_{\beta\mu}
+{{2}\over {15}}R_{\mu\alpha\beta\gamma}\nabla^\alpha{\cal R}^{\beta\gamma}
-{{7}\over {45}}R_{\mu\alpha}J^\alpha
\nonumber\\[10pt]
& &
+{2\over 5}\nabla_\alpha R_{\beta\mu}{\cal R}^{\beta\alpha}
-{{1}\over {15}}\nabla^\alpha R{\cal R}_{\alpha\mu}.
\end{eqnarray}

\noindent
The result for the coefficient $Z_{(4)}$ has more complicated form
\begin{eqnarray}\setcounter{equation}{157}
Z_{(4)}&=&Q_{(4)}
+2[{\cal R}^{\mu\nu},\nabla_\mu J_\nu]_+
+{{8}\over {9}}J_\mu J^\mu
+{{4}\over {3}}\nabla_\mu{\cal R}_{\alpha\beta}
\nabla^\mu{\cal R}^{\alpha\beta}
\nonumber\\[10pt]
& &
+6{\cal R}_{\mu\nu}{\cal R}^\nu_{\ \gamma}{\cal R}^{\gamma\mu}
+{{10}\over {3}}R^{\alpha\beta}{\cal R}^\mu_{\ \alpha}{\cal R}_{\mu\beta}
-R^{\mu\nu\alpha\beta}{\cal R}_{\mu\nu}{\cal R}_{\alpha\beta}
\nonumber\\[10pt]
& &
+\hat 1 \biggl\{{{3}\over {14}}\Square^2R
+{{1}\over {7}}R^{\mu\nu}\nabla_\mu\nabla_\nu R
-{{2}\over {21}}R^{\mu\nu}\Square R_{\mu\nu}
\nonumber\\[10pt]
& &
+{{4}\over {7}}R^{\alpha\ \beta}_{\ \mu\ \nu}
\nabla_\alpha\nabla_\beta R^{\mu\nu}
+{{4}\over {63}}\nabla_\mu R\nabla^\mu R
-{{1}\over {42}}\nabla_\mu R_{\alpha\beta}\nabla^\mu R^{\alpha\beta}
\nonumber\\[10pt]
& &
-{{1}\over {21}}\nabla_\mu R_{\alpha\beta}\nabla^\alpha R^{\beta\mu}
+{{3}\over {28}}\nabla_\mu R_{\alpha\beta\gamma\delta}
\nabla^\mu R^{\alpha\beta\gamma\delta}
+{{2}\over {189}}R_\beta^{\alpha}R_\gamma^{\beta}R_\alpha^{\gamma}
\nonumber\\[10pt]
& &
-{{2}\over {63}}R_{\alpha\beta}R^{\mu\nu}R^{\alpha\ \beta}_{\ \mu\ \nu}
+{{2}\over {9}}R_{\alpha\beta}R^\alpha_{\ \mu\nu\lambda}R^{\beta\mu\nu\lambda}
\nonumber\\[10pt]
& &
-{{16}\over {189}}R_{\alpha\beta}^{\ \ \ \mu\nu}
R_{\mu\nu}^{\ \ \ \sigma\rho}R_{\sigma\rho}^{\ \ \ \alpha\beta}
-{{88}\over {189}}R^{\alpha\ \beta}_{\ \mu\ \nu}
R^{\mu\ \nu}_{\ \sigma\ \rho}R^{\sigma\ \rho}_{\ \alpha\ \beta}\biggr\},
\end{eqnarray}
where
\begin{eqnarray}\setcounter{equation}{158}
Q_{(4)}&=&g^{\mu_1\mu_2}g^{\mu_3\mu_4}\nabla_{(\mu_1}\cdots\nabla_{\mu_4)}Q
\nonumber\\[10pt]
&=&
\Square^2 Q-{1\over 2}[{\cal R}^{\mu\nu},[{\cal R}_{\mu\nu},Q]]
-{{2}\over {3}}[J^\mu,\nabla_\mu Q]
\nonumber\\[10pt]
& &
+{{2}\over {3}}R^{\mu\nu}\nabla_\mu\nabla_\nu Q
+{{1}\over {3}}\nabla_\mu R\nabla^\mu Q .
\end{eqnarray}

{}From the fact that the quantities $U_1^{\mu\nu}$, $U_2^{\mu\nu\lambda}$
and $U_3^{\mu\nu\lambda\sigma}$, (2.138)--(2.140),  are equal to zero, (2.143), (2.146),
it follows that the quantities $\delta_3$ and $\delta_4$, (2.130), (2.131), are
equal to zero too:
$$
\delta_3=\delta_4=0.
\eqno(2.159)
$$
To calculate the De~Witt coefficient $[a_3]$, (2.128), it is sufficient to have
the formulas listed above. This coefficient is presented explicitly in the
paper [76]. (Let us note, to avoid misunderstanding, that our De~Witt
coefficients $a_k$, (1.43), differ from the coefficients $\tilde a_k$ 
used by the other
authors [26, 83, 84, 76] by a factor: $a_k=k!\tilde a_k$.)

Let us calculate the De~Witt coefficient $A_3$, (1.55), that determines 
the renormalized one--loop effective action (1.54) in the
four--dimensional physical space--time in the lowest 
nonvanishing approximation $\sim1/m^2$ [57, 58]. 
By integrating by parts and omitting the total derivatives we obtain from (2.128),
(2.141), (2.144), (2.145), (2.147) and (2.157)--(2.159)
\begin{eqnarray}\setcounter{equation}{160}
\lefteqn{A_3=\int d^n x\,g^{1/2}{\rm str}\Biggl\{P^3
+{{1}\over {30}}P\Bigl(R_{\mu\nu\alpha\beta}R^{\mu\nu\alpha\beta}
-R_{\mu\nu}R^{\mu\nu}+\Square R\Bigr)
}\qquad\qquad
\nonumber\\[10pt]
& &
+{{1}\over {2}}P{\cal R}_{\mu\nu}{\cal R}^{\mu\nu}
+{{1}\over {2}} P\Square P-{{1}\over {10}}J_\mu J^\mu
\nonumber\\[10pt]
& &
+{{1}\over {30}}
\left(2{\cal R}^\mu_{\ \nu}{\cal R}^\nu_{\ \alpha}{\cal R}^\alpha_{\ \mu}
-2R^\mu_\nu {\cal R}_{\mu\alpha}{\cal R}^{\alpha\nu}
+R^{\mu\nu\alpha\beta}{\cal R}_{\mu\nu}{\cal R}_{\alpha\beta}\right)
\nonumber\\[10pt]
& &
+\hat 1\Biggl[-{{1}\over {630}}R\Square R
+{{1}\over {140}}R_{\mu\nu}\Square R^{\mu\nu}
+{{1}\over {7560}}\bigg(
-64R^\mu_\nu R^\nu_\lambda R^\lambda_\mu
\nonumber\\[10pt]
& &
+48R^{\mu\nu}R_{\alpha\beta}R^{\alpha\ \beta}_{\ \mu\ \nu}
+6R_{\mu\nu}R^\mu_{\ \alpha\beta\gamma}R^{\nu\alpha\beta\gamma}
\nonumber\\[10pt]
& &
+17R_{\mu\nu}^{\ \ \ \alpha\beta}
R_{\alpha\beta}^{\ \ \ \sigma\rho}R_{\sigma\rho}^{\ \ \ \mu\nu}
-28R^{\mu \ \nu}_{\ \alpha \ \beta}R^{\alpha\ \beta}_{\ \sigma\ \rho}
R^{\sigma\ \rho}_{\ \mu \ \nu}\bigg)\Biggr]\Biggr\}.
\end{eqnarray}
The formula (2.160) is valid for any dimension of the space and for any fields.

To obtain the explicit expression for the coefficient $[a_4]$ one has to
substitute (2.141)--(2.159) in (2.129) as well as to calculate the quantity
$Z_{(6)}$, (2.133). To write down this quantity in a compact way we define the
following tensors constructed from the covariant derivatives of the curvature
tensor:
$$
I^{\alpha\beta}_{\ \ \ \gamma\mu_1\cdots\mu_n}=
\nabla_{(\mu_1}\cdots\nabla_{\mu_{n-1}}R^{\alpha\ \ \beta}_{\ |\gamma|\
\mu_n)},
$$
$$
K^{\alpha\beta}_{\ \ \ \mu_1\cdots\mu_n}=
\nabla_{(\mu_1}\cdots\nabla_{\mu_{n-2}}
R^{\alpha\ \ \beta}_{\ \mu_{n-1}\ \mu_n)},
$$
$$
L^{\alpha}_{\ \mu_1\cdots\mu_n}=
\nabla_{(\mu_1}\cdots\nabla_{\mu_{n-1}}R^{\alpha}_{\ \mu_n)},
\eqno(2.161)
$$
$$
M_{\mu_1\cdots\mu_n}=
\nabla_{(\mu_1}\cdots\nabla_{\mu_{n-2}}R_{\mu_{n-1}\mu_n)},
$$
$$
{\cal R}^\mu_{\ \mu_1\cdots\mu_n}
=\nabla_{(\mu_1}\cdots\nabla_{\mu_{n-1}}{\cal R}^\mu_{\ \mu_n)},
$$
and denote the contracted symmetrized covariant derivatives just by a
number in the brackets (analogously to (2.132)--(2.134). For example,
$$
I^{\alpha\beta}_{\ \ \ \gamma\mu(2)}
=g^{\mu_1\mu_2}I^{\alpha\beta}_{\ \ \ \gamma\mu\mu_1\mu_2},
$$
$$
K^{\alpha\beta}_{\ \ \  \mu\nu(4)}
=g^{\mu_1\mu_2}g^{\mu_3\mu_4}K^{\alpha\beta}_{\ \ \ \mu\nu\mu_1\cdots\mu_4},
$$
$$
L^{\alpha}_{\ \mu(4)}
=g^{\mu_1\mu_2}g^{\mu_3\mu_4}L^{\alpha}_{\ \mu\mu_1\cdots\mu_4},
\eqno(2.162)
$$
$$
M_{(8)}=g^{\mu_1\mu_2}\cdots g^{\mu_7\mu_8}M_{\mu_1\cdots\mu_8},
$$
$$
{\cal R}^\mu_{\ \nu(4)}
=g^{\mu_1\mu_2}g^{\mu_3\mu_4}{\cal R}^\mu_{\ \nu\mu_1\cdots\mu_4},
$$
etc.

In terms of introduced quantities, (2.161), and the notation (2.162) the quantity
$Z_{(6)}$, (2.133), takes the form
$$
Z_{(6)}=Z^M_{(6)}+\hat 1 Z^S_{(6)}.
\eqno(2.163)
$$
Here $Z^M_{(6)}$ is the matrix contribution
\begin{eqnarray}\setcounter{equation}{164}
Z^M_{(6)}&=&Q_{(6)}
+{{5}\over {2}}[{\cal R}^{\mu\nu},{\cal R}_{\mu\nu(4)}]_+
+{{32}\over {5}}[{\cal R}^{\mu\nu\alpha},{\cal R}_{\mu\nu\alpha(2)}]_+
\nonumber\\[10pt]
& &
-{{8}\over {5}}[J^\mu,{\cal R}_{\mu(4)}]_+
+{{9}\over {2}}{\cal R}^{\mu\nu\alpha\beta}{\cal R}_{\mu\nu\alpha\beta}
+{{27}\over {4}}{\cal R}^{\mu\nu}_{\ \ (2)}{\cal R}_{\mu\nu(2)}
\nonumber\\[10pt]
& &
+{{5}\over {4}}R^\mu_\nu[{\cal R}^{\nu\alpha},{\cal R}_{\mu\alpha(2)}]_+
+{{5}\over {2}}R_\mu^{\ \ \nu\alpha\beta}
[{\cal R}^{\mu\gamma},{\cal R}_{\alpha\beta\nu\gamma}]_+
\nonumber\\[10pt]
& &
+{{15}\over {8}}R^{\mu\nu\alpha\beta}
[{\cal R}_{\mu\nu},{\cal R}_{\alpha\beta(2)}]_+
+{{44}\over {15}}\nabla_\mu R_{\alpha\nu}[{\cal R}^{\mu\alpha},J^\nu]_+
\nonumber\\[10pt]
& &
+{{22}\over {5}}K^{\mu\nu\alpha\beta\gamma}
[{\cal R}_{\mu\gamma},{\cal R}_{\nu\alpha\beta}]_+
+{{22}\over {5}}K_{\mu\nu\alpha(2)}
[{\cal R}^{\mu\gamma,},{\cal R}^{\nu\alpha}_{\ \ \ \gamma}]_+
\nonumber\\[10pt]
& &
+{{64}\over {45}}R^\mu_{\ \nu}
{\cal R}_{\mu\alpha\beta}{\cal R}^{\nu\alpha\beta}
-{{16}\over {15}}R^{\mu\nu\alpha\beta}
[\nabla_\beta{\cal R}_{\mu\nu},J_\alpha]_+
\nonumber\\[10pt]
& &
+{{256}\over {45}}R_{\mu(\alpha|\nu|\beta)}
{\cal R}^{\mu\alpha}_{\ \ \ \gamma}{\cal R}^{\nu\beta\gamma}
+{{32}\over {45}}R^{\mu\nu}J_\mu J_\nu
\nonumber\\[10pt]
& &
+\left({{6}\over {5}}K_{\mu\nu(4)}
+{{17}\over {40}}R_{\mu\alpha\beta\gamma}R_\nu^{\ \ \alpha\beta\gamma}
+{{17}\over {60}}R_{\mu\alpha}R_\nu^\alpha\right)
{\cal R}^{\mu\sigma}{\cal R}^\nu_{\ \sigma}
\nonumber\\[10pt]
& &
+\biggl({{24}\over {5}}K_{\mu\nu\alpha\beta(2)}
+{{17}\over {40}}R_{\mu\gamma}R^\gamma_{\ \beta\nu\alpha}
+{{17}\over {40}}R_{\nu\gamma}R^\gamma_{\ \alpha\mu\beta}
+{{17}\over {30}}R_{\mu\nu\sigma\rho}R_{\beta\alpha}^{\ \ \ \ \sigma\rho}
\nonumber\\[10pt]
& &
+{{17}\over {60}}R_{\mu\sigma\alpha\rho}R^{\sigma\ \rho}_{\ \nu\ \beta}
+{{51}\over {80}}R_{\mu\beta\sigma\rho}R_{\nu\alpha}^{\ \ \ \sigma\rho}\biggr)
{\cal R}^{\mu\beta}{\cal R}^{\nu\alpha},
\end{eqnarray}
where
$$
Q_{(6)}=g^{\mu_1\mu_2}g^{\mu_3\mu_4}g^{\mu_5\mu_6}
\nabla_{(\mu_1}\cdots\nabla_{\mu_6)}Q,
$$
and $Z^S_{(6)}$ is the scalar contribution
\begin{eqnarray}\setcounter{equation}{165}
Z^S_{(6)}&=&{{7}\over {18}}M_{(8)}
+R^{\mu\nu}\left({{5}\over {18}}K_{\mu\nu(6)}
-{{5}\over {6}}L_{\mu\nu(4)}\right)
\nonumber\\[10pt]
& &
+{{20}\over {21}}R^{\mu\alpha\nu\beta}K_{\mu\nu\alpha\beta(4)}
-{{2}\over {5}}\nabla^\mu RL_{\mu(4)}
+\bigg({{5}\over {3}}K^{\mu\nu\alpha}_{\ \ \ \ (2)}
\nonumber\\[10pt]
& &
+M^{\mu\nu\alpha}\bigg)K_{\mu\nu\alpha(4)}
-{12\over 5}M^{\mu\nu\alpha}L_{\mu\nu\alpha(2)}
+{{20}\over {9}}K^{\mu\nu\alpha\beta\gamma}K_{\mu\nu\alpha\beta\gamma(2)}
\nonumber\\[10pt]
& &
+{{6}\over {25}}K^{\mu\nu}_{\ \ (4)}K_{\mu\nu(4)}
-{{27}\over {25}}M^{\mu\nu}_{\ \ \ (2)}M_{\mu\nu(2)}
+{{48}\over {25}}K^{\mu\nu\alpha\beta}_{\ \ \ \ \ (2)}K_{\mu\nu\alpha\beta(2)}
\nonumber\\[10pt]
& &
-{{18}\over {25}}M^{\mu\nu\alpha\beta}M_{\mu\nu\alpha\beta}
+{16\over 25}K^{\mu\nu\alpha\beta\gamma\rho}K_{\mu\nu\alpha\beta\gamma\rho}
\nonumber\\[10pt]
& &
+\biggl({{101}\over {450}}R^\mu_\alpha R^{\nu\alpha}
+{{68}\over {525}}R^{\mu}_{\ \alpha\beta\gamma}R^{\nu\alpha\beta\gamma}
-{{1}\over {5}}R^{\alpha\beta}
R^{\mu\ \nu}_{\ \alpha\ \beta}\biggr)K_{\mu\nu(4)}
\nonumber\\[10pt]
& &
+\biggl(-{{2}\over {5}}R^\mu_\alpha R^{\nu\alpha}
-{{6}\over {25}}R^{\mu}_{\ \alpha\beta\gamma}R^{\nu\alpha\beta\gamma}
+{{6}\over {25}}R^{\alpha\beta}
R^{\mu\ \nu}_{\ \alpha\ \beta}\biggr)M_{\mu\nu(2)}
\nonumber\\[10pt]
& &
+\left(-{1\over 6}R^{\mu\nu}R^{\alpha\beta}
-{1\over 3}R^{\lambda\alpha\nu\beta}\right)I_{(\alpha\beta)\mu\nu(2)}
-{1\over 3}R^{\mu\nu}
R^{\alpha\beta\gamma\delta}I_{\alpha\gamma\mu\nu\beta\delta}
\nonumber\\[10pt]
& &
+\left(-{2\over 5}R^{\mu\nu\alpha\beta}R^{\sigma\ \rho}_{\ \mu\ \alpha}
-R^\beta_\lambda R^{\lambda\sigma\nu\rho}\right)L_{\nu\beta\sigma\rho}
\nonumber\\[10pt]
& &
+\biggl(-{{6}\over {5}}R^\alpha_\rho R^{\rho\mu\beta\nu}
+{{2588}\over {1575}}R^\mu_\rho R^{\rho\alpha\nu\beta}
+{{4}\over {25}}R^{\mu\sigma\nu\rho}R^{\alpha\ \beta}_{\ \sigma\ \rho}
\nonumber\\[10pt]
& &
+{{1048}\over {1575}}R^{\mu\sigma\alpha\rho}R^{\nu\ \beta}_{\ \sigma\ \rho}
+{{962}\over {1575}}R^{\mu\alpha}_{\ \ \ \sigma\rho}
R^{\nu\beta\sigma\rho}\biggr)K_{\mu\nu\alpha\beta(2)}
\nonumber\\[10pt]
& &
+{{1088}\over {1575}}R^{\mu\alpha\beta\gamma}
R^{\nu\sigma\ \rho}_{\ \ \beta}K_{\mu\nu\alpha\gamma\sigma\rho}
+R^{\mu\nu}\biggl\{
-{4\over 45}\nabla^\alpha R I_{\mu\nu\alpha(2)}
\nonumber\\[10pt]
& &
+\left({7\over 10}M_{\mu\alpha\beta}
+{5\over 6}K_{\mu\alpha\beta(2)}\right)K_{\nu\ \ (2)}^{\ \ \alpha\beta}
-{1\over 3}K^{\sigma\rho}_{\ \ \ \mu(2)}I_{\sigma\rho\nu(2)}
\nonumber\\[10pt]
& &
-{2\over 3}I_{\sigma\rho\mu\nu\alpha}K^{\sigma\rho}_{\ \ \ \alpha(2)}
+{5\over 9}K_{\mu\alpha\beta\lambda\gamma}K_\nu^{\ \alpha\beta\lambda\gamma}
-{2\over 3}I_{\sigma\rho\mu\alpha\beta}K^{\sigma\rho\ \ \alpha\beta}_{\ \ \ \nu}
\nonumber\\[10pt]
& &
-{4\over 15}K_{\mu\nu\lambda\sigma\rho}M^{\lambda\sigma\rho}\biggr\}
+R^{\mu \ \ \nu}_{\ (\alpha\ \beta)}\biggl\{\bigg({7\over 5}
K_{\mu\sigma\rho}^{\ \ \ \ \alpha\beta}
-{9\over 10}K^{\alpha\beta}_{\ \ \ \mu\sigma\rho}
\nonumber\\[10pt]
& &
-{16\over 15}I^{\alpha\beta}_{\ \ \ \sigma\rho\mu}\bigg)M_\nu^{\ \sigma\rho}
+\biggl(-{7\over 10}K_{\rho\ \ \ (2)}^{\ \ \alpha\beta}
-{9\over 10}K_{\ \ \ \rho(2)}^{\alpha\beta}
\nonumber\\[10pt]
& &
-{4\over 15}I_{\ \ \ \rho(2)}^{\alpha\beta}\biggr)M_{\mu\nu}^{\ \ \ \rho}
-{8\over 45}\nabla^\rho RI_{\mu\nu\rho}^{\ \ \ \ \alpha\beta}
-{3\over 10}\nabla^\alpha RK_{\mu\nu \ \ (2)}^{\ \ \ \beta}
\nonumber\\[10pt]
& &
+{20\over 21}K_{\mu\rho\ \ (2)}^{\ \ \ \alpha}K_{\nu\ \ \ (2)}^{\ \ \rho\beta}
+{40\over 21}K_\mu^{\ \ \rho\alpha\beta\sigma}K_{\nu\rho\sigma(2)}
+{40\over 21}K_{\mu\rho}^{\ \ \ \alpha\lambda\gamma}
K_{\nu\ \ \ \lambda\gamma}^{\ \ \rho\beta}\biggr\}
\nonumber\\[10pt]
& &
-{{7}\over {450}}R^\mu_\nu R^\nu_\alpha R^\alpha_\beta R^\beta_\mu
+{{1}\over {90}}R_{\mu\alpha}R^{\alpha}_{\nu}
R^{\mu\ \nu}_{\ \sigma \ \rho}R^{\sigma\rho}
\nonumber\\[10pt]
& &
+{{817}\over {6300}}R_{\mu\alpha}R_{\nu}^{\alpha}
R^{\mu}_{\ \lambda\sigma\rho}R^{\nu\lambda\sigma\rho}
+{{391}\over {4725}}R_\mu^\nu R_\alpha^\beta
R^{\mu\ \alpha}_{\ \,\sigma\ \rho}R^{\sigma\ \rho}_{\ \nu\ \beta}
\nonumber\\[10pt]
& &
-{{2243}\over {9450}}R_{\mu\nu}R_{\alpha\beta}
R^{\mu\alpha}_{\ \ \ \sigma\rho}R^{\nu\beta\sigma\rho}
-{{1}\over {75}}R_{\mu\nu}R^{\alpha\beta}
R^{\mu\ \nu}_{\ \sigma\ \rho}R^{\sigma\ \rho}_{\ \alpha\ \beta}
\nonumber\\[10pt]
& &
-{16\over 4725}R^\nu_\mu R^{\mu\ \alpha}_{\ \sigma\ \rho}
R^{\sigma\ \rho}_{\ \lambda\ \gamma}R^{\lambda\ \gamma}_{\ \nu\ \alpha}
-{7\over 300}R_{\mu\nu}R^{\mu\ \nu}_{\ \alpha\ \beta}
R^\alpha_{\ \lambda\sigma\rho}R^{\beta\lambda\sigma\rho}
\nonumber\\[10pt]
& &
+{{8}\over {675}}R^\nu_\mu R^{\mu\lambda}_{\ \ \ \alpha\beta}
R^{\alpha\beta}_{\ \ \ \sigma\rho}R^{\sigma\rho}_{\ \ \ \nu\lambda}
+{{247}\over {9450}}R^{\mu\nu}_{\ \ \ \alpha\beta}
R^{\alpha\beta}_{\ \ \ \lambda\gamma}
R^{\lambda\gamma}_{\ \ \ \sigma\rho}R^{\sigma\rho}_{\ \ \ \mu\nu}
\nonumber\\[10pt]
& &
-{{32}\over {4725}}R^{\mu\ \nu}_{\ \alpha\ \beta}
R_{\ \lambda\ \gamma}^{\alpha\ \beta}
R^{\lambda\ \gamma}_{\ \rho\ \,\sigma}R_{\ \mu\ \nu}^{\rho\ \sigma}
+{{1}\over {105}}R^\mu_{\ \alpha\beta\gamma}R^{\nu\alpha\beta\gamma}
R_{\mu\lambda\sigma\rho}R_\nu^{\ \lambda\sigma\rho}
\nonumber\\[10pt]
& &
+{32\over 4725} R_{\mu\alpha\nu\beta}R^{\alpha\ \beta}_{\ \lambda\ \gamma}
R^{\mu\ \lambda}_{\ \,\sigma\ \rho}R^{\nu\sigma\gamma\rho}
-{232\over 4725}R^{\mu\nu}_{\ \ \sigma\rho}
R^{\alpha\beta\sigma\rho}R_{\mu\lambda\alpha\gamma}
R^{\lambda\ \gamma}_{\ \,\nu\ \beta}
\nonumber\\[10pt]
& &
+{299\over 4725}R^{\mu\nu}_{\ \ \sigma\rho}
R^{\alpha\beta\sigma\rho}R_{\mu\alpha\lambda\gamma}
R^{\ \ \ \lambda\gamma}_{\nu\beta}.
\end{eqnarray}

Further transformations of the expression for the quantity $Z_{(6)}$,
(2.163)--(2.165), in general
form appear to be pointless since they are very cumbersome and there are very
many independent invariants constructed from the curvatures and their covariant
derivatives. The set of invariants and the form of the result should be chosen
in accordance with the specific character of the considered problem. That is
why we present the result for the De~Witt coefficient $[a_4]$ by the set of
formulas (2.129), (2.141)--(2.159) and (2.161)--(2.165).

\section{Effective action of massive scalar, spinor and vector fields in
external gravitational field}

Let us illustrate the elaborated methods for the calculation of the one--loop
effective action on the example of real scalar, spinor and vector massive
quantum matter fields on a classical gravitational background in the
four--dimensional physical space--time. In this section we follow our papers
[171, 172].

The operator (1.22), (1.29) in this case has the form
\begin{equation}\setcounter{equation}{166}
\hat\Delta=\left\{\begin{array}{ll}
-\Square+\xi R+m^2, & (j=0),\\
\gamma^\mu\nabla_\mu+m, & (j=1/2),\\
-\delta d+m^2, & (j=1),
\end{array}
\right.
\end{equation}
where $m$ and $j$ are the mass and the spin of the field, $\xi$ is the coupling
constant of the scalar field with the gravitational field, $\gamma^\mu$ are the
Dirac matrices, $\gamma^{(\mu}\gamma^{\nu)}=\hat 1 g^{\mu\nu}$, $d$ is the
exterior derivative and $\delta$ is the operator of co--differentiation on
forms:
$(\delta\varphi)_{[\mu_1\cdots\mu_k]}
=\nabla^\mu\varphi_{[\mu\mu_1\cdots\mu_k]}$.

The commutator of covariant derivatives (3.13) (2.14) has the form
\begin{equation}\setcounter{equation}{167}
{\cal R}_{\mu\nu}=\left\{\begin{array}{ll}
0\\
{1\over 4}\gamma^\alpha\gamma^\beta R_{\alpha\beta\mu\nu}\ .\\
R^\alpha_{\ \beta\mu\nu}
\end{array}
\right.
\end{equation}

Using the equations
$$
\gamma^\mu\gamma^\nu\nabla_\mu\nabla_\nu
=\hat 1 \left(\Square-{1\over 4} R\right),
$$
$$
\left(d\delta+\delta d\right)\varphi^\mu
=\left(\delta^\mu_\nu\Square-R^\mu_\nu\right)\varphi^\nu,
$$
$$
\delta^2=\delta^2=0,
\eqno(2.168)
$$
one can express the Green functions of the operator $\hat \Delta$, (2.166), in
terms of the Green functions of the minimal operator, (1.30),
$$
\hat\Delta^{-1}=\Pi\left(-\Square-Q+m^2\right)^{-1},
\eqno(2.169)
$$
where
\begin{equation}\setcounter{equation}{170}
\Pi=\left\{\begin{array}{ll}
1\\
m-\gamma^\mu\nabla_\mu\ ,\\
1-{1\over m^2}d\delta
\end{array}
\right.
\qquad
Q=\left\{\begin{array}{ll}
-\xi R\\
-{1\over 4} \hat 1 R\ .\\
-R^\alpha_{\ \beta}
\end{array}
\right.
\end{equation}

Using the Schwinger--De~Witt representation,  (1.33),
$$
\Gamma_{(1)}=-{1\over 2 i}\log{\rm sdet}\hat\Delta
={1\over 2 i}\int\limits^\infty_0 {d s\over s}{\rm sdet}\exp(is\hat \Delta)
\eqno(2.171)
$$
and the equations
$$
{\rm tr}\exp(is\gamma^\mu\nabla_\mu)
={\rm tr}\exp(-is\gamma^\mu\nabla_\mu),
$$
$$
\exp(is \delta d)=\exp\{is(d\delta+\delta d)\}-\exp(is d\delta)+1,
\eqno(2.172)
$$
$$
{\rm tr}\exp(is d\delta)\Big\vert_{j=1}
={\rm tr}\exp(is\Square)\Big\vert_{j=0},
$$
we obtain up to non--essential infinite contributions $\sim \delta(0)$,
$$
\log\det(\gamma^\mu\nabla_\mu+m)
={1\over 2}\log\det\left\{\hat 1\left(-\Square+{1\over 4}R+m^2\right)\right\},
$$
\begin{equation}\setcounter{equation}{173}
\log\det(-\delta d+m^2)
=\log\det\left(-\Square \delta^\beta_\alpha
+R^\beta_\alpha+m^2\delta^\beta_\alpha\right)
-\log\det(-\Square+m^2)\Big\vert_{j=0}.
\end{equation}
\vskip10pt

Thus we have reduced the functional determinants of the operators (2.166) to the
functional determinants of the minimal operators of the form (1.30). By making
use of the formulas (2.171)--(2.173), (1.50) and (1.54) we obtain the asymptotic
expansion of the one--loop effective action in the inverse powers of the mass
$$
\Gamma_{(1){\rm ren}}={1\over 2(4\pi)^2}\sum\limits_{k\ge 3}
{B_k\over k(k-1)(k-2)m^{2(k-2)}},
\eqno(2.174)
$$
where
\begin{equation}\setcounter{equation}{175}
B_k=\left\{\begin{array}{ll}
A_k\\
{1\over 2} A_k\\
A_k\Big\vert_{(j=1)}-A_k\Big\vert_{(j=0,\xi=0)}
\end{array}
\right.
\end{equation}
and the coefficients $A_k$ are given by (1.55).

Let us stress that the coefficients $A_k$, (1.55), for the spinor field contain a
factor $(-1)$ in addition to the usual trace over the spinor indices according
to the definition of the supertrace (1.25).

Substituting the matrices ${\cal R}_{\mu\nu}$, (2.167), and $Q$, (2.170), in
(2.160) and using (2.175) we obtain the first coefficient, $B_3$, in the
asymptotic expansion (2.174)
\begin{eqnarray}\setcounter{equation}{176}
\lefteqn{B_3=\int d^4 x\,g^{1/2}\Biggl\{c_1R\Square R
+c_2R_{\mu\nu}\Square R^{\mu\nu}
+c_3R^3
+c_4RR_{\mu\nu}R^{\mu\nu}
}\qquad\qquad\qquad
\nonumber\\[10pt]
& &
+c_5RR_{\mu\nu\alpha\beta}R^{\mu\nu\alpha\beta}
+c_6R^\mu_\nu R^\nu_\lambda R^\lambda_\mu
+c_7R^{\mu\nu}R_{\alpha\beta}R^{\alpha\ \beta}_{\ \mu\ \nu}
\nonumber\\[10pt]
& &
+c_8R_{\mu\nu}R^\mu_{\ \alpha\beta\gamma}R^{\nu\alpha\beta\gamma}
+c_9R_{\mu\nu}^{\ \ \ \alpha\beta}
R_{\alpha\beta}^{\ \ \ \sigma\rho}R_{\sigma\rho}^{\ \ \ \mu\nu}
\nonumber\\[10pt]
& &
+c_{10}R^{\mu \ \nu}_{\ \alpha \ \beta}R^{\alpha\ \beta}_{\ \,\sigma\ \rho}
R^{\sigma\ \rho}_{\ \mu \ \nu}\Biggr\},
\end{eqnarray}
where the coefficients $c_i$ are given in the Table 1.

\begin{table}[h]
\begin{center}
{\bf Table 1}\\[4mm]
\begin{tabular}{|l|c|c|c|}
\hline
& Scalar field & Spinor field & Vector field \\
\hline
$c_{1}$  &  ${1\over 2}\xi^2-{1\over 5}\xi+{1\over 56}$
& $-{3 \over 280}$ & $-{27 \over 280}$ \\
$c_{2}$  &  ${1\over 140}$ & ${1 \over 28}$ & ${9 \over 28}$ \\
$c_{3}$  &  $\left({1\over 6}-\xi\right)^3$ & ${ 1\over 864}$
& $-{5 \over 72}$ \\
$c_{4}$  &  $-{1\over 30}\left({1\over 6}-\xi\right)$ & $-{1 \over 180}$
& ${31 \over 60}$ \\
$c_{5}$  &  ${1\over 30}\left({1\over 6}-\xi\right)$ & $-{7 \over 1440}$
& $-{1 \over 10}$ \\
$c_{6}$  &  $-{8 \over 945}$ & $-{25 \over 756}$ & $-{52 \over 63}$ \\
$c_{7}$  &  ${2 \over 315}$ & ${47 \over 1260}$ & $-{19 \over 105}$ \\
$c_{8}$  &  ${1 \over 1260}$ & ${19 \over 1260}$ & ${61 \over 140}$ \\
$c_{9}$  &  ${17 \over 7560}$ & ${29 \over 7560}$ & $-{67 \over 2520}$ \\
$c_{10}$  &  $-{1 \over 270}$ & $-{1 \over 108}$ & ${1 \over 18}$ \\[2mm]
\hline
\end{tabular}
\end{center}
\end{table}

The renormalized effective action (2.174) can be used to obtain the
renormalized matrix elements of the energy--momentum tensor of the quantum
matter fields in the external gravitational field [57, 58]
$$
\left<{\rm out, vac|T}_{\mu\nu}(x)|{\rm in, vac}\right>_{\rm ren}
=-\hbar 2g^{-1/2}{\delta \Gamma_{(1){\rm ren}}\over \delta g^{\mu\nu}(x)}
+O(\hbar^2).
\eqno(2.177)
$$

Such problems were intensively investigated last time [53, 54]. In particular,
in the papers [57, 58] the vacuum polarization of the quantum fields in the
external gravitational field of the black holes was investigated. In these
papers an expression for the renormalized one--loop effective action was
obtained that is similar to (2.174)--(2.176) but does not take into account the
terms, that do not contribute to the effective vacuum energy--momentum tensor
(2.177) when the background metric satisfies the vacuum Einstein equations, 
$R_{\mu\nu}=0$. Our result (2.176) is valid, however, in general case of
arbitrary background space. Moreover, using the results of Sect. 2.4 for the
De~Witt coefficient $[a_4]$, one can calculate the coefficient $A_4$ and,
therefore, the next term, $B_4$ in the asymptotic expansion of the effective
action (2.174) of order $1/m^4$. The technique for the calculation of the
De~Witt coefficients developed in this section is very algorithmic and can be
realized on computers (all the needed 
information is contained in the first three sections of the present chapter).
In this case one can also calculate the next terms of the expansion (2.174).

However, the effective action functional $\Gamma_{(1)}$ is, in general,
essentially non--local and contains an imaginary part. The asymptotic expansion
in the inverse powers of the mass (2.174) does not reflect these properties. It
describes good the effective action only in weak gravitational fields ($R\ll
m^2$). In strong
gravitational fields  ($R\gg m^2$), as well as for the massless matter fields,
the asymptotic expansion (2.174) becomes meaningless. In this case it is
necessary either to sum up some leading (in some approximation) terms or to use
from the very beginning the non--local methods for the Green function and the
effective action.

%
%
%
%
%
%
%
%
%
%
%
%
%
%
%
%
%
%
%
%
%

\chapter{Partial summation of the
semiclassical
 Schwinger--De~Witt expansion}
\markboth{\sc Chapter 3. Summation of Schwinger--De~Witt expansion}
{\sc Chapter 3. Summation of Schwinger--De~Witt expansion}
%
%
%
%
%
%
%
%
%
\section{Summation of asymptotic expansions}

The solution of the wave equation in external fields, (1.32), by means of the proper
time method, (1.33), turns out to be very convenient for investigation of many
general problems of the quantum field theory, especially for the analysis of
the ultraviolet behavior of Green functions, regularization and
renormalization. However, in practical calculations of concrete effects one
fails to use the proper time method directly and one is forced to use model
noncovariant methods.

In order to use the advantages of the covariant proper time method it is
necessary to sum up the asymptotic series (1.43) for the evolution function. 
In general case the exact summation is impossible. Therefore, one can try to
carry out the partial summation, i.e., to single out the leading (in some
approximation) terms and sum them up in the first line. On the one hand one can
limit oneself to a given order in external fields and sum up all
derivatives, on the other hand one can neglect the derivatives and sum up all
powers of external fields.

In this way we come across a certain difficulty. The point is that the
asymptotic series do not converge, in general. Therefore, in the paper [166] it
is proposed to give up the Schwinger--De~Witt representation (1.33), (1.49), and
treat it only as an auxiliary tool for the separation of the ultraviolet
divergences. It is stated there that the Schwinger--De~Witt representation,
(1.33), exists for a small class of spaces --- only when the semi--classical solution
is exact.

However, the divergence of the asymptotic series (1.43) does not mean at all
that one must give up the Schwinger--De~Witt representation (1.33). The matter
is, the transfer function $\Omega(s)$ is not analytical at the point $s=0$.
Therefore, it is natural that the direct summation of the power series in $s$,
(1.43), leads to divergences. In spite of this one can get a certain useful
information from the structure of the asymptotic (divergent) series.

Let us consider a physical quantity $G(\alpha)$ which is defined by an
asymptotic expansion of the perturbation theory in a parameter $\alpha$
$$
G(\alpha)=\sum\limits_{k\ge 0}c_k\alpha^k.
\eqno(3.1)
$$
The convergence radius of the series (3.1) is given by the expression [165]
$$
R=\left(\lim_{k\to\infty}\,{\rm sup}|c_k|^{1/k}\right)^{-1}.
\eqno(3.2)
$$

If $R\ne 0$ then in the disc $|\alpha|<R$ of the complex plane of the parameter
$\alpha$ the series (3.1) converges and defines an analytical function. If the
considered physical quantity $G(\alpha)$ is taken to be analytical function,
then outside the convergence disc of the series (3.1), $|\alpha|\ge R$, it
should be defined by analytical continuation. The function  $\tilde G(\alpha)$
obtained in this way will be certain to have singularities, the first one lying
on the circle $|\alpha|=R$. The analytical continuation through the boundary of
the convergence disc is impossible if all the points of the boundary (i.e., the
circle $|\alpha|=R$) are singular. In this case the physical quantity
$G(\alpha)$ appears to be meaningless for $|\alpha|\ge R$. If $R=0$, then the
series (3.1) diverges for any $\alpha$, i.e., the function $G(\alpha)$ is not
analytic in the point $\alpha=0$. In this case it is impossible to carry out
the summation and the analytical continuation. Nevertheless, one can gain an
impression of t
he exact quantity $G(\alpha)$ by making use of the Borel procedure for
summation of asymptotic (in general, divergent) series [167].

The idea consists in the following. One constructs a new series with better
convergence properties which reproduces the initial series by an integral
transform. Let us define the Borel function
$$
B(z)=\sum\limits_{k\ge 0}{c_k\over \Gamma(\mu k+\nu)}z^k,
\eqno(3.3)
$$
where $\mu$ and $\nu$ are some complex numbers (${\rm Re}\,\mu$, ${\rm Re}\,\nu
>0$).

The convergence radius $\tilde R$ of the series (3.3) equals
$$
\tilde R=\left(\lim_{k\to\infty}\,{\rm sup}\left\vert{c_k\over \Gamma(\mu
k+\nu)}\right\vert^{1/k}\right)^{-1}.
\eqno(3.4)
$$

Thus, when the coefficients $c_k$ of the series (3.1) rise not faster than
\break
$\exp(M k\,\log\, k)$,  
$M=\mbox{\rm const}$, then one can always choose $\mu$ in
such way, ${\rm Re}\,\mu\ge M$, that the convergence radius of the series (3.3)
will be not equal to zero $\tilde R\ne 0$, i.e., the Borel function $B(z)$ will
be analytical at the point $z=0$. Outside the convergence disc, $|z|\ge \tilde
R$, the Borel function is defined by analytical continuation.

\noindent
Let us define
$$
\tilde G(\alpha)=\int\limits_C d t\,t^{\nu-1}e^{-t}B(\alpha t^\mu),
\eqno(3.5)
$$
where the integration contour $C$ starts at the zero point and goes to infinity
in the right halfplane (${\rm Re}\, t\to +\infty$). The asymptotic expansion of
the function $\tilde G(\alpha)$ for $\alpha\to 0$ has the form (3.1).
Therefore, the function $\tilde G(\alpha)$, (3.5), (which is called the Borel sum
of the series (3.1)) can be considered as the true physical quantity
$G(\alpha)$.

The analytical properties of the Borel function $B(z)$ determine the
convergence properties of the initial series (3.1). So, if the initial series
(3.1) has a finite convergence radius $R\ne 0$, then from (3.4) it follows that
the Borel series (3.3) has an infinite convergence radius $\tilde R=\infty$
and, therefore, the Borel function $B(z)$ is an entire function (analytical in
any finite part of the complex plane).
In this case the function $\tilde G(\alpha)$, (3.5), is equal to the sum of the
initial series (3.1) for $|\alpha|<R$ and determines its analytical
continuation outside the convergence disc $|\alpha|\ge R$.
If the Borel function $B(z)$ has singularities in the finite part of the
complex plane, i.e., $\tilde R<\infty$, then from (3.4) it follows that the
series (3.1) has a convergence radius equal to zero $R=0$, and, therefore, the
physical quantity $G(\alpha)$ is not analytic at the point $\alpha=0$. At the
same time there always exist a region in the complex plane of the variable
$\alpha$ where the Borel sum $\tilde G(\alpha)$ is still well defined and can
be used for the analytical continuation to physical values of $\alpha$.
In this way different integration contours will give different functions
$\tilde G(\alpha)$. In this case one should choose the contour of integration
from some additional physical assumptions concerning the analytical properties
of the exact function $G(\alpha)$.

\section{Covariant methods for the investigation of  nonlocalities}

The De~Witt coefficients
$a_k$ have the background dimension $L^{-2k}$, where $L$ is the length
unit. Therefore, the standard Schwinger--De~Witt expansion, (1.43), (1.52) and (1.54),
is, in fact, an
expansion in the background dimension [77, 78]. In a given order in the
background dimension $L^{-2k}$ both the powers of
the background fields, $R^k$, as well as their 
derivatives, $\nabla^{2(k-1)}R$, are taken into account.
In order to investigate the nonlocalities it is convenient to reconstruct the
local Schwinger--De~Witt expansion in such a way that the expansion is
carried out in the background fields but its derivatives are taken
into account exactly from the very beginning. 
Doing this one can preserve the manifest
covariance using the methods developed in the Chap.~2.

Let us introduce instead of the Green function $G^A_{\ B}(x,y)$ of the operator
(1.30), whose upper index belongs to the tangent space in the point $x$ and the
lower one --- in the point $y$, a three--point Green function ${\cal G}^{A'}_{\
B'}(x,y|x')$, that depends on some additional fixed point $x'$,
$$
{\cal G}(x,y|x')={\cal
P}^{-1}(x,x')\Delta^{-1/2}(x,x')G(x,y)\Delta^{-1/2}(y,x'){\cal P}(y,x').
\eqno(3.6)
$$
This Green function is scalar at the points $x$ and $y$ and a matrix at the
point $x'$. In the following we will not exhibit the dependence of all
quantities on the fixed point $x'$.

The equation for the Green function ${\cal G}(x,y)$, (3.6), has the form (1.32)
$$
\left(F_x-\hat 1 m^2\right){\cal G}(x,y)=
-\hat 1g^{-1/2}(x)\Delta^{-1}(x)\delta(x,y),
\eqno(3.7)
$$
where $F_x$ is the operator (2.112) and $\hat 1=\delta^A_{\ B}$.

Let us single out in the operator $F_x$ the free part that is of zero order in
the background fields. Using (2.120) we have
$$
F_x=\hat 1\bar{\Square}_x+\tilde F_x,
\eqno(3.8)
$$
where
$$
\bar{\Square}_x=g^{\mu'\nu'}(x')\bar\nabla^x_{\mu'}\bar\nabla^x_{\nu'},
\eqno(3.9)
$$
$$
\tilde F_x=\hat 1\tilde X^{\mu'\nu'}(x)\bar\nabla^x_{\mu'}\bar\nabla^x_{\nu'}
+Y^{\mu'}(x)\bar\nabla^x_{\mu'}+Z(x),
\eqno(3.10)
$$
$$
\tilde X^{\mu'\nu'}=X^{\mu'\nu'}(x)-g^{\mu'\nu'}(x'),
\eqno(3.11)
$$
and the operators $\bar\nabla_{\mu'}$ and the quantities $X^{\mu'\nu'}$,
$Y^{\mu'}$ and $Z$ are defined by the formulas (2.41), (2.46), (2.122) and (2.123).
The operator $\tilde F$, (3.10), is of the first order in the background fields
and can be considered as a perturbation.

By introducing the free Green function ${\cal G}_0(x,y)$,
$$
\left(-\bar{\Square}_x+m^2\right){\cal G}_0(x,y)=\hat
1\Delta^{-1}(x)g^{-1/2}(x)\delta(x,y),
\eqno(3.12)
$$
and writing the equation (3.7) in the integral form
$$
{\cal G}(x,z)={\cal G}_0(x,z)
+\int d^n y\, g^{1/2}(y)\Delta(y)
{\cal G}_0(x,y)\tilde F_y{\cal G}(y,z),
\eqno(3.13)
$$
we obtain from (3.13) by means of direct iterations
\begin{eqnarray}\setcounter{equation}{14}
{\cal G}(x,z)&=&{\cal G}_0(x,z)
+\sum\limits_{k\ge 1}\int d^n y_1\, g^{1/2}(y_1)\Delta(y_1)
\cdots d^n y_k\, g^{1/2}(y_k)\Delta(y_k)
\nonumber\\[10pt]
& &\times
{\cal G}_0(x,y_1)\tilde F_{y_1}{\cal G}_0(y_1,y_2)
\cdots \tilde F_{y_k}{\cal G}_0(y_k,z).
\end{eqnarray}
Using the covariant Fourier integral (2.55) and the equations (2.56) and (2.57) we
obtain from (3.12) and (3.14) the momentum representations for the free Green
function
$$
{\cal G}_0(x,y)=\hat 1\int\,{d^n k^{\mu'}\over (2\pi)^n}\,g^{1/2}(x')
\exp\left\{ik_{\mu'}\left(\sigma^{\mu'}(y)
-\sigma^{\mu'}(x)\right)\right\}{1\over m^2+k^2},
\eqno(3.15)
$$
and for the full Green function
\begin{eqnarray}\setcounter{equation}{16}
{\cal G}(x,y)&=&\int\,{d^n k^{\mu'}\over (2\pi)^n}\,g^{1/2}(x')
{d^n p^{\mu'}\over (2\pi)^n}\,g^{1/2}(x')
\nonumber\\[10pt]
& &
\quad\times
\exp\left\{ip_{\mu'}\sigma^{\mu'}(y)
-ik_{\mu'}\sigma^{\mu'}(x)\right\}{\cal G}(k,p),
\end{eqnarray}
where
$$
{\cal G}(k,p)=(2\pi)^n g^{-1/2}(x'){\delta(k^{\mu'}-p^{\mu'})\over m^2+k^2}
+{\Pi(k,p)\over (m^2+k^2)(m^2+p^2)},
\eqno(3.17)
$$
\begin{eqnarray}\setcounter{equation}{18}
\Pi(k,p)&=&\tilde F(k,p)
+\sum\limits_{i\ge 1}\int\,{d^n q_1^{\mu'}\over (2\pi)^n}\,g^{1/2}(x')\cdots
{d^n q_i^{\mu'}\over (2\pi)^n}\,g^{1/2}(x') 
\nonumber\\[10pt]
& &
\times\tilde F(k,q_1){1\over m^2+q_1^2}\cdots
\tilde F(q_{i-1},q_i){1\over m^2+q_i^2}\tilde F(q_i,p),
\end{eqnarray}
$$
\tilde F(k,p)=-\hat 1\tilde
X^{\mu'\nu'}(k-p)p_{\mu'}p_{\nu'}-iY^{\mu'}(k-p)p_{\mu'}+Z(k-p),
\eqno(3.19)
$$
$$
k^2\equiv g_{\mu'\nu'}(x')k^{\mu'}k^{\nu'},
$$
and $\tilde X^{\mu'\nu'}(q)$, $Y^{\mu'}(q)$ and $Z(q)$ are the covariant Fourier
components, (2.55), of the coefficients $\tilde X^{\mu'\nu'}$,
$Y^{\mu'}$ and $Z$, (3.11), (2.46), (2.122), (2.123).

The formulas (3.15)--(3.19) reproduce the covariant generalization of the
usual diagrammatic technique. Therefore, one can apply good elaborated methods of the
Feynman momentum integrals. The Fourier components of the coefficient functions
$\tilde X^{\mu'\nu'}(q)$, $Y^{\mu'}(q)$ and $Z(q)$, can be expressed in terms of
the Fourier components of the external fields, $R^\mu_{\ \nu\alpha\beta}$,
${\cal R}_{\mu\nu}$ and $Q$, using the formulas obtained in Chap.~2.
As usual [48--50, 54] one should choose the contour of integration over $k_0$ in
the momentum integrals (3.15)--(3.18). Different ways of integration correspond
to different Green functions. For the causal (Feynman)  Green function one
should either assume $k^2\to k^2-i\varepsilon$ or go to the Euclidean sector of
the space--time [48--50].

Similarly, one can construct the kernels of any non--local operators of general
form, $f(\Square)$, where $f(z)$ is some function. In the zeroth approximation in
external fields we obtain
$$
f(\Square)(x,y)={\cal P}(x)\Delta^{1/2}(x)\bar f(\Square)(x,y)\Delta^{1/2}(y){\cal
P}^{-1}(y),
$$
$$
\bar f(\Square)(x,y)=\int\,{d^n k^{\mu'}\over (2\pi)^n}\,g^{1/2}(x')
\exp\left\{ik_{\mu'}\left(\sigma^{\mu'}(y)
-\sigma^{\mu'}(x)\right)\right\}f(-k^2).
\eqno(3.20)
$$

An important method for the investigation of the nonlocalities is the analysis
of the De~Witt coefficients $a_k$ and the partial summation of the asymptotic
series (1.43). In this case one should limit oneself to some order in
external fields and sum up all derivatives of external fields.
In order to get an effective expansion in external  fields it is convenient
to change a little the ``diagrammatic'' technique for the De~Witt coefficients
developed in Sect. 2.3. Although all the terms in the sum (2.117) have
equal background dimension, $L^{n-2k}$, they are of different order in
external fields. From the formula (2.117) it is not seen immediately what order
in external fields has each term of the sum (2.117), i.e., a single
diagram, since all the diagrams for the coefficient $<n|a_k>$ have $k$ blocks.
However, among these blocks, i.e., the matrix elements $<m|F|m+2>$, there are
dimensionless blocks that do not have any background dimension and are of zero
order in external fields. These are the blocks (matrix elements $<m|F|m+2>$)
with the number of outgoing lines equal to the number of
the incoming lines plus 2 (c.f. (2.124)). 
Therefore, one can order all the diagrams for the
De~Witt coefficients (i.e., the single terms of the sum (2.117)) in the
following way. The first diagram contains only one dimensionful block, all
others being dimensionless. The second class  contains all diagrams with two
dimensionful blocks, the third one --- three etc. The last diagram contains $k$
dimensionful blocks. To obtain the De~Witt coefficients in the first order in
external fields it is sufficient to restrict oneself to the first diagram.
To get the De~Witt coefficients in the second order in external fields it is
sufficient to restrict oneself to the first diagram and the set of diagrams
with two dimensionful blocks etc. 
This method is completely analogous to the separation of the
free part of the operator $F$, (3.8).
The dimensionful matrix elements $<m|F|n>$, (with $m\ge n$), of the operator
$F$, (2.120), are equal to the matrix elements of the operator $\tilde F$, (3.10).
When calculating the matrix elements (2.124) and (2.125) one can also neglect the
terms that do not contribute in the given order in external fields.

After such reconstruction (and making use of (2.124)) the formula (2.117) for
the De~Witt coefficients $<n|a_k>$ takes the form
\begin{eqnarray}\setcounter{equation}{21}
\lefteqn{<n|a_k>
=\sum\limits_{1\le N\le k;}\sum\limits_{1\le i_1<i_2<\cdots<i_{N-1}\le
k-1;}\sum\limits_{n_i}
\prod\limits_{1\le j\le N}
{{2i_j+n_j-1\choose i_{j-1}}\over {2i_j+n_j-1\choose i_j}}
}
\nonumber\\[10pt]
& &
\times <n;k-i_{N-1}-1|F|n_{N-1}><n_{N-1};i_{N-1}-i_{N-2}-1|F|n_{N-2}>
\nonumber\\[10pt]
& &
\times \cdots
<n_2;i_2-i_1-1|F|n_1><n_1;i_1-1|F|0>,
\end{eqnarray}
where the following notation is introduced
$$
<n;k|F|m>\equiv g^{\nu_1\nu_2}\cdots g^{\nu_{2k-1}\nu_{2k}}
<\nu_1 \cdots\nu_{n+2k}|F|\mu_1\cdots \mu_m>,
\eqno(3.22)
$$
$$
i_0\equiv 0,\qquad i_N\equiv k,\qquad n_N\equiv n,
$$
and the summation over $n_i$ is carried out in such limits that all matrix
elements are dimensionful, (i.e., for each $<n;k|F|m>$, (3.22), there holds:
$n+2k\ge m$):
$$
n_1+2(i_1-1)\ge 0, \quad n_2+2(i_2-i_1-1)\ge n_1, 
\quad\cdots,\quad  n+2(k-i_{N-1}-1)\ge
n_{N-1}.
$$

The formula (3.21) also enables to use the diagrammatic technique. However, for
analysing the general structure of the De~Witt coefficients and for the
partial summation it is no longer effective. Therefore, one should use the
analytic expression (3.21).

\section{Summation of the terms of first order in external fields}

Let us calculate the coincidence limit of the De~Witt coefficients
$[a_k]=<0|a_k>$ in the first order in external fields. Using the formula
(3.21) we obtain
$$
[a_k]={1\over {2k-1\choose k}}<0;k-1|F|0>+O(R^2),
\eqno(3.23)
$$
where $O(R^2)$ denotes all omitted terms of second order in external fields.

Using (3.22), (2.121)--(2.125) and (2.46) and the formulas of the Sect.~2.2 we
obtain from (3.23) up to quadratic terms
$$
[a_k]={k!(k-1)!\over (2k-1)!}\Square^{k-1}\left(Q+{k\over 2(2k+1)}\hat 1
R\right)+O(R^2), \qquad (k\ge 1).
\eqno(3.24)
$$
The expression (3.24) can be used for the calculation of the transfer function
$\Omega(s|x,x)$ in the first order in external fields. Substituting (3.24) in
(1.43), we obtain
$$
\Omega(s|x,x)=\hat 1+is\left(f_1(is\Square)Q+\hat 1f_2(is\Square)R\right)+O(R^2),
\eqno(3.25)
$$
where
$$
f_1(z)=\sum\limits_{k\ge 0}{k!\over (2k+1)!}z^k,
\eqno(3.26)
$$
$$
f_2(z)=\sum\limits_{k\ge 0}{(k+1)!(k+1)\over (2k+3)!}z^k.
\eqno(3.27)
$$
The power series (3.26) and (3.27) converge for any finite $z$ and hence define
entire functions.
One can sum up the series of the type (3.26) and (3.27) using the general formula
$$
{(k+l)!\over (2k+2l+1)!}={l!\over (2l)!}
\int\limits^1_0d\xi\,\xi^{2l}{1\over k!}
\left({1-\xi^2\over 4}\right)^k,
\eqno(3.28)
$$
that is easily obtained from the definition of the Euler beta--function [168].

Substituting (3.28) in (3.26) and (3.27) and summing over $k$ we obtain the
integral representations of the functions $f_1(z)$ and $f_2(z)$
$$
f_1(z)=\int\limits^1_0d\xi\,
\exp\left\{{1\over 4}\left(1-\xi^2\right)z\right\},
\eqno(3.29)
$$
$$
f_2(z)=\int\limits^1_0d\xi\,{1\over 4}(1-\xi^2)
\exp\left\{{1\over 4}\left(1-\xi^2\right)z\right\}.
\eqno(3.30)
$$
The kernels of the non--local operators $f_1(is\Square)$, $f_2(is\Square)$ should be
understood in terms of covariant momentum expansions (3.20).

Using the obtained transfer function, (3.25), one can easily obtain the Green
function at coinciding points, $G(x,x)$, in the first order in external
fields. Substituting (3.25) in (1.49) and supposing ${\rm Im}\, m^2<0$ (for the
causal Green function), we obtain after the integration over the proper time in
the $n$--dimensional space
\begin{eqnarray}\setcounter{equation}{31}
G(x,x)&=&i(4\pi)^{-n/2}\Biggl\{\Gamma\left(1-{n\over 2}\right)\hat 1 m^{n-2}
\nonumber\\[10pt]
& &
+\Gamma\left(2-{n\over 2}\right)m^{n-4}
\left[\tilde F_1\left(-{\Square\over 4m^2}\right)Q
+\hat 1\tilde F_2\left(-{\Square\over 4m^2}\right)R\right]\Biggr\}
\nonumber\\[10pt]
& &
+O(R^2),
\end{eqnarray}
where $\Gamma(z)$ is the Euler gamma--function, and
$$
\tilde F_1(z)=\int\limits^1_0d\xi\,\left[1+(1-\xi^2)z\right]^{{n\over 2}-2},
\eqno(3.32)
$$
$$
\tilde F_2(z)=\int\limits^1_0d\xi\,{1\over 4}(1-\xi^2)
\left[1+(1-\xi^2)z\right]^{{n\over 2}-2}.
\eqno(3.33)
$$

By expanding in the dimension $n$ in the neighborhood of the point $n=4$ and
subtracting the pole $1/(n-4)$ we obtain the renormalized Green function,
$G_{\rm ren}(x,x)$, in the physical four--dimensional space--time (1.51), (1.52),
up to terms of second order in external fields
$$
G_{\rm ren}(x,x)={i\over (4\pi)^2}\left\{F_1\left(-{\Square\over 4m^2}\right)Q
+\hat 1 F_2\left(-{\Square\over 4m^2}\right)R\right)\Biggr\}
+O(R^2),
\eqno(3.34)
$$
where
$$
F_1(z)=2-J(z),
\eqno(3.35)
$$
$$
F_2(z)={1\over 18}\left(5-{3\over z}\right)
-{1\over 6}\left(1-{1\over 2 z}\right)J(z),
\eqno(3.36)
$$
$$
J(z)=2(1+z)\int\limits^1_0{d\xi\over 1+(1-\xi^2)z}.
\eqno(3.37)
$$

The formfactors $F_1(z)$, (3.35), and $F_2(z)$, (3.36), are normalized according
the conditions
$$
F_1(0)=F_2(0)=0.
\eqno(3.38)
$$
The integral (3.37) determines an analytical single--valued function in the
complex plane $z$ with a cut along the negative part of the real axis from $-1$
to $-\infty$:
$$
J(z)=2\sqrt{1+{1\over z}}\log\,\left(\sqrt{z+1}+\sqrt z\right), \qquad
|\arg(z+1)|<\pi,
\eqno(3.39)
$$
$$
J(x\pm i\varepsilon)=2\sqrt{1+{1\over x}}
\log\,\left(\sqrt{-x-1}+\sqrt{-x}\right)\pm i\pi\sqrt{1+{1\over x}}, \qquad
(x<-1).
\eqno(3.40)
$$

Thus we obtained the non--local expression for the Green function at coinciding
points, (3.34). It reproduces the local Schwinger--De~Witt expansion, (1.52), up
to quadratic terms in eternal fields by expanding in inverse powers of
the mass. The power series determining the formfactors $F_1(z)$, (3.35), and
$F_2(z)$, (3.36), converge in the region $|z|<1$. For $z=-1$ there is a threshold
singularity --- the branching point. Outside the circle $|z|\ge 1$ the
formfactors are defined by the analytic continuation. The boundary conditions
for the formfactors fix uniquely the ambiguity in the Green function (3.34).
For the causal Green function (${\rm Im} \, m^2<0$) the lower bank of the cut
is the physical one. Therefore, with account of (3.40),
the imaginary parts of the formfactors (3.35) and
(3.36) in the pseudo--Euclidean region above the threshold $z=x-i\varepsilon$,
$x<-1$, equal
$$
{\rm Im}F_1(x-i\varepsilon)=\pi\sqrt{1+{1\over x}},
\eqno(3.41)
$$
$$
{\rm Im}F_2(x-i\varepsilon)={\pi\over 6}
\left(1-{1\over 2 x}\right)\sqrt{1+{1\over x}}.
\eqno(3.42)
$$
The ultraviolet asymptotics $|z|\to \infty$ of the formfactors (3.35) and (3.36)
have the form
$$
F_1(z)\Bigg\vert_{|z|\to\infty}=-\log\,(4z)+2
+O\left({1\over z}\log\, z\right),
\eqno(3.43)
$$
$$
F_2(z)\Bigg\vert_{|z|\to\infty}={1\over 6}\left(-\log\,(4z)+{5\over 3}\right)
+O\left({1\over z}\log\, z\right).
\eqno(3.44)
$$

Let us consider the case of the massless field, $m^2=0$. Taking the limit
$m^2\to 0$ in (3.31) for ${\rm Re}\,n>2$ we obtain the Green function of the
massless field at coinciding points in the $n$--dimensional space in the first
order in external fields
\begin{eqnarray}\setcounter{equation}{46}
G(x,x)&=&i(4\pi)^{-n/2}{\Gamma\left(2-{n\over 2}\right)
\left(\Gamma\left({n\over 2}-1\right)\right)^2\over \Gamma(n-2)}
\nonumber\\[10pt]
& &
\times (-\Square)^{{n\over 2}-2}\left(Q+{n-2\over 4(n-1)}\hat 1 R\right)
+O(R^2).
\end{eqnarray}

The formula (3.46) determines the analytic function of the dimension $n$ in the
region $2<{\rm Re}\,n<4$. After the analytic continuation there appear poles in
the points $n=4,6,8,\dots$, that reflect the ultraviolet divergences, and a
pole in the point $n=2$ reflecting the infrared divergence. In the
two--dimensional space eq. (3.46) gives
$$
G(x,x)={i\over 4\pi}\left\{-2\left({2\over n-2}+{\bf C}
+\log{-\Square\over 4\pi\mu^2}\right){1\over \Square}Q
-\hat 1{1\over \Square}R\right\}+O(R^2).
\eqno(3.47)
$$
For even dimensions, $n=2N$, ($N\ge 2$), from (3.46) we obtain 
\begin{eqnarray}\setcounter{equation}{48}
G(x,x)&=&{i\over (4\pi)^N}\cdot{(N-2)!\over (2N-3)!}\Biggl\{
-\Biggl({2\over n-2N}+\Psi(N-1)
\nonumber\\[10pt]
& &
-2\Psi(2N-2)+\log\,{-\Square\over 4\pi\mu^2}\Biggr)
\Square^{N-2}\left(Q
+{N-1\over 2(2N-1)}\hat 1 R\right)
\nonumber\\[10pt]
& &
-{1\over 2(2N-1)^2}\Square^{N-2}R\Biggr\}+O(R^2),
\end{eqnarray}
where
$$
\Psi(k)=-{\bf C}+\sum\limits_{1\le l \le k-1}{1\over l}.
\eqno(3.49)
$$

In particular, in the physical four--dimensional space--time, $n\to 4$, we have
from (3.48)
\begin{eqnarray}\setcounter{equation}{50}
G(x,x)&=&{i\over (4\pi)^2}\Biggl\{-\left({2\over n-4}+{\bf C}-2
+\log\,{-\Square\over 4\pi\mu^2}\right)
\left(Q+{1\over 6}\hat 1 R\right)
\nonumber\\[10pt]
& &
\qquad\ \ \,
-{1\over 18}\hat 1 R\Biggr\}+O(R^2).
\end{eqnarray}

The renormalized Green function of the massless field can be obtained by using the
ultraviolet asymptotics of the formfactors (3.43) and (3.44) and substituting the
renormalization parameter instead of the mass, $m^2\to \mu^2$. This reduces simply to a
change of the normalization of the formfactors (3.38).

Let us stress that in the massless case the divergence of the coincidence limit
of the Green function in the first order in external fields, (3.48), (3.50),
is proportional to the De~Witt coefficient $[a_{{n\over 2}-1}]$, (3.24).
Therefore, in the conformally invariant case [54, 53],
$$
Q=-{n-2\over 4(n-1)}\hat 1 R,
\eqno(3.51)
$$
the linear part of the coefficient $[a_{{n\over 2}-1}]$, (3.24), is equal to zero
and the Green function at coinciding points is finite in the first order in
external fields, (3.47)--(3.50). In odd dimensions, $n=2N+1$, the Green function,
(3.31), (3.46), is finite.

\section{Summation of the terms of second order in external fields}

Let us calculate the coincidence limit of the De~Witt coefficients
$[a_k]=<0|a_k>$ in the second order in external fields. From the formula
(3.21) we have up to cubic terms in external fields
\begin{eqnarray}\setcounter{equation}{52}
[a_k]&=&{1\over {2k-1\choose k}}<0;k-1|F|0>
\nonumber\\[10pt]
& &
+\sum\limits_{1\le i\le k-1;}\,\sum\limits_{0\le n_i\le 2(k-i-1)}
{{2k-1\choose i}\over {2k-1\choose k}{2i+n_i-1\choose i}}
\nonumber\\[10pt]
& &
\times<0;k-i-1|F|n_i><n_i;i-1|F|0>
+O(R^3),
\end{eqnarray}
where $O(R^3)$ denotes all omitted terms of third order in external fields.

The total number of terms, that are quadratic in external fields and
contain
arbitrary number of derivatives, is infinite. Therefore, we will also neglect
the total derivatives and the trace--free terms in (3.52) assuming to use
the results for the calculation of the coefficients $A_k$, (1.55), and the
one--loop effective action, (1.50), (1.54).

The number of the terms quadratic in external fields, in the
coefficients $A_k$ is finite. Let us write the general form of the
coefficients
$A_k$ up to the terms of the third order in external fields
\begin{eqnarray}
\setcounter{equation}{53}
A_k&=&\int d^nx\,g^{1/2}{\rm str}{1\over 2}\Biggl\{\alpha_kQ\Square^{k-2}Q
-2\beta_kJ_\mu\Square^{k-3}J^\mu
+\gamma_k Q\Square^{k-2}R
\nonumber\\[10pt]
& &
\qquad\qquad
+\hat 1\left(\delta_kR_{\mu\nu}\Square^{k-2}R^{\mu\nu}
+\varepsilon_kR\Square^{k-2}R\right)+O(R^3)\Biggr\},
\end{eqnarray}
where $J_\mu=\nabla_\alpha{\cal R}^\alpha_{\ \mu}$.

All other quadratic invariants can be reduced to those written above up
 to third order terms using the integration by parts and the Bianci identity.
For example,
$$
\int dx^n\,g^{1/2}{\rm str}\,{\cal R}_{\mu\nu}\Square^k{\cal R}^{\mu\nu}
=-2\int dx^n\,g^{1/2}{\rm str}\,\left\{J_\mu\Square^{k-1}J^\mu
+O(R^3)\right\},
$$
$$
\int dx^n\,g^{1/2}R_{\mu\nu\alpha\beta}\Square^kR^{\mu\nu\alpha\beta}
=\int dx^n\,g^{1/2}\left\{\left(4R_{\mu\nu}\Square^kR^{\mu\nu}-R\Square^kR\right)
+O(R^3)\right\}.
\eqno(3.54)
$$

Using (3.22), (2.121)--(2.125) and (2.46) and the formulas of the Sect. 2.2
one can calculate the coefficients $\alpha_k$, $\beta_k$, $\gamma_k$,
$\delta_k$ and $\varepsilon_k$ from (3.52) and (1.55). Omitting the
intermediate computations we present the result
$$
\alpha_k={k!(k-2)!\over (2k-3)!},
\eqno(3.55)
$$
$$
\beta_k={k!(k-1)!\over (2k-1)!},
\eqno(3.56)
$$
$$
\gamma_k=2(k-1){k!(k-1)!\over (2k-1)!},
\eqno(3.57)
$$
$$
\delta_k=2{k!k!\over (2k+1)!},
\eqno(3.58)
$$
$$
\varepsilon_k=(k^2-k-1){k!k!\over (2k+1)!}.
\eqno(3.59)
$$

Using the obtained coefficients, (3.55)--(3.59), we calculate the trace of the
transfer function $\Omega(s)$, (1.43),
$$
\int d^nx\,g^{1/2}{\rm str}\,\Omega(s|x,x)=\sum\limits_{k\ge 0}{(is)^k\over
k!}A_k.
\eqno(3.60)
$$
Substituting (3.53) and (3.55)--(3.59) in (3.60) we obtain
\begin{eqnarray}
\setcounter{equation}{61}
\lefteqn{\int d^nx\,g^{1/2}{\rm str}\,\Omega(s|x,x)
=\int d^nx\,g^{1/2}{\rm str}\Biggl\{\hat 1
+is\left(Q+{1\over 6}\hat 1 R\right)
}\nonumber\\[10pt]
& &
+{(is)^2\over 2}\biggl[Qf_1(is\Square)Q
+2J_\mu f_3(is\Square){1\over -\Square}J^\mu
+2Qf_2(is\Square)R
\nonumber\\[10pt]
& &
+\hat 1\left(R_{\mu\nu}f_4(is\Square)R^{\mu\nu}
+Rf_5(is\Square)R\right)\biggr]+O(R^3)\Biggr\},
\end{eqnarray}
where
$$
f_3(z)=\sum\limits_{k\ge 0}{(k+1)!\over (2k+3)!}z^k,
\eqno(3.62)
$$
$$
f_4(z)=\sum\limits_{k\ge 0}2{(k+2)!\over (2k+5)!}z^k,
\eqno(3.63)
$$
$$
f_5(z)=\sum\limits_{k\ge 0}{(k+2)!\over (2k+5)!}(k^2+3k+1)z^k,
\eqno(3.64)
$$
and $f_1(z)$ and $f_2(z)$ are given by the formulas (3.26), (3.27), (3.29) and
(3.30).

The series (3.62)--(3.64) converge for any finite $z$ and define in the same
manner as (3.26) and (3.27) entire functions. The summation of the series (3.62)
and (3.63) can be performed by means of the formula (3.28). Substituting (3.28) in
(3.62) and (3.63) and summing over $k$ we obtain the integral representation of
the functions $f_3(z)$ and $f_4(z)$:
$$
f_3(z)=\int\limits_0^1d\xi\,{1\over 2}\xi^2
\exp\left\{{1\over 4}(1-\xi^2)z\right\},
\eqno(3.65)
$$
$$
f_4(z)=\int\limits_0^1d\xi\,{1\over 6}\xi^4
\exp\left\{{1\over 4}(1-\xi^2)z\right\}.
\eqno(3.65)
$$
Noting that from (3.64) it follows
$$
f_5(z)={1\over 16}f_1(z)-{1\over 4}f_3(z)-{1\over 8}f_4(z),
\eqno(3.67)
$$
we obtain from the formulas (3.29), (3.65) and (3.66)
$$
f_5(z)=\int\limits_0^1d\xi\,{1\over 8}\left({1\over 2}-\xi^2
-{1\over 6}\xi^4\right)
\exp\left\{{1\over 4}(1-\xi^2)z\right\}.
\eqno(3.68)
$$
Using (3.61) one can calculate the one--loop effective action up to cubic terms
in external fields. Substituting (3.61) in (1.50) and assuming ${\rm
Im}\,m^2<0$, after integration over the proper time we obtain in the
$n$--dimensional space
\begin{eqnarray}
\setcounter{equation}{69}
\lefteqn{
\hskip-30pt
\Gamma_{(1)}={1\over 2(4\pi)^{n/2}}\int d^n x\, g^{1/2}{\rm str}\,\Biggl\{
\Gamma\left(-{n\over 2}\right)\hat 1 m^n
+\Gamma\left(1-{n\over 2}\right)m^{n-2}\left(Q
+{1\over 6}\hat 1 R\right)
}\nonumber\\[10pt]
& &
+{1\over 2}\Gamma\left(2-{n\over 2}\right) m^{n-4}\Biggl[
Q\tilde F_1\left(-{\Square\over 4m^2}\right)Q
\nonumber\\[10pt]
& &
+2J_\mu{1\over -\Square}\tilde F_3\left(-{\Square\over 4m^2}\right)J^\mu
+2Q\tilde F_2\left(-{\Square\over 4m^2}\right)R
\nonumber\\[10pt]
& &
+\hat 1\left(R_{\mu\nu}\tilde F_4\left(-{\Square\over 4m^2}\right)
R^{\mu\nu}
+R\tilde F_5\left(-{\Square\over 4m^2}\right)R\right)\Biggr]+O(R^3)\Biggr\},
\end{eqnarray}
where
$$
\tilde F_3(z)=\int\limits_0^1 d\xi\,{1\over 2}
\xi^2\left[1+(1-\xi^2)z\right]^{{n\over 2}-2},
\eqno(3.70)
$$
$$
\tilde F_4(z)=\int\limits_0^1 d\xi\,{1\over 6}
\xi^4\left[1+(1-\xi^2)z\right]^{{n\over 2}-2},
\eqno(3.71)
$$
$$
\tilde F_5(z)=\int\limits_0^1 d\xi\,{1\over 8}\left({1\over 2}-\xi^2
-{1\over 6}\xi^4\right)
\left[1+(1-\xi^2)z\right]^{{n\over 2}-2},
\eqno(3.72)
$$
and $\tilde F_1(z)$ and $\tilde F_2(z)$ are given by the formulas (3.32) and (3.33).

Subtracting the pole in the dimension $1/(n-4)$, we obtain the renormalized
effective action in the physical four--dimensional space--time, (1.53), (1.54), up
to terms of third order in external fields
\begin{eqnarray}
\setcounter{equation}{73}
\Gamma_{(1){\rm ren}}&=&{1\over 2(4\pi)^2}\int d^4x\, g^{1/2}{\rm str}{1\over 2}
\Biggl\{QF_1\left(-{\Square\over 4m^2}\right)Q
\nonumber\\[10pt]
& &
+2J_\mu{1\over -\Square}\tilde F_3\left(-{\Square\over 4m^2}\right)J^\mu
+2Q\tilde F_2\left(-{\Square\over 4m^2}\right)R
\nonumber\\[10pt]
& &
+\hat 1\left(R_{\mu\nu}\tilde F_4\left(-{\Square\over 4m^2}\right)
R^{\mu\nu}
+R\tilde F_5\left(-{\Square\over 4m^2}\right)R\right)\Biggr]
+O(R^3)\Biggl\},
\nonumber\\[10pt]
& &
\end{eqnarray}
where
$$
F_3(z)={4\over 9}+{1\over 3z}-{1\over 6}\left(1+{1\over z}\right)J(z),
\eqno(3.74)
$$
$$
F_4(z)={23\over 225}+{7\over 45z}+{1\over 15z^2}
-{1\over 30}\left(1+{1\over z}\right)^2J(z),
\eqno(3.75)
$$
$$
F_5(z)={1\over 900}-{37\over 360z}-{1\over 120 z^2}
-{1\over 60}\left(1-{3\over z}-{1\over 4z^2}\right)J(z),
\eqno(3.76)
$$
and $F_1(z)$, $F_2(z)$ and $J(z)$ are given by the formulas (3.35)--(3.37).

The formfactors $F_3(z)$, (3.74), $F_4(z)$, (3.75), and $F_5(z)$, (3.76), are
normalized by the conditions
$$
F_3(0)=F_4(0)=F_5(0)=0.
\eqno(3.77)
$$
The normalization conditions of the formfactors (3.38) and (3.77) correspond to
the normalization of the effective action
$$
\Gamma_{(1){\rm ren}}\Bigg\vert_{m^2\to \infty}=0.
\eqno(3.78)
$$

Thus, by means of the partial summation of the local Schwinger--De~Witt
expansion, (1.43), (3.60), we obtained a non--local expression for the effective
action up to terms of third order in external fields, (3.73). Although the
power series, that define the formfactors, i.e., the power series for the
function $J(z)$, (3.37), converge only in the region $|z|<1$, $z=-1$ being the
threshold branch point, the expressions (3.35)--(3.37) and (3.74)--(3.76) are valid
for any $z$. That means that the proper time method automatically does the
analytical continuation in the ultraviolet region $|z|\to \infty$. 
All the
ambiguity, which arises by the partial summation of the asymptotic expansion
(3.60), reduces to the arbitrariness in the boundary conditions for the
formfactors. Specifying the causal boundary conditions leads to the
single--valued expression for the effective action. Using the prescription
$m^2\to m^2-i\varepsilon$ and the equation (3.40) we obtain the imaginary parts
of the formfactors (3.74)--(3.76) in the 
pseudo--Euclidean region ($z=x-i\varepsilon$) above the threshold ($x<-1$)
$$
{\rm Im}\, F_3(x-i\varepsilon)={\pi\over 6}\left(1+{1\over x}\right)
\sqrt{1+{1\over x}},
\eqno(3.79)
$$
$$
{\rm Im}\, F_4(x-i\varepsilon)={\pi\over 30}\left(1+{1\over x}\right)^2
\sqrt{1+{1\over x}},
\eqno(3.80)
$$
$$
{\rm Im}\, F_5(x-i\varepsilon)={\pi\over 60}\left(1
-{3\over x}-{1\over 4x^2}\right)
\sqrt{1+{1\over x}}.
\eqno(3.81)
$$

The imaginary parts of all formfactors, (3.41), (3.42), (3.79)--(3.81), are
positive. This ensures the fulfillment of the important condition
$$
{\rm Im}\,\Gamma_{(1){\rm ren}}>0.
\eqno(3.82)
$$

\noindent
The ultraviolet asymptotics $|z|\to\infty$ of the formfactors (3.74)--(3.76)
have the form
$$
F_3(z)\Big\vert_{|z|\to\infty}={1\over 6}\left(-\log\, (4z)+{8\over 3}\right)
+O\left({1\over z}\log\, z\right),
\eqno(3.83)
$$
$$
F_4(z)\Big\vert_{|z|\to\infty}={1\over 30}\left(-\log\, (4z)+{46\over
15}\right)
+O\left({1\over z}\log\, z\right),
\eqno(3.84)
$$
$$
F_5(z)\Big\vert_{|z|\to\infty}={1\over 60}\left(-\log\, (4z)+{1\over 15}\right)
+O\left({1\over z}\log\, z\right).
\eqno(3.85)
$$

Let us consider the case of the massless field. Taking the limit $m^2\to 0$ in
(3.69) in the region ${\rm Re}\, n>2$ we obtain the one--loop effective action
for the massless field in the $n$--dimensional space up to terms of third order
in external fields
\begin{eqnarray}
\setcounter{equation}{86}
\Gamma_{(1)}&=&{1\over 2(4\pi)^{n/2}}\int d^nx\, g^{1/2}{\rm str}
{\Gamma\left(2-{n\over 2}\right)
\left(\Gamma\left({n\over 2}-1\right)\right)^2
\over \Gamma(n-2)}
\Biggl\{{1\over 2}Q\left(-\Square\right)^{{n\over 2}-2}Q
\nonumber\\[10pt]
& &
+{1\over 2(n-1)}J_\mu\left(-\Square\right)^{{n\over 2}-3}
J^\mu
+{n-2\over 4(n-1)}Q\left(-\Square\right)^{{n\over 2}-2}R
\nonumber\\[10pt]
& &
+{1\over 4(n^2-1)}\hat 1\left[R_{\mu\nu}\left(-\Square\right)^{{n\over 2}-2}
R^{\mu\nu}
+{1\over 8}(n^2-2n-4)R\left(-\Square\right)^{{n\over 2}-2}R\right]
\nonumber\\[10pt]
& &
+O(R^3)\Biggr\}.
\end{eqnarray}

The formula (3.86) defines an analytical function of the dimension $n$ in the
region $2<{\rm Re}\,n<4$. The analytical continuation leads to the poles in the
points $n=2, 4, 6, \dots$. In odd dimensions the expression (3.86) is finite
and directly defines the effective action. In even dimensions the expression
(3.86) is proportional to the De~Witt coefficient $A_{n/2}$, (3.53).

Separating the pole $1/(n-4)$ in (3.86) we obtain the effective action for the
massless field in the physical four--dimensional space--time up to terms $O(R^3)$
\begin{eqnarray}
\setcounter{equation}{87}
\Gamma_{(1)}&=&{1\over 2(4\pi)^2}\int d^nx\,g^{1/2}{\rm str}\,
\Biggl\{{1\over 2}Q\left(-{2\over n-4}-{\bf C}+2
-\log\,{-\Square\over 4\pi\mu^2}\right)Q
\nonumber\\[10pt]
& &
+{1\over 6}J_\mu{1\over-\Square}\left(-{2\over n-4}-{\bf C}+{8\over 3}
-\log\,{-\Square\over 4\pi\mu^2}\right)J^\mu
\nonumber\\[10pt]
& &
+{1\over 6}Q\left(-{2\over n-4}-{\bf C}+{5\over 3}
-\log\,{-\Square\over 4\pi\mu^2}\right)R
\nonumber\\[10pt]
& &
+\hat 1\Biggl[{1\over 60}R_{\mu\nu}\left(-{2\over n-4}-{\bf C}+{46\over 15}
-\log\,{-\Square\over 4\pi\mu^2}\right)R^{\mu\nu}
\nonumber\\[10pt]
& &
+{1\over 120}R\left(-{2\over n-4}-{\bf C}+{1\over 15}
-\log\,{-\Square\over 4\pi\mu^2}\right)R\Biggr]+O(R^3)\Biggr\}.
\end{eqnarray}

An analogous expression was obtained in the paper [45]. However, in that paper
the coefficients $\alpha_k$, $\beta_k$, $\gamma_k$, $\delta_k$, and $\varepsilon_k$,
(3.55)--(3.59), in the De~Witt coefficient $A_k$, (3.53), were not calculated.
That is why in the paper [45] it was assumed additionally that
the power series in the proper time for the functions $f_i(z)$, (3.26), (3.27) and
(3.62)--(3.64), converge as well as the proper time integral (1.50) in the upper
limit does. Besides, in [45] only divergent and logarithmic,
$\log(-\Square)$, terms in (3.87) were calculated explicitly. The complete result
(3.87) together with the finite constants is obtained in present work.

The renormalized effective action for the massless field can be obtained using
the ultraviolet asymptotics of the formfactors (3.43), (3.44) and (3.83)--(3.85)
and substituting the renormalization parameter instead of the mass, $m^2\to\mu^2$.
This reduces just to a renormalization of the formfactors (3.38) and (3.77).
Although in the massless case the finite terms in (3.87) are absorbed by the
renormalization ambiguity, the finite terms in the ultraviolet asymptotics
$m^2\to 0$ of the formfactors with fixed normalization of the effective action
(3.78)
are essential.

Let us consider the massless field in the two--dimensional space. In this case
there are no ultraviolet divergences; instead, the pole in the dimension
$1/(n-2)$ in (3.86) reflects the non--local infrared divergences
\begin{eqnarray}
\setcounter{equation}{88}
\Gamma_{(1)}&=&{1\over 4\pi(n-2)}\int d^nx\,g^{1/2}{\rm str}\,
\left(Q{1\over-\Square}Q+J_\mu{1\over \Square^2}J^\mu\right)
\nonumber\\[10pt]
& &
+{1\over 2(4\pi)}\int d^2x\,g^{1/2}{\rm str}\,
\Biggl\{Q\left(\log\,{-\Square\over 4\pi\mu^2}
+{\bf C}\right){1\over-\Square}Q
\nonumber\\[10pt]
& &
+J_\mu\left(\log\,{-\Square\over 4\pi\mu^2}
+{\bf C}-2\right){1\over \Square^2}J^\mu
+Q{1\over-\Square}R
\nonumber\\[10pt]
& &
+{1\over 12}\hat 1R{1\over-\Square}R+O(R^3)\Biggr\},
\end{eqnarray}
where we made use of the fact that
any two--dimensional
space satisfies identically the Einstein equations
$$
R_{\mu\nu}={1\over 2}g_{\mu\nu}R.
\eqno(3.89)
$$
 For the scalar field,
$J_\mu\equiv\nabla^\alpha{\cal R}_{\alpha\mu}=0$, in the conformally invariant
case (3.51),  $Q=0$, the effective action of the massless field in the
two--dimensional
space is finite up to cubic terms in external fields, (3.88), and has the
form
$$
\Gamma_{(1)}={1\over 24(4\pi)}\int d^2x\,g^{1/2}
\left\{R{1\over -\Square}R+O(R^3)\right\}.
\eqno(3.90)
$$
On the other hand, any two--dimensional space is conformally flat. Therefore,
any functional of the metric $g_{\mu\nu}$ is uniquely determined by the trace
of
 the functional
derivative. Hence, the effective action of the massless conformally invariant field
in
the two--dimensional space can be obtained by the integration of the
conformal anomaly [45, 54, 95]. The exact answer has the form (3.90) without
the third order terms $O(R^3)$, i.e., they vanish in the conformally invariant case.
When the conformal invariance is absent the infrared divergences $Q\Square^{-1}Q$
and $J_\mu\Square^{-2}J^\mu$ as well as the finite terms of higher orders
$O(R^3)$, (3.88), appear.

\section{Summation of the terms without covariant de\-ri\-va\-ti\-ves of
external fields}

In the Sects. 3.3 and 3.4 we summed up the linear and quadratic terms 
in external
fields in the asymptotic expansion of the transfer function $\Omega(s)$
in the powers of the proper time, (1.43). We showed that the corresponding
series (3.26), (3.27) and (3.62)--(3.64) 
converge for any value of the proper time and
define the entire functions (3.29), (3.30) and (3.65)--(3.68).

In the present section we are going to sum up those terms in the asymptotic
expansion of the transfer function at coinciding points, (1.43),
$$
\Omega(s|x,x)=\sum\limits_{k\ge 0}{(is)^k\over k!}[a_k],
\eqno(3.91)
$$
that do not contain the covariant derivatives of the external field. The
summation of the linear and quadratic terms determines the high--energy
asymptotics of the effective action ($\Square R\gg R^2$), whereas the terms
without derivatives determine its low-energy asymptotics ($\Square R\ll R^2$).
We limit ourselves, for simplicity, to the scalar case, i.e., we set
${\cal R}_{\mu\nu}=0$.

The De~Witt coefficients in the coinciding
points $[a_k]$ have the following general form
$$
[a_k]=\sum\limits_{0\le l\le k}Q^l
\underbrace{R^{. . . .}\cdots R_{. . . . }}_{k-l}
+O(\nabla R),
\eqno(3.92)
$$
where the sum contains a finite number of local terms constructed from the
curvature tensor and the potential term $Q$ and $O(\nabla R)$ denotes a finite
number of local terms constructed from the curvature tensor, the potential term
$Q$ and their covariant derivatives, so that in the case of covariantly
constant
curvature tensor and the potential term,
$$
\nabla_\mu R_{\alpha\beta\gamma\delta}=0, \qquad
\nabla_\mu Q=0,
\eqno(3.93)
$$
these terms vanish, $O(\nabla R)=0$.

Thus, to find the terms without covariant derivatives in the De~Witt
coefficients
$[a_k]$, (3.92), it is sufficient to restrict oneself to the constant potential
term, $Q={\rm const}$, and symmetric spaces, (3.93). Let us note that all
symmetric
spaces in two-- three-- and four--dimensional cases are characterized by one
dimensionful constant [164]. Therefore, the terms without derivatives in (3.92)
can be expressed only in terms of the scalar curvature $R$
$$
[a_k]=\sum\limits_{0\le l\le k}c_{k,l} R^{k-l}Q^l
+O(\nabla R),
\eqno(3.94)
$$
where $c_{k,l}$ are numerical coefficients.

On the other hand, from the definition of the De~Witt coefficients as the
coefficients of the asymptotic expansion of the transfer function, (1.43), it is
easy to get the dependence of the De~Witt coefficients on the potential term
$Q$
$$
[a_k]=\sum\limits_{0\le l\le k} {k\choose l}c_{k-l}R^{k-l}Q^l
+O(\nabla R),
\eqno(3.95)
$$
where
$$
c_k\equiv c_{k,0}=R^{-k}[a_k]\Big\vert_{Q=0}
$$
are dimensionless De~Witt coefficients computed for $Q=0$.

In order to calculate the coefficients $c_k$ it is sufficient to consider
the case of De Sitter space, (2.104), with the positive curvature
$$
R^\mu_{\ \alpha\nu\beta}={R\over n(n-1)}\left(\delta^\mu_\nu g_{\alpha\beta}
-\delta^\mu_\beta g_{\alpha\nu}\right),
\qquad R={\rm const} >0,
\eqno(3.96)
$$
since in other symmetric spaces, (3.93), they are the same as in the De Sitter
space.

The De~Witt coefficients in De Sitter space can be calculated in the lines of
the
method developed in the Chap. 2. However, it is more convenient to find them
by
means of the asymptotic expansion of the Green function in the coinciding
points.
It was obtained in the paper [80] in any $n$--dimensional De Sitter space
$$
G(x,x)={i\over (4\pi)^{n/2}}\left({R\over n(n-1)}\right)^{{n\over 2}-1}
\Gamma\left(1-{n\over 2}\right)\Phi_n(\beta),
\eqno(3.97a)
$$
$$
\Phi_n(\beta)={\Gamma\left({n-1\over 2}+i\beta\right)
\Gamma\left({n-1\over 2}-i\beta\right)\over
\Gamma\left({1\over 2}+i\beta\right)\Gamma\left({1\over 2}-i\beta\right)},
\eqno(3.97b)
$$
where
$$
\beta^2={n(n-1)\over R}\left(m^2-Q-{n-1\over 4n}R\right).
\eqno(3.98)
$$

The asymptotic expansion of the function (3.97b) in the inverse
powers of $\beta^2$ has the form
$$
\Phi_n(\beta)=\sum\limits_{k\ge0} d_k(n)\beta^{2\left({n\over 2}-1-k\right)}.
\eqno(3.99)
$$
The coefficients $d_k(n)$ are determined by using the asymptotic expansion
of the gamma--function [168] from the relation
$$
\sum\limits_{k\ge 0} d_k(n)z^k=\exp\left\{
\sum\limits_{k\ge 1}{(-1)^{k+1}\over k(2k+1)}
B_{2k+1}\left({n-1\over 2}\right)z^k\right\},
\eqno(3.100)
$$
$B_k(x)$ being the Bernoulli polynomials [168], and have the form
$$
d_0(n)=1,
$$
$$
d_k(n)=\sum\limits_{1\le l\le k}{(-1)^{k+l}\over l!}
\sum\limits_{{\scriptstyle k_1,\dots,k_l\ge 1\atop
\scriptstyle k_1+\cdots+k_l=k}}
{B_{2k_1+1}\left({n-1\over 2}\right)\over k_1(2k_1+1)}\cdots
{B_{2k_l+1}\left({n-1\over 2}\right)\over k_l(2k_l+1)}.
\eqno(3.101)
$$

On the other hand, using the Schwinger--De~Witt presentation of the Green
function, (1.49), the proper time expansion of the transfer function, (3.91), and
the
equation (3.95), we obtain the asymptotic Schwinger--De~Witt expansion
 for the Green function
$$
G(x,x)={i\over (4\pi)^{n/2}}\left\{{R\over n(n-1)}\right\}^{{n\over 2}-1}
\sum\limits_{k\ge 0}{\Gamma\left(k+1-{n\over 2}\right)\over k!}
b_k(n)\beta^{2\left({n\over 2}-1-k\right)},
\eqno(3.102)
$$
where
$$
b_k(n)= \left\{{R\over n(n-1)}\right\}^{-k}[a_k]\Big\vert_{Q=-{n-1\over 4n}R}
\eqno(3.103)
$$
are the dimensionless De~Witt coefficients calculated for $Q=-{n-1\over 4n}R$.

The total De~Witt coefficients for arbitrary $Q$ are expressed in terms of the
coefficients $b_k$, (3.103), according to (3.95):
$$
[a_k]=\sum\limits_{0\le l\le k}{k\choose l}b_{k-l}(n)
\left({R\over n(n-1)}\right)^{k-l}\left(Q+{n-1\over 4n}R\right)^l
+O(\nabla R).
\eqno(3.104)
$$

Comparing (3.97), (3.99) and (3.102) we obtain the coefficients $b_k(n)$
$$
b_k(n)=(-1)^kk!{\Gamma\left({n\over 2}-k\right)\over
\Gamma\left({n\over 2}\right)}d_k(n).
\eqno(3.105)
$$
To get the coefficients $b_k(n)$ for integer values of the dimension $n$ one
has to take the limit in (3.105). Thereby it is important to take into account
the dependence of the coefficients
$d_k(n)$, (3.101), on the dimension $n$. From the definition of these
coefficients,
(3.97b), (3.99), one can show that in integer dimensions ($n=2, 3, \dots$) they
vanish for $k\ge [n/2]$
$$
d_k(n)=0, \qquad \left(n=2,3,4\dots;\ k\ge \left[{n\over 2}\right]\right).
\eqno(3.106)
$$

Let us consider first the case of odd dimension ($n=3, 5, 7, \dots$).
In this case the gamma--function in (3.105) does not have any poles and the
formula
(3.105) immediately gives the coefficients $b_k(n)$. From (3.106) it follows
that only first $(n-1)/2$ coefficients $b_k(n)$ do not vanish, i.e.,
$$
b_k(n)=0, \qquad \left(n=3, 5, 7, \dots;\
k\ge {n-1\over 2}\right).
\eqno(3.107)
$$
This is the consequence of the finiteness of the Green function (3.97) in odd
dimension.

In even dimensions ($n=2, 4, 6,\dots$) and for $k\le {n\over 2}-1$ the expression
(3.105) is also single--valued and immediately defines the coefficients
$b_k(n)$.
For $k\ge {n\over 2}$ there appear poles in the gamma--function that are
suppressed
by the zeros of the coefficients $d_k(n)$, (3.106). Using the definition of the
coefficients $d_k(n)$, (3.100), (3.101), and the properties of the Bernoulli
 polynomials [168], we obtain the coefficients $b_k(n)$ in even dimension
($n=2, 4, 6, \dots$) for $k\ge n/2$
$$
b_k(n)={(-1)^{k-{n\over 2}}k!\over \Gamma\left({n\over 2}\right)
\Gamma\left(k+1-{n\over 2}\right)}
\sum\limits_{0\le l\le {n\over 2}-1}
{(-1)^l\over k-l}\left(1-2^{1+2l-2k}\right)B_{2k-2l}d_l(n),
\eqno(3.108)
$$
$$
\left(n=2, 4, 6, \dots;\ k\ge {n\over 2}\right),
$$
where $B_l=B_l(0)$ are the Bernoulli numbers and the coefficients $d_l(n)$, ($l\le
n/2-1$),
are calculated by means of the formula (3.101).

Let us list the dimensionless De~Witt coefficients $b_k(n)$, (3.103), in two--
three-- and four--dimensional (physical) space--time. Substituting $n=2$ and $n=4$
in the formulas (3.101), (3.105) and (3.108) we obtain
$$
b_k(2)=(-1)^{k-1}\left(1-2^{1-2k}\right)B_{2k}, \qquad (k\ge 0),
\eqno(3.109)
$$
\begin{equation}\setcounter{equation}{110}
\left\{
\begin{array}{ll}
b_0(4)=1,\\
b_k(4)=(-1)^k\left\{(k-1)\left(1-2^{1-2k}\right)B_{2k}
-{k\over 4}\left(1-2^{3-2k}\right)B_{2k-2}\right\}, \ (k\ge 1).
\end{array}
\right.
\end{equation}

The expression for the coefficients $b_k(n)$ in even dimension for $k\ge n/2$,
(3.108), can be written in a more convenient and compact form.
Using the integral representation of the Bernoulli numbers [168]
$$
B_{2k}=(-1)^{k+1}{4k\over 1-2^{1-2k}}
\int\limits_0^\infty{dt\over e^{2\pi t}+1}t^{2k-1}
\eqno(3.111)
$$
and the definition of the coefficients $d_k(n)$, (3.97b), (3.99)--(3.101), we
obtain from (3.108)
\begin{equation}
\setcounter{equation}{112}
b_k(n)={4(-1)^{{n\over 2}-1}k!\over\Gamma\left({n\over 2}\right)
\Gamma\left(k+1-{n\over 2}\right)}
\int\limits_0^\infty{dt\, t\over e^{2\pi t}+1}
{\Gamma\left({n-1\over 2}+it\right)
\Gamma\left({n-1\over 2}-it\right)\over
\Gamma\left({1\over 2}+it\right)\Gamma\left({1\over 2}-it\right)}t^{2k-n},
\end{equation}
$$
\left(n=2, 4, 6, \dots;\ k\ge {n\over 2}\right).
$$

Thus we have obtained all the terms in the De~Witt coefficients $[a_k]$, that do
not
contain covariant derivatives, (3.104), where the coefficients $b_k(n)$ are
given
by the formulas (3.105), (3.101) and (3.108)--(3.112).

Using the obtained De~Witt coefficients, (3.104), one can calculate the transfer
function
in the coinciding points, (3.91). Substituting (3.104) in (3.91) and summing the
powers of the potential term we obtain
$$
\Omega(s|x,x)=\exp\left\{is\left(Q+{n-1\over 4n}R\right)\right\}
\omega\left({isR\over n(n-1)}\right)+O(\nabla R),
\eqno(3.113)
$$
where
$$
\omega(z)=\sum\limits_{k\ge 0}{z^k\over k!}b_k(n).
\eqno(3.114)
$$
Let us divide the series (3.114) in two parts
$$
\omega(z)=\omega_1(z)+\omega_2(z),
\eqno(3.115)
$$
where
$$
\omega_1(z)=\sum\limits_{0\le k\le \left[{n\over 2}\right]-1}
{z^k\over k!}b_k(n),
\eqno(3.116)
$$
$$
\omega_2(z)=\sum\limits_{k\ge \left[{n\over 2}\right]}
{z^k\over k!}b_k(n).
\eqno(3.117)
$$
The first part $\omega_1(z)$, (3.116), is a polynomial. Using the expression
(3.105) for coefficients $b_k$ and the definition of the coefficients $d_k$,
(3.97), (3.99), one can write it in the integral form
$$
\omega_1(z)=2{(-z)^{{n\over 2}-1}\over \Gamma\left({n\over 2}\right)}
\int\limits_0^\infty dt\,t\exp\,(-t^2)
{\Gamma\left({n-1\over 2}+i{t\over\sqrt{-z}}\right)
\Gamma\left({n-1\over 2}-i{t\over\sqrt{-z}}\right)\over
\Gamma\left({1\over 2}+i{t\over\sqrt{-z}}\right)
\Gamma\left({1\over 2}-i{t\over\sqrt{-z}}\right)}.
\eqno(3.118)
$$

The second part $\omega_2(z)$, (3.117), is an asymptotic series. In odd dimension
all coefficients of this series are equal to zero, (3.107). Therefore, in this
case $\omega_2(z)=0$ up to an arbitrary non--analytical function in the vicinity like
$\exp({\rm const}/z)$.

{}From the obtained expressions (3.108)--(3.112) one can get
the asymptotics of the coefficients $b_k(n)$ as $k\to\infty$ in even dimensions
$$
b_k(n)\Big\vert_{k\to\infty}=4{(-1)^{{n\over 2}-1}\over
\Gamma\left({n\over 2}\right)}
{k!(2k-1)!\over\Gamma\left(k+1-{n\over 2}\right)}(2\pi)^{-2k},
\qquad(n=2, 4, 6,\dots).
\eqno(3.119)
$$
Herefrom it is immediately seen that in this case the asymptotic series
(3.117)
diverges for any $z\ne 0$, i.e., its convergence radius is equal to zero and the
point $z=0$ is a singular point of the function $\omega_2(z)$. This makes the
asymptotic series of the terms without covariant derivatives of external
fields very different from the corresponding series of linear and quadratic 
terms  in
external fields, (3.26), (3.27), (3.62)--(3.64), which converge for any $z$
and define entire functions, (3.29), (3.30), (3.65)--(3.68).

Based on the divergence of the asymptotic series (3.117), it is concluded
in the paper [166] that the Schwinger--De~Witt representation of the Green
function is meaningless. However, the asymptotic divergent series (3.117)
can be used to obtain quite certain idea about the function $\omega_2(z)$.
To do this one has to make use of the methods for summation of asymptotic
series
discussed in Sect. 3.1.

Let us define the function $\omega_2(z)$ according to the formulas (3.5) and (3.3):
$$
\omega_2(z)=\int\limits_0^\infty dy\, y^{\nu-1}e^{-y}B(zy^\mu),
\eqno(3.120)
$$
where
$$
B(z)=\sum\limits_{k\ge {n\over 2}} {z^k\over k!}{b_k\over \Gamma(\mu k+\nu)},
\eqno(3.121)
$$
and $\mu$ and $\nu$ are some complex numbers, such that
${\rm Re}\,\left(\mu{n\over 2}+\nu\right)>0.$ The series (3.121) for the Borel
 function $B(z)$ converges in case ${\rm Re}\,\mu>1$ for any $z$ and in case
${\rm Re}\,\mu=1$ for $|z|<\pi^2$. This can be seen easily using the formulas
(3.4) and (3.119). Let us substitute the integral representation of the
coefficients $b_k$, (3.112), in (3.121)
and change the order of integration and summation. We get
$$
B(z)=-4{(-z)^{{n\over 2}}\over\Gamma\left({n\over 2}\right)}
\int\limits_0^\infty{dt\, t\over e^{2\pi t}+1}
{\Gamma\left({n-1\over 2}+it\right)
\Gamma\left({n-1\over 2}-it\right)\over
\Gamma\left({1\over 2}+it\right)\Gamma\left({1\over 2}-it\right)}H(zt^2),
\eqno(3.122)
$$
where
$$
H(z)=\sum\limits_{k\ge 0}{z^k\over k!\Gamma\left(\mu k+\nu+{n\over 2}\right)}.
\eqno(3.123)
$$

The series (3.123) converges for any $z$ and defines an entire function.
For example, for $\mu=1$, $\nu=1-{n\over 2}$ it reduces to the Bessel
function of zeroth order
$$
H(z)\Big\vert_{\mu=1,\,\nu=1-{n\over 2}}=J_0\left(2\sqrt{-z}\right).
\eqno(3.124)
$$
One can show that the integral (3.122) converges for sure on the negative part
of the real axis, ${\rm Re}\,z<0$, ${\rm Im}\,z=0$. Therefore, for such $z$ it
gives the analytical continuation of the series (3.121). The whole Borel
function $B(z)$ on the whole complex plane can be obtained by the analytic
continuation of the integral (3.122).

Let us substitute now the expression (3.122) for the Borel function in the
integral (3.120) and change the order of integration over $t$ and over $y$.
Integrating over $y$ and summing over $k$ we obtain
$$
\omega_2(z)=-4{(-z)^{{n\over 2}}\over\Gamma\left({n\over 2}\right)}
\int\limits_0^\infty{dt\, t\over e^{2\pi t}+1}
{\Gamma\left({n-1\over 2}+it\right)
\Gamma\left({n-1\over 2}-it\right)\over
\Gamma\left({1\over 2}+it\right)\Gamma\left({1\over 2}-it\right)}\exp(zt^2).
\eqno(3.125)
$$
This integral converges in the region ${\rm Re}\, z\le 0$. In the other
part of the complex plane the function $\omega_2(z)$ is defined by analytical
continuation. In this way we obtain a single--valued analytical function in
the complex plane with a cut along the positive part of the real axis from $0$
to $\infty$. Thus the point $z=0$ is a singular (branch) point
of the function $\omega_2(z)$. It is this fact that makes the power series
over $z$, (3.117), to diverge for any $z\ne 0$.

In the region ${\rm Re}\, z<0$ one can obtain analogous expression for the
total function $\omega(z)$, (3.115). Changing the integration variable
$t\to t\sqrt{-z}$ in (3.118) and adding it to (3.125) we obtain
$$
\omega(z)=2{(-z)^{{n\over 2}}\over\Gamma\left({n\over 2}\right)}
\int\limits_0^\infty dt\, t \tanh\,(\pi t)
{\Gamma\left({n-1\over 2}+it\right)
\Gamma\left({n-1\over 2}-it\right)\over
\Gamma\left({1\over 2}+it\right)\Gamma\left({1\over 2}-it\right)}\exp(zt^2).
\eqno(3.126)
$$

Thus, we summed up the divergent asymptotic series of the terms without
covariant derivatives of the external field in the transfer function, (3.113),
(3.114), (3.126).

Using the obtained transfer function, (3.113), (3.118), (3.126), one can calculate
the one--loop effective action, (1.50), up to terms with covariant derivatives
$O(\nabla R)$. In order to obtain the renormalized effective action
$\Gamma_{(1){\rm ren}}$ with the normalization condition (3.78) it is
sufficient
to subtract from the transfer function $\Omega(s)$, (3.113), the potentially
divergent terms and integrate over the proper time. As the result we obtain in
two--dimensional space
\begin{eqnarray}
\setcounter{equation}{127}
\Gamma_{(1){\rm ren}}\Bigg\vert_{n=2}
&=&{1\over 2(4\pi)}\int d^2x\,g^{1/2}\Biggl\{\left(
m^2-Q-{1\over 6}R\right)\log\,{m^2-Q-{1\over 8}R\over m^2}
\nonumber\\[10pt]
& &
+Q+{1\over 8}R-2R\chi_1\left(2{m^2-Q\over R}-{1\over 4}\right)
+O(\nabla R)\Biggr\},
\end{eqnarray}
and in physical four--dimensional space--time
\begin{eqnarray}
\setcounter{equation}{128}
\Gamma_{(1){\rm ren}}\Bigg\vert_{n=4}
&=&{1\over 2(4\pi)^2}\int d^4x\,g^{1/2}\Biggl\{-{1\over 2}\biggl[
m^4-2m^2\left(Q+{1\over 6}R\right)
\nonumber\\[10pt]
& &
+Q^2+{1\over 3}QR+{29\over 1080}R^2\biggr]
\log\,{m^2-Q-{9\over 48}R\over m^2}
\nonumber\\[10pt]
& &
-{1\over 2}m^2\left(Q+{9\over 48}R\right)+{5\over 4}Q^2+{43\over 96}QR
\nonumber\\[10pt]
& &
+{41\over 1024}R^2+{1\over 36}R^2
\chi_2\left(12{m^2-Q\over R}-{9\over 4}\right)
+O(\nabla R)\Biggr\},
\nonumber\\[10pt]
& &
\end{eqnarray}
where
$$
\chi_1(z)=\int\limits_0^\infty{dt\, t\over e^{2\pi t}+1}
\log\,\left(1-{t^2\over z}\right),
\eqno(3.129)
$$
$$
\chi_2(z)=\int\limits_0^\infty{dt\, t\over e^{2\pi t}+1}
\left(t^2+{1\over 4}\right)
\log\,\left(1-{t^2\over z}\right).
\eqno(3.130)
$$

In odd dimensions the effective action is finite.
Substituting (3.118), (3.113) in (1.50) and integrating over the proper time
we obtain, for example, in three--dimensional space
$$
\Gamma_{(1)}\Bigg\vert_{n=3}={1\over 3(4\pi)}
\int d^3x\,g^{1/2}\left\{\left(m^2-Q-{1\over 6}R\right)^{3/2}
+O(\nabla R)\right\}.
\eqno(3.131)
$$

In higher dimensions the terms without covariant derivatives of the curvature
tensor, (3.92),  no longer reduce only to powers of the scalar curvature
$R$, (3.94), --- they also contain invariants of the Ricci and Weyl tensors.

Thus via the partial summation of the covariantly constant terms of the 
Schwinger--De~Witt 
asymptotic expansion (an expansion in background dimension) we
obtained a non--analytical  (in the background field expression) for the effective
action up to terms with covariant derivatives, (3.127)--(3.131).
Although the corresponding asymptotic series diverge, the expressions
(3.129) and (3.130) are defined in the whole complex plane $z$ with the cut along
the positive part of the real axis. There appears a natural arbitrariness
connected with the possibility to choose the different banks of the cut.

%
%
%
%
%
%
%
%
%
%
%
%
%
%
%
%
%
%
%

\chapter{Higher--derivative quantum
gravity}
\markboth{\sc Chapter 4. Higher--derivative quantum gravity}{\sc Chapter 4.
Higher--derivative quantum gravity}
%
%
%
%
%
%
%
%
%
%
%
%
\section{Quantization of gauge field theories. 
Uni\-que effective action}

Let $M=\{\varphi^i\}$ be the configuration space of a boson gauge field and
$S(\varphi)$ be a classical action functional that is invariant with respect to
local gauge transformations 
$$
\delta \varphi^i=R^i_{\ \alpha}(\varphi)\xi^\alpha,
\eqno(4.1)
$$
forming the gauge group $G$.
Here $\xi^\alpha$ are the group parameters and $R^i_{\ \alpha}(\varphi)$ are the
local generators of the gauge transformations that form a closed Lie algebra
$$
[\hat \RR_\alpha, \hat \RR_\beta]=C^\gamma_{\ \alpha\beta}\hat \RR_\gamma,
\eqno(4.2)
$$
where
$$
\hat \RR_\alpha\equiv R^i_{\ \alpha}(\varphi){\delta\over \delta\varphi^i}
\eqno(4.3)
$$
and $C^\gamma_{\ \alpha\beta}$ are the structure constants of the gauge group
satisfying the Jacobi identity
$$
C^\alpha_{\ \mu[\beta}C^\mu_{\ \gamma\delta]}=0.
\eqno(4.4)
$$
The classical equations of motion determined by the action functional
$S(\varphi)$ have the form
$$
\varepsilon_i(\varphi)=0,
\eqno(4.5)
$$
where $\varepsilon_i\equiv S_{,i}$ is the `extremal' of the action. The
equation (4.5) defines the `mass shell' in the quantum perturbation theory.

The equations (4.5) are not independent. The gauge invariance of the action
leads to the first type constraints between the dynamical variables that are
expressed through the Noether identities
$$
\hat \RR_\alpha S=R^i_{\ \alpha}\varepsilon_i=0.
\eqno(4.6)
$$

The physical dynamical variables are the group orbits (the classes of gauge
equivalent field configurations), and the physical configuration space is the
space of orbits ${\cal M}=M/G$. To have coordinates on the orbit space one has
to put some suuplementary 
gauge conditions (a set of constraints) that isolate in the space
$M$ a subspace ${\cal M}'$, which intersects each orbit only in one point. Each
orbit is represented then by the point in which it intersects the given
subspace ${\cal M}'$. If one reparametrizes the initial configuration space by
the new variables, $M=\{I^A, \chi_\mu\}$, where $I^A$ are the physical
gauge-\-invariant variables enumerating the orbits, ${\cal M}=\{I^A\}$, and
$\chi_\mu$ are group variables that enumerate the points on each orbit,
$G=\{\chi_\mu\}$, then the gauge conditions can be written in the form
$$
\chi_\mu(\varphi)=\theta_\mu,
\eqno(4.6a)
$$
where $\theta_\mu$ are some constants,  i.e., $\theta_{\mu,i}=0$. 
Thus we obtain the
coordinates on the physical space ${\cal M}'=\{I^A, \theta_\mu\}$.

The group variables $\chi_\mu$ transform under the action of the gauge group
analogously to (4.1)
$$
\delta \chi_\mu=Q_{\mu\alpha}(\varphi)\xi^\alpha,
\eqno(4.7)
$$
where
$$
Q_{\mu\alpha}(\varphi)=\chi_{\mu,i}(\varphi)R^i_{\ \alpha}(\varphi), \qquad
(\det Q\ne 0),
\eqno(4.8)
$$
are the generators of gauge transformations of group variables. They form a
representation of the Lie algebra of the gauge group
$$
[\hat\QQ_\alpha, \hat\QQ_\beta]=C^\gamma_{\ \alpha\beta}\hat\QQ_\gamma,
\eqno(4.9)
$$
where
$$
\hat\QQ_\alpha\equiv Q_{\mu\alpha}{\delta\over \delta\chi_\mu} .
$$

All physical gauge--invariant quantities (in particular, the action
$S(\varphi)$) are expressed only in terms of the gauge--invariant variables $I^A$
and do not depend on the group variables $\chi_\mu$
$$
S(\varphi)=\bar S(I^A(\varphi)).
\eqno(4.10)
$$
The action $\bar S(I^A)$ is an usual non--gauge--invariant action. Therefore one
can quantize the theory in the variables $I^A$, and then go back to the initial
field variables $\varphi^i$. The functional integral for the standard effective
action takes the form [45]
\begin{eqnarray}
\setcounter{equation}{11}
\exp\left\{{i\over \hbar}\Gamma(\Phi)\right\}&=&\int d\varphi {\cal
M}(\varphi)\delta(\chi_\mu(\varphi)-\theta_\mu)\det Q(\varphi)
\nonumber\\[10pt]
& &
\times \exp\left\{{i\over
\hbar}\left[S(\varphi)-(\varphi^i-\Phi^i)\Gamma_{,i}(\Phi)\right]\right\},
\end{eqnarray}
where ${\cal M}(\varphi)$ is a local measure. 
The measure ${\cal M}(\varphi)$  is gauge invariant up
to the terms $R^i_{\ \alpha,i}$ and $C^\mu_{\ \alpha\mu}$ that are proportional to
derivatives of the $\delta$--function at coinciding points $\delta(0)$ and that
should vanish by the regularization. The exact form of the measure ${\cal
M}(\varphi)$ must be determined by the canonical quantization of the theory. In
most practically important cases 
${\cal M}(\varphi)=1+\delta(0)(\dots)$ [45]. Therefore,
below we will simply omit the measure ${\cal M}(\varphi)$.

Since the effective action (4.11) must not depend on the arbitrary constants
$\theta_\mu$, one can integrate over them with a Gaussian weight. As result we
get
\begin{eqnarray}
\setcounter{equation}{12}
\exp\left\{{i\over \hbar}\Gamma(\Phi)\right\}&=&\int d\varphi \det Q(\varphi)
(\det H)^{1/2}
\nonumber\\[10pt]
& &
\times \exp\left\{{i\over \hbar}\left[S(\varphi)
-{1\over 2}\chi_\mu(\varphi)H^{\mu\nu}\chi_\nu(\varphi)
-(\varphi^i-\Phi^i)\Gamma_{,i}(\Phi)\right]\right\},
\nonumber\\[10pt]
& &
\end{eqnarray}
where $H^{\mu\nu}$ is a non--degenerate matrix ($\det H\ne 0$), that does not
depend on quantum field ($\delta H^{\mu\nu}/\delta\varphi^i=0$). The
determinants of the operators $Q$ and $H$ in (4.12) can be also represented as
result of integration over the anti--commuting variables, so called
Faddeev-Popov  [15, 14] and Nielsen-Kallosh  [31, 32] ``ghosts''.

Using the equation (4.12) we find the one--loop effective action
$$
\Gamma=S+\hbar \Gamma_{(1)}+O(\hbar^2),
\eqno(4.13)
$$
$$
 \Gamma_{(1)}=-{1\over 2i}\log {\det \Delta\over \det H(\det F)^2},
 \eqno(4.14)
$$
where
$$
\Delta_{ik}=-S_{,ik}+\chi_{\mu i}H^{\mu\nu}\chi_{\nu k},
\qquad \chi_{\mu i}\equiv \chi_{\mu, i}(\varphi)\Big\vert_{\varphi=\Phi},
\eqno(4.15)
$$
$$
F=Q(\varphi)\Big\vert_{\varphi=\Phi}.
\eqno(4.16)
$$

On the mass shell, $\Gamma_{,i}=0$, the standard effective action, (4.12), and
therefore the $S$--matrix too, does not depend neither on the gauge (i.e., on the
choice of arbitrary functions $\chi_\mu$ and $H^{\mu\nu}$) nor on the
parametrization of the quantum field [34, 35, 45]. However, off mass shell the
background field as well as the Green functions and the effective action
crucially depend both on the gauge fixing and on the parametrization of the
quantum field. Moreover, the effective action is not, generally speaking, a
gauge--invariant functional of the background field because the usual procedure of
the gauge fixing of the quantum field automatically fixes the gauge of the
background field.

The gauge invariance of the effective action off mass shell can be preserved by
using the De~Witt's background field gauge. In this gauge the functions
$\chi_\mu(\varphi)$ and the matrix $H^{\mu\nu}$ depend parametrically on the
background field [14, 34]
$$
\chi_\mu(\varphi)=\chi_\mu(\varphi,\Phi),\qquad
\chi_\mu(\Phi,\Phi)=0, \qquad H^{\mu\nu}=H^{\mu\nu}(\Phi),
$$
and are covariant with respect to simultaneous gauge transformations of the
quantum $\varphi$ and background $\Phi$ fields, i.e., they form the adjoint
representation of the gauge group
$$
\hat\RR_\alpha(\varphi)\chi_\mu(\varphi,\Phi)
+\hat\RR_\alpha(\Phi)\chi_\mu(\varphi,\Phi)
+C^\nu_{\ \mu\alpha}\chi_\nu(\varphi, \Phi)=0,
$$
$$
\hat\RR_\alpha(\Phi)H^{\mu\nu}(\Phi)
-C^\mu_{\ \beta\alpha}H^{\beta\nu}(\Phi)
-C^\nu_{\ \beta\alpha}H^{\mu\beta}(\Phi)=0.
\eqno(4.17)
$$

In the background field gauge method one fixes the gauge of the quantum field but not
the gauge of the background field. In this case the standard effective action
(4.12) will be gauge invariant functional of the background field provided the
generators of gauge transformations $R^i_{\ \alpha}(\varphi)$ are linear in the
fields, i.e., $R^i_{\ \alpha, kn}(\varphi)=0$ [34].

However, the off--shell effective action still depends parametrically on the
choice of the gauge (i.e., on the functions $\chi_\mu$, and $H^{\mu, \nu}$), as
well as on the parametrization of the quantum field $\varphi$. As a matter of
fact this is the same problem, since the change of the gauge is in essence a
reparametrization of the physical space of orbits ${\cal M}'$ [45].
Consequently we do not have any unique effective action off mass shell.

The off--shell effective action can give much more information about quantum
processes than just the $S$--matrix. It should give the effective equations for
the background field with all quantum corrections, $\Gamma_{,i}(\Phi)=0$.

A possible solution of the problem of constructing the off--shell effective
action was proposed by Vilkovisky [45, 46] both for usual and the gauge field
theories.

The condition of the reparametrization invariance of the functional
$S(\varphi)$ means that $S(\varphi)$ should be a scalar on the configuration
space $M$ which is treated as a manifold with coordinates $\varphi^i$. The
non-covariance of the effective action $\Gamma(\varphi)$ can be traced to the
source term $(\varphi^i-\Phi^i)\Gamma_{,i}(\Phi)$ in (4.12) as the difference
of coordinates $(\varphi^i-\Phi^i)$ is not a geometric object. One can achieve
covariance if one replaces this difference by a two--point function,
$\sigma^i(\varphi,\Phi)$, that transforms like a vector with respect to
transformations of the background field $\Phi$ and like a scalar with respect to
transformations of the quantum field. This quantity can be constructed by
introducing a symmetric gauge--invariant connection $\Gamma^i_{\ mn}(\varphi)$
on $M$ and by identifying $\sigma^i(\varphi,\Phi)$ with the tangent vector at
the point $\Phi$ to the geodesic connecting the points $\varphi$ and $\Phi$.
More precisely, $\sigma^i(\varphi,\Phi)$ is
 defined to be the solution of the equation
$$
\sigma^k\nabla_k\sigma^i=\sigma^i,
\eqno(4.18)
$$
where
$$
\nabla_k\sigma^i={\delta\over\delta\Phi^k}\sigma^i
+\Gamma^i_{\ km}(\Phi)\sigma^m,
\eqno(4.19)
$$
with the boundary condition
$$
\sigma^i\Big\vert_{\varphi=\Phi}=0.
\eqno(4.20)
$$
Thus we obtain a ``unique effective action'' $\tilde \Gamma(\Phi)$
according to Vilkovisky
\begin{eqnarray}
\setcounter{equation}{21}
\lefteqn{\exp\left\{{i\over \hbar}\tilde\Gamma(\Phi)\right\}=\int d\varphi
(\det H(\Phi))^{1/2}\det Q(\varphi, \Phi)
}\nonumber\\[10pt]
& &
\times \exp\left\{{i\over \hbar}\left[S(\varphi)
-{1\over 2}\chi_\mu(\varphi,\Phi)H^{\mu\nu}(\Phi)\chi_\nu(\varphi,\Phi)
+\sigma^i(\varphi,\Phi)\tilde\Gamma_{,i}(\Phi)\right]\right\},
\nonumber\\[10pt]
& &
\end{eqnarray}
Since the integrand in (4.21) is a scalar on $M$, the equation (4.21) defines a
re\-pa\-ra\-met\-ri\-za\-tion--\-invariant effective action. Let us note, that
such a modification corresponds to the definition of the background field
$\Phi$ in a manifestly reparametrization--invariant way,
$$
<\sigma^i(\varphi,\Phi)>=0,
\eqno(4.22)
$$
instead of the usual definition $\Phi^i=<\varphi^i>$.

The construction of the perturbation theory is performed by the change of the
variables in the functional integral, $\varphi^i\to \sigma^i(\varphi,\Phi)$, and
by expanding all the functionals in the covariant Taylor series
$$
S(\varphi)=\sum_{k\ge 0}{(-1)^k\over k!}\sigma^{i_1}\cdots\sigma^{i_k}
\left[\nabla_{(i_1}\cdots\nabla_{i_k)}S(\varphi)\right]
\Big\vert_{\varphi=\Phi}.
\eqno(4.23)
$$
The diagrammatic technique for the usual effective action results from the
substitution of the covariant functional derivatives $\nabla_i$
 instead of the usual ones in the expression for the standard effective action
(up to the terms $\sim \delta(0)$ that are caused by the Jacobian of the change of
variables).

In particular, in the one--loop approximation, (4.13)--(4.16), 
$$
\tilde\Gamma_{(1)}=-{1\over 2i}\log {\det \tilde\Delta\over \det H(\det F)^2},
 \eqno(4.24)
$$
where
$$
\tilde\Delta_{ik}=-\nabla_i\nabla_k S+\chi_{\mu i}H^{\mu\nu}\chi_{\nu k}.
\eqno(4.25)
$$

To construct the connection on the physical configuration space ${\cal M}'$ let
us introduce, first of all, a non-degenerate gauge--invariant metric
$E_{ik}(\varphi)$ in the initial configuration space $M$ that satisfies the
Killing equations
$$
{\cal D}_m R_{n\alpha}+{\cal D}_n R_{m\alpha}=0,
\eqno(4.26)
$$
where $R_{m\alpha}=E_{mk}R^k_{\ \alpha}$ and ${\cal D}_m $ means the covariant
derivative with Christoffel connection of the metric $E_{ik}$
$$
\Bigl\{ ^i_{jk}\Bigr\}=\frac{1}{2}E^{-1 im}(E_{mj,k}+E_{mk,j}-E_{jk,m}).
\eqno(4.27)
$$
The metric $E_{ik}(\varphi)$ must ensure the non-degeneracy of the matrix
$$
N_{\mu\nu}=R^i_{\ \mu}E_{ik}R^k_{\ \nu}, \qquad
(\det N\ne 0).
\eqno(4.28)
$$
This enables one to define the De~Witt projector [14, 33]
$$
{\Pi_\perp}^m_{\ \,i}=\delta^m_{\ i}-R^m_{\ \alpha}B^\alpha_{\ i},
\qquad \Pi_\perp^2=\Pi_\perp,
\eqno(4.29)
$$
where
$$
B^\mu_{\ n}=N^{-1\; \mu\alpha}R^i_{\ \alpha}E_{in}.
\eqno(4.30)
$$

The projector $\Pi_\perp$ projects every orbit to one point,
$$
{\Pi_\perp}^m_{\ \,i}R^i_{\ \alpha}=0
\eqno(4.31)
$$
and is orthogonal to the generators $R^k_{\ \alpha}$ in the metric $E_{ik}$
$$
R^k_{\ \beta}E_{km}{\Pi_\perp}^m_{\ \,i}=0.
\eqno(4.32)
$$
Therefore, the subspace $\Pi_\perp M$ is the space of the orbits.

The natural physical conditions lead to the following form of the connection on the
configuration space [45, 46]
$$
\Gamma^i_{\ mn}=\left\{ ^i_{\ mn}\right\}+T^i_{\ mn},
\eqno(4.33)
$$
where
$$
T^i_{\ mn}=-2B^\mu_{(m}{\cal D}_{n)}R^i_{\ \mu}
+B^\alpha_{\ (m}B^\beta_{\ n)}R^k_{\ \beta}{\cal D}_kR^i_{\ \alpha}.
\eqno(4.34)
$$
The main property of the connection (4.33) is
$$
\nabla_n R^i_{\ \alpha}\propto R^i_{\ \alpha}.
\eqno(4.35)
$$
It means that the transformation of the quantity $\sigma^i$ under the gauge 
transformations of both the quantum and
background field  is proportional
to $R^i_{\ \alpha}(\Phi)$, and, therefore, ensures the gauge invariance of the
term $\sigma^i\tilde\Gamma_{,i}$ in (4.21). This leads then to the
reparametrization invariance and the independence of the unique
effective action $\tilde \Gamma(\Phi)$ on gauge fixing off mass shell.

On the other hand, it is obvious that on mass shell the unique effective action
coincides with the standard one,
$$
\tilde\Gamma(\Phi)\Big\vert_{\rm on-shell}
=\Gamma(\Phi)\Big\vert_{\rm on-shell},
$$
and leads to the standard $S$--matrix [45].

To calculate the unique effective action one can choose any gauge. The result
will be the same. It is convenient to use the orthogonal gauge, i.e., simply put
the non--physical group variables equal to zero,
$$
\chi_\mu(\varphi, \Phi)=R^i_{\ \mu}(\Phi)E_{ik}(\Phi)\sigma^k(\varphi,\Phi)=0,
\eqno(4.36)
$$
i.e., $\sigma^i=\sigma^i_\perp$, where $\sigma^i_\perp={\Pi_\perp}^i_{\
n}\sigma^n$.
The non-metric part of the connection $T^i_{\ mn}$ satisfies the equation
$$
{\Pi_\perp}^m_{\ k}T^i_{\ mn}{\Pi_\perp}^n_{\ j}=0.
\eqno(4.37)
$$
Using this equation one can show that it does not contribute to the quantity
$\sigma^i_\perp$. Therefore in the orthogonal gauge (4.36) the quantity $T^i_{
\ mn}$ in the connection (4.33) can be omitted. As a result we obtain for the
unique effective action the equation
\begin{eqnarray}
\setcounter{equation}{38}
\exp\left\{{i\over \hbar}\tilde\Gamma(\Phi)\right\}&=&\int d\varphi
\delta\left(R_{i\mu}(\Phi)\sigma^i(\varphi,\Phi)\right)
\det \bar Q(\varphi, \Phi)
\nonumber\\[10pt]
& &
\times \exp\left\{{i\over \hbar}\left[S(\varphi)
+\sigma^i(\varphi,\Phi)\tilde\Gamma_{,i}(\Phi)\right]\right\},
\end{eqnarray}
where
$$
\bar Q_{\mu\nu}(\varphi,\Phi)=R_{i\mu}(\Phi)R^i_{\ \nu}(\varphi)
$$

The change of variables $\varphi \to \sigma^i(\varphi, \Phi)$ and covariant
expansions of all functionals of the form
$$
S(\varphi)=\sum_{k\ge 0}{(-1)^k\over k!}\sigma^{i_1}\cdots\sigma^{i_k}
\left[{\cal D}_{(i_1}\cdots{\cal
D}_{i_k)}S(\varphi)\right]\Big\vert_{\varphi=\Phi}
\eqno(4.39)
$$
lead (up to terms $\sim\delta(0)$) to the standard perturbation theory with simple
replacement of usual functional derivatives by the covariant ones with the
Christoffel connection, ${\cal D}_i$. In particular, in the one--loop
approximation we have
$$
\tilde\Gamma_{(1)}=-{1\over 2i}\log {\det \tilde\Delta_\perp\over (\det N)^2},
 \eqno(4.40)
$$
where
$$
\tilde\Delta_\perp=\Pi_\perp\tilde\Delta\Pi_\perp,
$$
$$
\tilde\Delta_{ik}=-{\cal D}_i{\cal D}_k S=-S_{,ik}
+\left\{ ^j_{\ ik}\right\}S_{,j}
\eqno(4.41)
$$

\section{One--loop divergences of higher--\-de\-ri\-va\-tive quan\-tum
gravity}

The theory of gravity with higher derivatives as well as the Einstein gravity and any
other metric theory of gravity is a non--Abelian gauge theory with the group of
diffeomorphisms $G$ of the space--time as the gauge group.
The complete configuration space $M$ is the space of all pseudo--Riemannian
metrics on the space--time, and the physical configuration space, the space of
orbits ${\cal M}$, is the space of geometries on the space--time.

We will parametrize the gravitational field by the metric tensor of the
space--time
$$
\varphi^i=g_{\mu\nu}(x), \qquad i\equiv (\mu\nu, x).
\eqno(4.42)
$$
The parameters of the gauge transformation are the components of the
infinitesimal vector of the coordinate  transformation
$$
x^\mu\to x^\mu-\xi^\mu(x),
$$
$$
\xi^\mu=\xi^\mu(x), \qquad \mu\equiv (\mu,x).
\eqno(4.43)
$$
The local generators of the gauge transformations in the parametrization (4.42)
are linear in the fields, i.e., $R^i_{\ \alpha,kn}=0$, and have the form
$$
R^i_{\ \alpha}=2\nabla_{(\mu}g_{\nu)\alpha}\delta(x,y), \qquad
i\equiv(\mu\nu,x), \qquad \alpha\equiv(\alpha,y).
\eqno(4.44)
$$
Here and below, when writing the kernels of the differential operators, all the
derivatives act on the first argument of the $\delta$--function.

The local metric tensor of  the configuration  space, that satisfies the
Killing equations (4.26), has the form
$$
E_{ik}=g^{1/2}E^{\mu\nu,\alpha\beta}\delta(x,y), \qquad
i\equiv(\mu\nu,x), \qquad k\equiv(\alpha\beta,y),
$$
$$
E^{\mu\nu,\alpha\beta}=g^{\mu(\alpha}g^{\beta)\nu}
-{1\over 4}(1+\kappa)g^{\mu\nu}g^{\alpha\beta},
\eqno(4.45)
$$
where $\kappa\ne 0$ is a numerical parameter. In the four--dimensional
pseudo--Euclidean space--time it is equal to the determinant of the $10\times 10$
matrix:
$$
\kappa=\det\left(g^{1/2}E^{\mu\nu,\alpha\beta}\right).
\eqno(4.46)
$$
The Christoffel symbols of the metric $E_{ik}$ have the form
$$
\left\{ ^i_{jk}\right\}
=\left\{ ^{\mu\nu,\alpha\beta}_{\lambda\rho}\right\}
\delta(x,y)\delta(x,z),
$$
$$
i\equiv(\lambda\rho,x),\qquad j\equiv(\mu\nu,y),\qquad k\equiv(\alpha\beta,z),
$$
$$
\left\{ ^{\mu\nu,\alpha\beta}_{\lambda\rho}\right\}=
-\delta^{(\mu}_{(\lambda}g^{\nu)(\alpha}\delta^{\beta)}_{\rho)}
+{1\over 4}\left(g^{\mu\nu}\delta^{\alpha\beta}_{\lambda\rho}
+g^{\alpha\beta}\delta^{\mu\nu}_{\lambda\rho}
+\kappa^{-1}E^{\mu\nu,\alpha\beta}g_{\lambda\rho}\right),
$$
$$
\delta^{\mu\nu}_{\alpha\beta}\equiv
\delta^{\mu}_{(\alpha}\delta^{\nu}_{\beta)}.
\eqno(4.47)
$$
The matrix $N_{\mu\nu}(\varphi)$, (4.28), is a second order differential operator
of the form
$$
N_{\mu\nu}(\varphi)=2g^{1/2}\left\{-g_{\mu\nu}\Square
+{1\over 2}(\kappa-1)\nabla_\mu\nabla_\nu-R_{\mu\nu}\right\}\delta(x,y).
\eqno(4.48)
$$
The condition that the operator $N_{\mu\nu}$, (4.48), should be non--degenerate in
the perturbation theory on the flat background,
$\det N\big\vert_{R=0}\ne 0$, imposes a constraint on the parameter of the
metric, $\kappa\ne 3$.

Let us write the classical action with quadratic terms in the curvature of
general type
$$
S(\varphi)=-\int d^4 x g^{1/2}\Biggl\{\epsilon R^*R^*+{1\over 2f^2}C^2
-{1\over 6\nu^2}R^2-{1\over k^2}R+2{\lambda\over k^4}
\Biggr\},
\eqno(4.49)
$$
where $R^*R^*\equiv\frac{1}{4}\varepsilon^{\mu\nu\alpha\beta}
\varepsilon_{\lambda\rho\gamma\delta}R^{\lambda\rho}_{\ \ \mu\nu}
R^{\gamma\delta}_{\  \ \alpha\beta}$ is the topological term,
$\varepsilon_{\mu\nu\alpha\beta}$ is the antisymmetric Levi-Civita tensor,
$C^2\equiv C^{\mu\nu\alpha\beta}C_{\mu\nu\alpha\beta}$ is the square of the
Weyl tensor,
$\epsilon$ is the topological coupling constant, $f^2$ ---  the Weyl one, $\nu^2$ ---
the conformal one, $k^2$ --- the Einstein one, and $\lambda=\Lambda k^2$ is the
dimensionless cosmological constant.
Here and below we omit the surface terms $\sim \Square R$ that do not contribute
neither to the equations of motions nor to the $S$--matrix.

The extremal of the classical action, (4.5), and the Noether identities, (4.6),
have the form
\begin{eqnarray}
\setcounter{equation}{50}
\varepsilon_i(\varphi)&=&\varepsilon^{\mu\nu}\equiv{\delta S\over \delta
g_{\mu\nu}}
=-g^{1/2}\Biggl\{{1\over k^2}\left(R^{\mu\nu}
-{1\over 2}g^{\mu\nu}R+\Lambda g^{\mu\nu}\right)
\nonumber\\[10pt]
& &
+{1\over f^2}\Biggl[{2\over 3}(1+\omega)R(R^{\mu\nu}-{1\over 4}g^{\mu\nu}R)
+{1\over 2}g^{\mu\nu}R_{\alpha\beta}R^{\alpha\beta}
-2R^{\alpha\beta}R^{\mu\ \nu}_{\ \alpha \ \beta}
\nonumber\\[10pt]
& &
+{1\over 3}(1-2\omega)\nabla^\mu\nabla^\nu R-\Square R^{\mu\nu}
+{1\over 6}(1+4\omega)g^{\mu\nu}\Square R\Biggr]
\Biggr\},
\end{eqnarray}
$$
\nabla_\mu\varepsilon^{\mu\nu}=0,
\eqno(4.51)
$$
where $\omega\equiv f^2/(2\nu^2)$.

Let us calculate the one--loop divergences of the standard effective action
(4.14). From (4.14) we have up to the terms $\sim\delta(0)$ 
$$
\Gamma^{\rm div}_{(1)}=-{1\over 2i}\left(\log\det\Delta\big\vert^{\rm div}
-\log\det H\big\vert^{\rm div}
-2\log\det F\big\vert^{\rm div}\right).
\eqno(4.52)
$$
The second variation of the action (4.49) has the form
\begin{eqnarray}
\setcounter{equation}{53}
-S_{,ik}&=&\Biggl\{\hat F_{(0)\lambda\rho\sigma\tau}
\nabla^\lambda\nabla^\rho\nabla^\sigma\nabla^\tau
+\nabla^\rho\hat F_{(2)\rho\sigma}\nabla^\sigma
\nonumber\\[10pt]
& &
+\hat F_{(3)\rho}\nabla^\rho+\nabla^\rho\hat F_{(3)\rho}
+\hat F_{(4)}\Biggr\}g^{1/2}\delta(x,y),
\end{eqnarray}
where $\hat F_{(0)\lambda\rho\sigma\tau}, \hat F_{(2)\rho\sigma},
\hat F_{(3)\rho}$ and $\hat F_{(4)}$ mean the tensor $10\times 10$ matrices
\ ($\hat F_{(0)\lambda\rho\sigma\tau}$
$=F^{\mu\nu,\alpha\beta}_{(0)\lambda\rho\sigma\tau}$, etc.).
They satisfy the following symmetry relations
$$
\hat F^T_{(2)\rho\sigma}=\hat F_{(2)\rho\sigma},\qquad
\hat F^T_{(3)\rho}=-\hat F_{(3)\rho}, \qquad
\hat F^T_{(4)}=\hat F_{(4)},
$$
$$
\hat F_{(2)\rho\sigma}=\hat F_{(2)\sigma\rho},
\eqno(4.54)
$$
with the symbol `T' meaning the transposition, and have the form
\begin{eqnarray*}
F^{\mu\nu,\alpha\beta}_{(0)\lambda\rho\sigma\tau}
&=&{1\over 2f^2}\Biggl\{g_{\lambda\rho}g_{\sigma\tau}
\left(g^{\mu(\alpha}g^{\beta)\nu}-{1+4\omega\over
3}g^{\mu\nu}g^{\alpha\beta}\right)
\nonumber\\[10pt]
& &
+{1+4\omega\over 3}\left(g^{\mu\nu}\delta^{\alpha\beta}_{\sigma\tau}
g_{\lambda\rho}+g^{\alpha\beta}\delta^{\mu\nu}_{\lambda\rho}g_{\sigma\tau}
\right)
\nonumber\\[10pt]
& &
+{2\over 3}(1-2\omega)\delta^{\mu\nu}_{\lambda\rho}
\delta^{\alpha\beta}_{\sigma\tau}
-2\delta^{(\nu}_\lambda g^{\mu)(\alpha}\delta^{\beta)}_\tau g_{\sigma\rho}
\Biggr\},
\end{eqnarray*}
\begin{eqnarray*}
F^{\mu\nu,\alpha\beta}_{(2)\rho\sigma}&=&{1\over 2f^2}\Biggl\{
2R_{\rho\sigma}\left(g^{\mu(\alpha}g^{\beta)\nu}
-{1\over 2}g^{\mu\nu}g^{\alpha\beta}\right)
+4g_{\rho\sigma}R^{(\mu|\alpha|\nu)\beta}
\nonumber\\[10pt]
& &
+2\delta^{(\mu}_{(\sigma}R^{\nu)(\alpha}\delta^{\beta)}_{\rho)}
+2\left(g^{\alpha\beta}\delta^{(\mu}_{(\rho}R^{\nu)}_{\sigma)}
+g^{\mu\nu}\delta^{(\alpha}_{(\rho}R^{\beta)}_{\sigma)}\right)
\nonumber\\[10pt]
& &
-4\left(R^{(\mu}_{(\rho}g^{\nu)(\alpha}\delta^{\beta)}_{\sigma)}
+R^{(\alpha}_{(\rho}g^{\beta)(\mu}\delta^{\nu)}_{\sigma)}\right)
\nonumber\\[10pt]
& &
+{4\over 3}(1+\omega)\left[
R^{\mu\nu}\left(\delta^{\alpha\beta}_{\rho\sigma}
-g^{\alpha\beta}g_{\rho\sigma}\right)
+R^{\alpha\beta}\left(\delta^{\mu\nu}_{\rho\sigma}
-g^{\mu\nu}g_{\rho\sigma}\right)
\right]
\nonumber\\[10pt]
& &
+\left(m^2_2+{2\over 3}(1+\omega)R\right)
\Bigl(-g_{\rho\sigma}g^{\alpha(\mu}g^{\nu)\beta}
+g_{\rho\sigma}g^{\alpha\beta}g^{\mu\nu}
\nonumber\\[10pt]
& &
-g^{\mu\nu}\delta^{\alpha\beta}_{\rho\sigma}
-g^{\alpha\beta}\delta^{\mu\nu}_{\rho\sigma}
+2\delta^{(\beta}_{(\sigma}g^{\alpha)(\mu}\delta^{\nu)}_{\rho)}
\Bigr)\Biggl\},
\end{eqnarray*}
\begin{eqnarray*}
F^{\mu\nu,\alpha\beta}_{(3)\rho}&=&{1\over 2f^2}\Biggl\{
{1\over 2}\left(g^{\alpha\beta}\nabla^{(\mu}R^{\nu)}_\rho
-g^{\mu\nu}\nabla^{(\alpha}R^{\beta)}_\rho\right)
\nonumber\\[10pt]
& &
+{1\over 3}(1-2\omega)\left(g^{\mu\nu}\nabla_\rho R^{\alpha\beta}
-g^{\alpha\beta}\nabla_\rho R^{\mu\nu}\right)
+g^{(\beta(\mu}\nabla^{\nu)}R^{\alpha)}_\rho
\nonumber\\[10pt]
& &
-g^{(\mu(\beta}\nabla^{\alpha)}R^{\nu)}_\rho
+{1\over 12}(1+4\omega)\left(
g^{\alpha\beta}\delta^{(\mu}_\rho\nabla^{\nu)}R
-g^{\mu\nu}\delta^{(\alpha}_\rho\nabla^{\beta)}R\right)
\nonumber\\[10pt]
& &
+{1\over 6}(1-2\omega)\left(
\delta^{(\alpha}_\rho g^{\beta)(\mu}\nabla^{\nu)}R
-\delta^{(\mu}_\rho g^{\nu)(\alpha}\nabla^{\beta)}R\right)
\nonumber\\[10pt]
& &
+{2\over 3}(2-\omega)\left(
\delta^{(\beta}_\rho\nabla^{\alpha)}R^{\mu\nu}
-\delta^{(\mu}_\rho\nabla^{\nu)}R^{\alpha\beta}\right)
\nonumber\\[10pt]
& &
+{3\over 2}\left(\nabla^{(\beta}R^{\alpha)(\mu}\delta^{\nu)}_\rho
-\nabla^{(\mu}R^{\nu)(\alpha}\delta^{\beta)}_\rho\right)
\Biggr\},
\end{eqnarray*}
\begin{eqnarray}
\setcounter{equation}{55}
F^{\mu\nu,\alpha\beta}_{(4)}&=&{1\over 2f^2}\Biggl\{
-{1\over 2}\left(g^{\alpha\beta}R^\mu_\rho R^{\nu\rho}
+g^{\mu\nu}R^\alpha_\rho R^{\beta\rho}\right)
+4R^{(\mu \  \nu)}_{\ \ \rho \ \sigma}R^{\rho\alpha\sigma\beta}
\nonumber\\[10pt]
& &
\nonumber\\[10pt]
& &
-\left(g^{\mu(\alpha}g^{\beta)\nu}-{1\over 2}g^{\mu\nu}g^{\alpha\beta}\right)
\Bigl[R_{\rho\sigma}R^{\rho\sigma}-{1\over 3}(1+\omega)R^2
\nonumber\\[10pt]
& &
+{1\over 3}(1+4\omega)\Square R
-m_2^2(R-2\Lambda)\Bigr]
\nonumber\\[10pt]
& &
-{3\over 2}R^{\rho\sigma}\left(g^{\alpha\beta}R^{\mu\ \nu}_{\ \rho \ \sigma}
+g^{\mu\nu}R^{\alpha\ \beta}_{\ \rho \ \sigma}\right)
+R^{\alpha(\mu}R^{\nu)\beta}
\nonumber\\[10pt]
& &
+\left(m_2^2+{2\over 3}(1+\omega)R\right)\Bigl(R^{\mu\nu}g^{\alpha\beta}
+R^{\alpha\beta}g^{\mu\nu}
\nonumber\\[10pt]
& &
-3g^{(\alpha(\mu}R^{\nu)\beta)}
-R^{(\mu|\alpha|\nu)\beta}\Bigr)
\nonumber\\[10pt]
& &
+6g^{(\beta(\nu}R^{\mu)\ \alpha)}_{\ \; \rho \ \ \sigma}R^{\rho\sigma}
+{1\over 2}\left(R^{(\alpha\ \beta)(\mu}_{\ \ \rho}R^{\nu)\rho}
+R^{(\mu\ \nu)(\alpha}_{\ \ \rho}R^{\beta)\rho}
\right)
\nonumber\\[10pt]
& &
-{4\over 3}(1+\omega)R^{\mu\nu}R^{\alpha\beta}
-\omega\left(g^{\alpha\beta}\nabla^\mu\nabla^\nu R
+g^{\mu\nu}\nabla^\alpha\nabla^\beta R
\right)
\nonumber\\[10pt]
& &
+2\Square R^{(\mu|\alpha|\nu)\beta}
-{4\over 3}(1-2\omega)g^{(\beta(\nu}\nabla^{\mu)}\nabla^{\alpha)}R
\nonumber\\[10pt]
& &
+2\Square R^{(\mu(\alpha}g^{\beta)\nu)}
+{2\over 3}(1+\omega)\Bigl(
\nabla^{(\alpha}\nabla^{\beta)}R^{\mu\nu}
\nonumber\\[10pt]
& &
+\nabla^{(\mu}\nabla^{\nu)}R^{\alpha\beta}
-g^{\alpha\beta}\Square R^{\mu\nu}-g^{\mu\nu}\Square R^{\alpha\beta}
\Bigr)\Biggr\},
\end{eqnarray}
where $m_2^2=f^2/k^2$.

Next let us choose the most general linear covariant De~Witt gauge condition
[33]
$$
\chi_\mu(\varphi,\Phi)=R^k_{\ \mu}(\Phi)E_{ki}(\Phi)h^i,
\eqno(4.56)
$$
where $h^i=\varphi^i-\Phi^i$. In usual notation this condition reads
$$
\chi_\mu=-2g^{1/2}\nabla_\nu\left\{h^\nu_\mu
-\frac{1}{4}(1+\kappa)\delta^\nu_\mu h\right\},
\eqno(4.57)
$$
where raising and lowering the indices as well as the covariant derivative
are defined by means of the background metric $g_{\mu\nu}$ and $h\equiv
h^\mu_\mu$. The ghost operator $F$, (4.8), (4.16), in the gauge (4.56) is equal
to the operator $N$, (4.48),
$$
F_{\mu\nu}=N_{\mu\nu}=2g^{1/2}\left\{-g_{\mu\nu}\Square
+\frac{1}{2}(\kappa-1)\nabla_\mu\nabla_\nu-R_{\mu\nu}\right\}\delta(x,y).
\eqno(4.57a)
$$

It is obvious that for the operator $\Delta$, (4.15), to be nondegenerate in
the flat space--time it is necessary to choose the matrix $H$ as a second order
differential operator
$$
H^{\mu\nu}={1\over 4\alpha^2}g^{-1/2}\left\{-g^{\mu\nu}\Square
+\beta\nabla^\mu\nabla^\nu+R^{\mu\nu}+P^{\mu\nu}\right\}\delta(x,y),
\eqno(4.58)
$$
where $\alpha$ and $\beta$ are numerical constants, $P^{\mu\nu}$ is an
arbitrary symmetric tensor, e.g.
$$
P^{\mu\nu}=p_1R^{\mu\nu}+g^{\mu\nu}\left(p_2R+p_3{1\over k^2}\right),
\eqno(4.59)
$$
$p_1$, $p_2$ and $p_3$ being some arbitrary numerical constants. Such form of
the operator $H$ does not increase the order of the operator $\Delta$, (4.15),
and preserves its locality. Thus we obtain a very wide six--parameter ($\kappa$,
$\alpha$, $\beta$, $p_1$, $p_2$, $p_3$) class of gauges. In particular, the
harmonic De~Donder--Fock--Landau gauge,
$$
\nabla_\nu\left(h^\nu_\mu-\frac{1}{2}\delta^\nu_\mu h\right)=0,
\eqno(4.60)
$$
corresponds to $\kappa=1$ and $\alpha=0$. For $\alpha=0$ the dependence on other
parameters disappear.

It is most convenient to choose the ``minimal'' gauge
$$
\alpha^2=\alpha_0^2\equiv f^2, \qquad
\beta=\beta_0\equiv\frac{1}{3}(1-2\omega), \qquad
\kappa=\kappa_0\equiv{3\omega\over 1+\omega},
\eqno(4.61)
$$
which makes the operator $\Delta$, (4.15), diagonal in leading derivatives, i.e.,
it takes the form
$$
\Delta_{ik}={1\over 2f^2}\left\{\hat E_{(0)}\Square^2
+\nabla^\rho\hat D_{\rho\sigma}\nabla^\sigma
+\nabla^\rho\hat V_{\rho} + \hat V_{\rho}\nabla^\rho
+\hat P\right\}\delta(x,y),
\eqno(4.62)
$$
where
$$
\hat E_{(0)}=E^{\mu\nu,\alpha\beta}(\kappa=\kappa_0), \qquad
\hat D_{\rho\sigma}=D_{\rho\sigma}^{\ \mu\nu,\alpha\beta},
$$
$$
\hat V_\rho=V_\rho^{\ \mu\nu,\alpha\beta}, \qquad
\hat P=P^{\mu\nu,\alpha\beta},
$$
are some tensor matrices that satisfy the same symmetry conditions, (4.54). They
have the following form
\begin{eqnarray*}
D^{\mu\nu,\alpha\beta}_{\rho\sigma}&=&{1\over 2f^2}\Biggl\{
2R_{\rho\sigma}\left[g^{\alpha(\mu}g^{\nu)\beta}
-{1\over 2}\left(1+{p_1\over 16}
\left({1+4\omega\over 1+ \omega}\right)^2\right)
g^{\mu\nu}g^{\alpha\beta}\right]
\\[10pt]
& &
+2\left(1+{p_1\over 4}
\left({1+4\omega\over 1+ \omega}\right)\right)
\left(g^{\alpha\beta}R^{(\mu}_{(\rho}\delta^{\nu)}_{\sigma)}
+g^{\mu\nu}R^{(\alpha}_{(\rho}\delta^{\beta)}_{\sigma)}\right)
+4g_{\rho\sigma}R^{(\mu|\alpha|\nu)\beta}
\\[10pt]
& &
-2p_1\delta^{(\mu}_{(\rho}R^{\nu)(\alpha}\delta^{\beta)}_{\sigma)}
-4\left(\delta^{(\alpha}_{(\rho}g^{\beta)(\nu}R^{\mu)}_{\sigma)}
+\delta^{(\mu}_{(\sigma}g^{\nu)(\beta}R^{\alpha)}_{\rho)}\right)
\\[10pt]
& &
+{4\over 3}(1+\omega)\left[
R^{\mu\nu}\left(\delta^{\alpha\beta}_{\rho\sigma}
-g^{\alpha\beta}g_{\rho\sigma}\right)
+R^{\alpha\beta}\left(\delta^{\mu\nu}_{\rho\sigma}
-g^{\mu\nu}g_{\rho\sigma}\right)
\right]
\\[10pt]
& &
+\left(m^2_2+{2\over 3}(1+\omega)R\right)
\Bigl(-g_{\rho\sigma}g^{\alpha(\mu}g^{\nu)\beta}
+g_{\rho\sigma}g^{\mu\nu}g^{\alpha\beta}
-g^{\mu\nu}\delta^{\alpha\beta}_{\rho\sigma}
\\[10pt]
& &
-\delta^{\mu\nu}_{\rho\sigma}g^{\alpha\beta}
+2\delta^{(\beta}_{(\sigma}g^{\alpha)(\mu}\delta^{\nu)}_{\rho)}
\Bigr)
+\left(p_2R+p_3{1\over k^2}\right)\Biggl[
-2\delta^{(\beta}_{(\sigma}g^{\alpha)(\mu}\delta^{\nu)}_{\rho)}
\\[10pt]
& &
+{1\over 2}\cdot{1+4\omega\over
1+\omega}\left(g^{\mu\nu}\delta^{\alpha\beta}_{\rho\sigma}
+g^{\alpha\beta}\delta^{\mu\nu}_{\rho\sigma}\right)
-{1\over 8}\left({1+4\omega\over 1+\omega}\right)^2
g_{\rho\sigma}g^{\mu\nu}g^{\alpha\beta}
\Biggr]\Biggl\},
\end{eqnarray*}

\begin{eqnarray*}
V^{\mu\nu,\alpha\beta}_{\rho}&=&{1\over 2f^2}\Biggl\{
\left({1\over 2}+{p_1\over 8}\cdot{1+4\omega\over 1+\omega}\right)
\left(g^{\alpha\beta}\nabla^{(\mu}R^{\nu)}_\rho
-g^{\mu\nu}\nabla^{(\alpha}R^{\beta)}_\rho\right)
\\[10pt]
& &
+{1\over 3}(1-2\omega)\left(g^{\mu\nu}\nabla_\rho R^{\alpha\beta}
-g^{\alpha\beta}\nabla_\rho R^{\mu\nu}\right)
\\[10pt]
& &
+g^{(\beta(\mu}\nabla^{\nu)}R^{\alpha)}_\rho
-g^{(\mu(\beta}\nabla^{\alpha)}R^{\nu)}_\rho
\\[10pt]
& &
+{1\over 12}(1+4\omega)\left(1-{3\over 4}\cdot{p_1\over 1+\omega}\right)
\left(
g^{\alpha\beta}\delta^{(\mu}_\rho\nabla^{\nu)}R
-g^{\mu\nu}\delta^{(\alpha}_\rho\nabla^{\beta)}R\right)
\\[10pt]
& &
+\left({1-2\omega\over 6}-{p_2\over 2}\right)
\left(
\delta^{(\alpha}_\rho g^{\beta)(\mu}\nabla^{\nu)}R
-\delta^{(\mu}_\rho g^{\nu)(\alpha}\nabla^{\beta)}R\right)
\\[10pt]
& &
+{2\over 3}(2-\omega)\left(
\delta^{(\alpha}_\rho\nabla^{\beta)}R^{\mu\nu}
-\delta^{(\mu}_\rho\nabla^{\nu)}R^{\alpha\beta}\right)
\\[10pt]
& &
+\left(2+{p_1\over 2}\right)
\left(\nabla^{(\alpha}R^{\beta)(\mu}\delta^{\nu)}_\rho
-\nabla^{(\mu}R^{\nu)(\alpha}\delta^{\beta)}_\rho\right)
\Biggr\},
\end{eqnarray*}

\begin{eqnarray}
\setcounter{equation}{63}
P^{\mu\nu,\alpha\beta}&=&{1\over 2f^2}\Biggl\{
-{1\over 2}\left(1+{p_1\over 4}\cdot{1+4\omega\over 1+\omega}\right)
\left(g^{\alpha\beta}R^\mu_\rho R^{\nu\rho}
+g^{\mu\nu}R^\alpha_\rho R^{\beta\rho}
\right)
\nonumber\\[10pt]
& &
-\left(g^{\mu(\alpha}g^{\beta)\nu}-{1\over 2}g^{\mu\nu}g^{\alpha\beta}\right)
\Biggl[R_{\rho\sigma}R^{\rho\sigma}
-{1\over 3}(1+\omega)R^2
\nonumber\\[10pt]
& &
+{1\over 3}(1+4\omega)\Square R
-m_2^2(R-2\Lambda)\Biggr]
\nonumber\\[10pt]
& &
-{3\over 2}\left(1-{p_1\over 12}\cdot{1+4\omega\over 1+\omega}\right)
R^{\rho\sigma}\left(g^{\alpha\beta}R^{\mu\ \nu}_{\ \rho \ \sigma}
+g^{\mu\nu}R^{\alpha\ \beta}_{\ \rho \ \sigma}\right)
\nonumber\\[10pt]
& &
+(2+p_1)R^{\alpha(\mu}R^{\nu)\beta}
-{1\over 2}p_1\left(R^{\rho(\mu}R^{\nu)(\alpha\ \beta)}_{\ \ \ \ \rho}
+R^{\rho(\alpha}R^{\beta)(\mu\ \nu)}_{\ \ \ \ \rho}
\right)
\nonumber\\[10pt]
& &
+6g^{(\beta(\nu}R^{\mu)\ \alpha)}_{\ \ \rho \ \ \sigma}R^{\rho\sigma}
-{4\over 3}(1+\omega)R^{\mu\nu}R^{\alpha\beta}
+4R^{(\mu\ \nu)\sigma}_{\ \ \rho}R^{\rho(\alpha\ \beta)}_{\ \ \ \sigma}
\nonumber\\[10pt]
& &
+\left(m_2^2+{2\over 3}(1+\omega)R\right)
\Biggl(R^{\mu\nu}g^{\alpha\beta}
+R^{\alpha\beta}g^{\mu\nu}
\nonumber\\[10pt]
& &
-3g^{(\alpha(\mu}R^{\nu)\beta)}
-R^{(\mu|\alpha|\nu)\beta}\Biggr)
\nonumber\\[10pt]
& &
+\left(p_2R+p_3{1\over k^2}\right)
\left(g^{(\alpha(\mu}R^{\nu)\beta)}
-R^{(\mu|\alpha|\nu)\beta}\right)
+2\Square R^{(\mu|\alpha|\nu)\beta}
\nonumber\\[10pt]
& &
-\omega\left(g^{\alpha\beta}\nabla^\mu\nabla^\nu R
+g^{\mu\nu}\nabla^\alpha\nabla^\beta R
\right)
+2\Square R^{(\mu(\alpha}g^{\beta)\nu)}
\nonumber\\[10pt]
& &
+{2\over 3}(1+\omega)
\Big(-g^{\alpha\beta}\Square R^{\mu\nu}
-g^{\mu\nu}\Square R^{\alpha\beta}
+\nabla^{(\alpha}\nabla^{\beta)}R^{\mu\nu}
\nonumber\\[10pt]
& &
+\nabla^{(\mu}\nabla^{\nu)}R^{\alpha\beta}\Big)
-{4\over 3}(1-2\omega)g^{(\beta(\nu}\nabla^{\mu)}\nabla^{\alpha)}R
\Biggr\}.
\end{eqnarray}

The divergences of the determinants of the operators $H$, (4.58), $F$, (4.57a),
and $\Delta$, (4.62), can be calculated by means of the algorithms for the
non-minimal vector operator of second order and the minimal tensor operator of
forth order. These algorithms were obtained first in [152--154] and confirmed in
[77] by using the generalized Schwinger--De~Witt technique. In the dimensional
regularization  up to the terms $\sim \Square R$ they have the form
\begin{eqnarray}
\setcounter{equation}{64}
\lefteqn{\log \det \left\{-\Square\delta^\mu_\nu+\beta\nabla^\mu\nabla_\nu
+R^\mu_\nu+P^\mu_\nu\right\}\Big\vert^{\rm div}
}\qquad
\nonumber\\[10pt]
& &
=i{2\over (n-4)(4\pi)^2}\int d^4 x g^{1/2}\Biggl\{
-{8\over 45}R^*R^*+{7\over 60}C^2+{1\over 36}R^2
\nonumber\\[10pt]
& &
+{1\over 6}(\xi+6)R_{\mu\nu}P^{\mu\nu}-{1\over 12}(\xi+2)RP
+{1\over 48}\xi^2P^2
\nonumber\\[10pt]
& &
+{1\over 24}(\xi^2+6\xi+12)P_{\mu\nu}P^{\mu\nu}\Biggr\},
\end{eqnarray}
where
$$
\xi\equiv{\beta\over 1-\beta}, \qquad P\equiv P^\mu_{\ \mu},
$$
and
\begin{eqnarray}
\setcounter{equation}{65}
\lefteqn{
\log\det\left\{\hat E\Square^2
+\nabla^\mu\hat D_{\mu\nu}\nabla^\nu
+\nabla^\mu\hat V_{\mu} + \hat V_{\mu}\nabla^\mu
+\hat P\right\}\Big\vert^{\rm div}
}\qquad
\nonumber\\[10pt]
& &
=i{2\over (n-4)(4\pi)^2}\int d^4 x g^{1/2}{\rm tr}\Biggl\{
\hat 1\left(-{1\over 180}R^*R^*+{1\over 60}C^2+{1\over 36}R^2
\right)
\nonumber\\[10pt]
& &
+{1\over 6}\hat{\cal R}_{\mu\nu}\hat{\cal R}^{\mu\nu}
-\hat E^{-1}\hat P+{1\over 12}R\hat E^{-1}\hat D
-{1\over 6}R_{\mu\nu}\hat E^{-1}\hat D^{\mu\nu}
\nonumber\\[10pt]
& &
+{1\over 48}\hat E^{-1}\hat D\hat E^{-1}\hat D
+{1\over 24}\hat E^{-1}\hat D_{\mu\nu}\hat E^{-1}\hat D^{\mu\nu}\Biggr\},
\end{eqnarray}
where
$$
\hat D\equiv \hat D^\mu_{\ \mu},
$$
$\hat E$, $\hat D^{\mu\nu}$, $\hat V^\mu$ and $\hat P$ are the tensor matrices
($\hat E= E^{AB}$, $\hat E^{-1}=E^{-1}_{AB}$, $\hat D^{\mu\nu}=D^{AB\mu\nu}$
etc.), $\hat 1=\delta^A_B$, $\hat {\cal R}_{\mu\nu}={\cal R}^A_{\ B\mu\nu}$ is the
commutator of covariant derivatives of the tensor field
$$
[\nabla_\mu,\nabla_\nu]h^A={\cal R}^A_{\ B\mu\nu}h^B,
\eqno(4.66)
$$
`tr' means the matrix trace and $n$ is the dimension of the spacetime.

In our case, $h^A=h_{\mu\nu}$,
$$
{\cal R}^A_{\ B\mu\nu}={\cal R}^{\ \ \alpha\beta}_{\gamma\delta\ \ ,\mu\nu}
=-2\delta^{(\alpha}_{(\gamma}R^{\beta)}_{\ \ \delta)\mu\nu}.
\eqno(4.67)
$$

Using eqs. (4.64) and (4.65) we obtain in the minimal gauge (4.61)
\begin{eqnarray}
\setcounter{equation}{68}
\lefteqn{\log\det H\Big\vert^{\rm div}
=i{2\over (n-4)(4\pi)^2}\int d^4 x g^{1/2}\Biggl\{
-{8\over 45}R^*R^*
+{7\over 60}C^2}\qquad\qquad
\nonumber\\[10pt]
& &
+{1\over 36}R^2
+{13+10\omega\over 12(1+\omega)}R_{\mu\nu}P^{\mu\nu}
-{5+2\omega\over 24(1+\omega)}RP
\nonumber\\[10pt]
& &
+{1\over 192}\left({1-2\omega\over 1+\omega}\right)^2P^2
+{28\omega^2+80\omega+61\over 96(\omega+1)^2}P_{\mu\nu}P^{\mu\nu}\Biggr\},
\nonumber\\[10pt]
& &
\end{eqnarray}

\begin{eqnarray}
\log\det F\Big\vert^{\rm div}
&=&i{2\over (n-4)(4\pi)^2}\int d^4 x g^{1/2}\Biggl\{
-{1\over 540}(20\omega^2+100\omega+41)R^*R^*
\nonumber\\[10pt]
& &
+{1\over 135}(5\omega^2+25\omega+2)C^2
+{1\over 81}(5\omega^2+16\omega+20)R^2\Biggr\},
\nonumber\\[10pt]
& &
\end{eqnarray}

\begin{eqnarray}
\log\det \Delta\Big\vert^{\rm div}
&=&i{2\over (n-4)(4\pi)^2}\int d^4 x g^{1/2}\Biggl\{
-{1\over 54}(4\omega^2+20\omega+253)R^*R^*
\nonumber\\[10pt]
& &
+{1\over 54}(4\omega^2+20\omega+367)C^2
+{1\over 162}(200\omega^2+334\omega+107)R^2
\nonumber\\[10pt]
& &
+{1\over 6}\left(40\omega-26-{3\over \omega}\right){f^2\over k^2}R
\nonumber\\[10pt]
& &
+{1\over k^4}\left[{4\over 3}\lambda(14f^2+\nu^2)+{1\over 2}(5f^4+\nu^4)\right]
-{13+10\omega\over 12(1+\omega)}R_{\mu\nu}P^{\mu\nu}
\nonumber\\[10pt]
& &
+{5+2\omega\over 24(1+\omega)}RP
-{1\over 192}\left(1-2\omega\over 1+\omega\right)^2P^2
\nonumber\\[10pt]
& &
-{28\omega^2+80\omega+61\over 96(\omega+1)^2}P_{\mu\nu}P^{\mu\nu}
\Biggr\}.
\end{eqnarray}

Substituting the obtained expressions (4.68)--(4.70) in (4.52) we get the
divergences of the standard one--loop effective action off mass shell
\begin{eqnarray}
\Gamma_{(1)}^{\rm div}&=&
-{1\over (n-4)(4\pi)^2}\int d^4 x g^{1/2}
\Biggl\{\beta_1R^*R^*+\beta_2C^2
\nonumber\\[10pt]
& &
+\beta_3R^2+\beta_4{1\over k^4}+\gamma{1\over k^2}(R-4\Lambda)\Biggr\},
\end{eqnarray}
where
$$
\beta_1=-{196\over 45},
$$
$$
\beta_2={133\over 20},
\eqno(4.72)
$$
$$
\beta_3={5\over 18}{f^4\over \nu^4}+{5\over 6}{f^2\over \nu^2}+{5\over 36},
\eqno(4.73)
$$
$$
\beta_4={1\over 2}(5f^4+\nu^4)
+{2\over 3}\lambda\left(10{f^4\over \nu^2}+15f^2-\nu^2\right),
\eqno(4.74)
$$
$$
\gamma={5\over 3}{f^4\over \nu^2}-{13\over 6}f^2-{1\over 2}\nu^2.
\eqno(4.75)
$$

Therefrom it is immediately seen that the gauge fixing tensor, $P^{\mu\nu}$,
(4.59) does not enter the result. In the next section we will calculate the
divergences of the effective action in arbitrary gauge and will show that the
tensor $P^{\mu\nu}$ does not contribute in the divergences in general case too.
If one puts $P_{\mu\nu}=0$ then the divergences of the operator $H$ do not
depend on the gauge fixing parameters at all.

Our result for divergences, (4.71)--(4.75), does not coincide with the results of
the papers [152--155] in the coefficient $\beta_3$, (4.73). Namely, the last term
in (4.73) is equal to $5/36$ instead of the incorrect value $-1/36$ obtained in
[152--155]. We will check our result, (4.73), by means of completely independent
computation on the De Sitter background in Sect. 4.5.

\section{Off--shell one--loop divergences of  the
stan\-dard effective action in arbitrary
gau\-ge and the  divergences of the unique
effective action}

Let us study now the dependence of the obtained result for the divergences of the
standard effective action, (4.71)--(4.75), on the choice of the gauge. Let us
consider the variation of the one--loop effective action (4.1) with respect to
variation of the gauge condition (i.e., the functions $\chi_\mu$ and
$H^{\mu\nu}$)
\begin{eqnarray}
\setcounter{equation}{76}
\delta\Gamma_{(1)}&=&-{1\over i}\Biggl\{\left(
\Delta^{-1\; ik}\chi_{\mu k}H^{\mu\nu}
-R^i_{\ \alpha}F^{-1\; \alpha\nu}\right)\delta\chi_{\nu i}
\nonumber\\[10pt]
& &
+{1\over 2}\left(
\chi_{\mu i}\Delta^{-1\; ik}\chi_{\nu k}-H^{-1}_{\mu\nu}
\right)\delta H^{\mu\nu}\Biggr\}.
\end{eqnarray}
Using the Ward identities
$$
\Delta^{-1\; ik}\chi_{\mu k}H^{\mu\nu}-R^i_{\ \alpha}F^{-1\; \alpha \nu}
=-\Delta^{-1\; ik}\varepsilon_jR^j_{\ \alpha,k}F^{-1\; \alpha\nu},
\eqno(4.77)
$$
\begin{eqnarray}
\setcounter{equation}{78}
\chi_{\mu i}\Delta^{-1\; ik}\chi_{\nu k}-H^{-1}_{\ \mu\nu}&=&
-H^{-1}_{\ \mu\alpha}F^{-1\; \beta\alpha}\varepsilon_jR^j_{\ \beta,k}
\nonumber\\[10pt]
& &
\times\left(R^k_{\ \gamma}-\Delta^{-1\; kn}\varepsilon_mR^m_{\ \gamma,n}
\right)F^{-1\; \gamma\delta}H^{-1}_{\ \delta\nu},
\end{eqnarray}
that follow from the Noether identity, (4.6), one can derive from (4.76)
\begin{eqnarray}
\lefteqn{\delta\Gamma_{(1)}=-{1\over i}\Biggl\{
-\Delta^{-1\; ik}\varepsilon_jR^j_{\ \alpha,k}F^{-1\; \alpha\nu}
\delta\chi_{\nu i}}
\nonumber\\[10pt]
& &
+{1\over 2}F^{-1\; \beta\alpha}\varepsilon_jR^j_{\ \beta,k}
\left(R^k_{\ \gamma}-\Delta^{-1\; kn}\varepsilon_mR^m_{\ \gamma,n}
\right)F^{-1\; \gamma\delta}\delta (H^{-1}_{\ \delta\alpha})\Biggr\}.
\end{eqnarray}
Herefrom, it follows, in particular, that the one--loop effective action on
mass shell, $\varepsilon=0$, (4.5), does not depend on the gauge,
$$
\delta\Gamma\Big\vert_{\rm on-shell}=0.
\eqno(4.80)
$$

Since the effective action on the mass shell is well defined, it is analytical
in background fields in the neighborhood of the mass shell (4.5). Therefore, it
can be expanded in powers of the extremal [45]. As the extremal has the
background dimension (in our case, (4.50), equal to four in mass units), this
expansion will be, in fact, an expansion in the background dimension. It is
obvious, that to calculate the divergences of the effective action it is
sufficient to limit oneself to the terms of background dimension not greater
than four. Thus one can obtain the divergences by taking into account only
linear terms in the extremal. Moreover, from the dimensional grounds it follows
that only the trace of the extremal (4.50),
$$
\varepsilon\equiv g_{\mu\nu}\varepsilon^{\mu\nu}=g^{1/2}\left\{
{1\over k^2}(R-4\Lambda)-{1\over \nu^2}\Square R\right\},
\eqno(4.81)
$$
contributes to the divergences. Therefore, a dependence of the divergent part
of the effective action on the gauge parameters appears only in the
$\gamma$--coefficient, (4.75). The other, $\beta$--coefficients, (4.72)--(4.74), do
not depend on the gauge.

So, from (4.79) we obtain the variation of the one--loop effective action with
respect to the variation of the gauge
\begin{eqnarray}
\setcounter{equation}{82}
\delta\Gamma_{(1)}^{\rm div}
&=&{1\over i}\Biggl\{
\varepsilon_jR^j_{\ \alpha,k}\Delta^{-1\; ki}F^{-1\; \alpha\beta}
\delta\chi_{\beta i}
\nonumber\\[10pt]
& &
-{1\over 2}\varepsilon_jR^j_{\ \alpha,k}
R^k_{\ \beta}F^{-1\; \beta\gamma}F^{-1\; \alpha\delta}\delta (H^{-1}_{\
\gamma\delta})\Biggr\}\Bigg\vert^{\rm div}.
\end{eqnarray}

Herefrom one can obtain the divergences of the effective action in any gauge.
To do this one has, first, to fix some form of the gauge condition with
arbitrary parameters, then, to calculate the divergences of the effective action
for some convenient choice of the gauge parameters and, finally, to integrate the
equation (4.82) over the gauge parameters.

Let us restrict ourselves to the covariant De~Witt gauge, (4.56), with arbitrary
gauge parameters $\alpha$, $\beta$ and $P^{\mu\nu}$. Since the coefficient at
the variation $\delta H^{-1}$ in (4.82) does not depend on the matrix $H$ (i.e.,
on the gauge parameters $\alpha$, $\beta$ and $P^{\mu\nu}$), one can integrate
over the operator $H$ immediately. Thus we obtain the divergences of the
effective action in arbitrary gauge
\begin{eqnarray}
\setcounter{equation}{83}
\Gamma_{(1)}^{\rm div}(\kappa,\alpha,\beta,P)
&=&\Gamma_{(1)}^{\rm div}(\kappa_0,\alpha_0,\beta_0)
+{1\over i}\Biggl\{\int\limits_{\kappa_0}^{\kappa}d \kappa\; U_1^{\rm
div}(\kappa)
\nonumber\\[10pt]
& &
-{1\over 2}\biggl[U_2^{\rm div}(\kappa,\alpha,\beta,P)
-U_2^{\rm div}(\kappa,\alpha_0,\beta_0,P)\biggr]\Biggr\},
\end{eqnarray}
where $\Gamma_{(1)}^{\rm div}(\kappa_0,\alpha_0,\beta_0)$ is the divergent part
of the effective action in the minimal gauge, (4.71)--(4.75),
$$
U_1(\kappa)=\varepsilon_jR^j_{\ \alpha,k}
\Delta^{-1\; ki}(\kappa,\alpha_0,\beta_0,P)F^{-1\; \alpha\beta}(\kappa)R^n_{\
\beta}E'_{in},
\eqno(4.84)
$$
$$
E'_{in}={d\over d\kappa}E_{in}=-{1\over
4}g^{1/2}g^{\mu\nu}g^{\alpha\beta}\delta(x,y),
$$
$$
U_2(\kappa,\alpha,\beta,P)
=\varepsilon_jR^j_{\ \alpha,k}R^k_{\ \beta}F^{-1\; \beta\gamma}(\kappa)
F^{-1\; \alpha\delta}(\kappa)
H^{-1}_{\ \gamma\delta}(\alpha,\beta,P)\Bigg\vert^{\rm div}.
\eqno(4.85)
$$

To calculate the quantities $U_1$, (4.84), and $U_2$, (4.85), one has to find the
ghost propagators $F^{-1}(\kappa)$ and $H^{-1}_{\
\gamma\delta}(\alpha,\beta,P)$ for arbitrary $\kappa$, $\alpha$, $\beta$ and
$P$ and the gravitational propagator $\Delta^{-1}(\kappa, \alpha_0,\beta_0)$
for arbitrary parameter $\kappa$ and minimal values of other parameters,
$\alpha_0$ and $\beta_0$. The whole background dimension that causes the
divergences is contained in the extremal $\varepsilon_i$. So, when calculating
the divergences of the quantities $U_1$, (4.84), and $U_2$, (4.85), one can take
all propagators to be free, i.e., the background quantities (the spacetime
curvature, the commutator of covariant derivatives etc.) and the mass terms can be
neglected. This is so because together with any dimensionful terms there appear
automatically a Green function $\Square^{-1}$ and the whole term becomes finite.
Therefore, in particular, the gauge fixing tensor $P^{\mu\nu}$ does not
contribute in the divergences of the effective action at all. 
In the minimal gauge, (4.61), we have shown this by explicit calculation in
previous section.

Using the explicit forms of the operators $F(\kappa)$, (4.57a), 
$H(\alpha,\beta)$, (4.58), and
$\Delta(\kappa,\alpha_0,\beta_0)$, (4.15), (4.53), we find the free Green
functions of these operators
$$
F^{-1\; \mu\nu}(\kappa)={1\over 2}\left(-g^{\mu\nu}\Square
+{\kappa-1\over \kappa-3}\nabla^\mu\nabla^\nu
\right)\Square^{-2}g^{-1/2}\delta(x,y),
\eqno(4.86)
$$
$$
H^{-1}_{\mu\nu}(\alpha,\beta)=4\alpha^2\left(-g_{\mu\nu}\Square
-{\beta\over 1-\beta}\nabla_\mu\nabla_\nu
\right)\Square^{-2}g^{1/2}\delta(x,y),
\eqno(4.87)
$$
\begin{eqnarray}
\setcounter{equation}{88}
\lefteqn{\Delta^{-1\; ik}(\kappa,\alpha_0,\beta_0)=2f^2\Biggl\{
E^{-1}_{(0)\mu\nu,\alpha\beta}\Square^2
+\rho_2\nabla_\mu\nabla_\nu\nabla_\alpha\nabla_\beta
}\qquad
\nonumber\\[10pt]
& &
+\rho_1\left(g_{\mu\nu}\nabla_\alpha\nabla_\beta
+g_{\alpha\beta}\nabla_\mu\nabla_\nu\right)\Square\Biggr\}
\Square^{-4}g^{-1/2}\delta(x,y),
\end{eqnarray}
where
$$
E^{-1}_{(0)\mu\nu,\alpha\beta}\equiv
E^{-1}_{\mu\nu,\alpha\beta}\Big\vert_{\kappa=\kappa_0}
=g_{\mu(\alpha}g_{\beta)\nu}
-{1+4\omega\over 12\omega}g_{\mu\nu}g_{\alpha\beta},
\eqno(4.89)
$$
$$
\rho_1={1\over 3\omega}\left(\omega+1+{3\over \kappa-3}\right),
\eqno(4.90)
$$
$$
\rho_2=-{4\over 3\omega(\omega+1)}\left(\omega+1+{3\over \kappa-3}\right)^2.
\eqno(4.91)
$$
Let us note that in the minimal gauge, (4.61), $\rho_1=\rho_2=0$ and, therefore,
$\Delta^{-1\; ik}_{(0)}$$=2f^2E^{-1}_{(0)\mu\nu,\alpha\beta}$
$\Square^{-2}g^{-1/2}$$\delta(x,y)$.

Substituting the free propagators, (4.86)--(4.88), in (4.84) and (4.85) we obtain
the divergences of the quantities $U_1$ and $U_2$
\begin{eqnarray}
\setcounter{equation}{92}
U_1^{\rm div}&=&{f^2\over \kappa-3}\int d^4 x\;\varepsilon^{\mu\nu}
\Biggl\{(\rho_2+6\rho_1-2\kappa_0^{-1})\nabla_\mu\nabla_\nu
\Square^{-3}g^{-1/2}\delta(x,y)\Bigg\vert_{y=x}^{\rm div}
\nonumber\\[10pt]
& &
-\left(\rho_1-{1+\omega\over 3\omega}\right)g_{\mu\nu}
\Square^{-2}g^{-1/2}\delta(x,y)\Bigg\vert_{y=x}^{\rm div}\Biggr\},
\end{eqnarray}
$$
U_2^{\rm div}=2\alpha^2\left(3+{4\over (\kappa-3)^2(1-\beta)}\right)
\int d^4 x\;\varepsilon^{\mu\nu}\nabla_\mu\nabla_\nu
\Square^{-3}g^{-1/2}\delta(x,y)\Bigg\vert_{y=x}^{\rm div}.
\eqno(4.93)
$$

Using the divergences of the coincidence limits of the Green functions and
their derivatives in the dimensional regularization
$$
\Square^{-2}g^{-1/2}\delta(x,y)\Bigg\vert_{y=x}^{\rm div}
=-i{2\over (n-4)(4\pi)^2},
$$
$$
\nabla_\mu\nabla_\nu
\Square^{-3}g^{-1/2}\delta(x,y)\Bigg\vert_{y=x}^{\rm div}
=-i{2\over (n-4)(4\pi)^2}{1\over 4}g_{\mu\nu},
\eqno(4.94)
$$
and the explicit form of the extremal $\varepsilon^{\mu\nu}$, (4.50), we obtain
$$
U_1^{\rm div}=i{1\over (n-4)(4\pi)^2}{6\nu^2\over (\kappa-3)^2}
\left(1+{2\over (\omega+1)(\kappa-3)}\right)\int d^4 x\;g^{1/2}{1\over k^2}
(R-4\Lambda),
$$
$$
U_2^{\rm div}=-i{1\over (n-4)(4\pi)^2}\alpha^2
\left(3+{4\over (\kappa-3)^2(1-\beta)}\right)\int d^4 x\;g^{1/2}{1\over k^2}
(R-4\Lambda).
\eqno(4.95)
$$
Substituting these expressions in (4.83) and integrating over $\kappa$ we
obtain finally
\begin{eqnarray*}
\Gamma_{(1)}^{\rm div}(\kappa,\alpha,\beta,P)
&=&\Gamma_{(1)}^{\rm div}(\kappa_0,\alpha_0,\beta_0)
\nonumber\\[10pt]
& &
-{1\over (n-4)(4\pi)^2}\int d^4 x\;g^{1/2}\Delta\gamma(\kappa,\alpha,\beta)
{1\over k^2}(R-4\Lambda),
\end{eqnarray*}
where
$$
\Delta\gamma(\kappa,\alpha,\beta)
={13\over 6}f^2+{4\over 3}\nu^2-{3\over 2}\alpha^2
-{2\alpha^2\over (\kappa-3)^2(1-\beta)}
+{6\nu^2(\kappa-2)\over (\kappa-3)^2}.
\eqno(4.96)
$$

Thus the off--shell divergences of the effective action in arbitrary gauge have
the same form, (4.71), where the coefficients $\beta_1$, $\beta_2$, $\beta_3$ and
$\beta_4$ do not depend on the gauge and are given by the expressions
(4.72)--(4.74), and the $\gamma$--coefficient reads
$$
\gamma(\kappa,\alpha,\beta)={5\over 3}{f^4\over \nu^2}+{5\over 6}\nu^2
-{3\over 2}\alpha^2
-{2\alpha^2\over (\kappa-3)^2(1-\beta)}
+{6\nu^2(\kappa-2)\over (\kappa-3)^2}.
\eqno(4.97)
$$
In particular, in the harmonic gauge, (4.60), ($\kappa=1$ and $\alpha=0$), we have
$$
\gamma(1,0,\beta)={5\over 3}{f^4\over \nu^2}-{2\over 3}\nu^2.
\eqno(4.98)
$$

The dependence of the divergences on the parametrization of the quantum field
also exhibits only in the $\gamma$--coefficient. Rather than to study this
dependence, let us calculate the divergences of the unique effective action
$\tilde\Gamma$, (4.24), that does not depend neither on the gauge nor on the
parametrization of the quantum field.

\noindent
{}From (4.24) we have
$$
\tilde\Gamma^{\rm div}_{(1)}=-{1\over
2i}\left(\log\det\tilde\Delta\big\vert^{\rm div}
-\log\det H\big\vert^{\rm div}
-2\log\det F\big\vert^{\rm div}\right).
\eqno(4.99)
$$
The unique effective action $\tilde\Gamma$, (4.24), (4.99), differs from the
usual one, (4.14), (4.52), only by the operator $\tilde\Delta$, (4.25). It is
obtained from the operator $\Delta$, (4.15), by substituting the covariant
functional derivatives instead of the usual ones:
$$
S_{,ik} \to \nabla_i\nabla_k S={\cal D}_i{\cal D}_kS-T^j_{\ ik}\varepsilon_j
=S_{,ik}-\Gamma^j_{\ ik}\varepsilon_j,
$$
$$
\tilde\Delta_{ik}=\tilde\Delta^{\rm loc}_{ik}+T^j_{\ ik}\varepsilon_j
=\Delta_{ik}+\Gamma^j_{\ ik}\varepsilon_j,
\eqno(4.100)
$$
where
$$
\tilde\Delta^{\rm loc}_{ik}
=-{\cal D}_i{\cal D}_kS+\chi_{\mu i}H^{\mu\nu}\chi_{\nu k}.
\eqno(4.101)
$$

Since the non--metric part of the connection $T^j_{\ ik}$, (4.34), is non--local,
the operator $\tilde\Delta$, (4.100), is an integro--differential one. The
calculation of the determinants of such operators offers a serious problem.
However, as the non--local part of the operator $\tilde\Delta$, (4.100), is
proportional to the extremal $\varepsilon_i$, it exhibits only off mass shell.
Therefore, the calculation of the determinant of the operator $\tilde\Delta$,
(4.100), can be based on the expansion in the non--local part, $T^j_{\
ik}\varepsilon_j$. To calculate the divergences it is again sufficient to limit
oneself only to linear terms
$$
\log\det\tilde\Delta\Big\vert^{\rm div}
=\log\det\tilde\Delta_{\rm loc}\Big\vert^{\rm div}
+\tilde\Delta^{-1\; mn}_{\rm loc}T^i_{\ mn}\varepsilon_i\Big\vert^{\rm div}.
\eqno(4.102)
$$

To calculate this expression one can choose any gauge, because the answer for
the unique effective action does not depend on the gauge. Let us choose the
De~Witt gauge, (4.56),
$$
\chi_{\mu i}=R^k_{\ \mu}E_{ki}, \qquad
F_{\mu\nu}=N_{\mu\nu}.
\eqno(4.103)
$$

Using the Ward identities for the Green function of the operator
$\tilde\Delta_{\rm loc}$, (4.101), in De~Witt gauge we get (up to terms
proportional to the extremal)
$$
B^\alpha_{\ i}\tilde\Delta_{\rm loc}^{-1\; ik}
=N^{-1\; \alpha\mu}H^{-1}_{\mu\nu}N^{-1\;\nu\beta}R^k_{\ \beta}+O(\varepsilon),
$$
$$
B^\alpha_{\ i}\tilde\Delta_{\rm loc}^{-1\; ik}B^\beta_{\ k}=N^{-1\;
\alpha\mu}H^{-1}_{\mu\nu}N^{-1\;\nu\beta}+O(\varepsilon).
\eqno(4.104)
$$
Using the explicit form of $T^i_{\ mn}$, (4.34), in (4.102) and the eqs. (4.104)
we obtain
$$
\log\det\tilde\Delta\Big\vert^{\rm div}
=\log\det\tilde\Delta_{\rm loc}\Big\vert^{\rm div}-U_3^{\rm div},
$$
where
$$
U_3=\varepsilon_j{\cal D}_kR^j_{\ \alpha}R^k_{\ \beta}N^{-1\;
\alpha\mu}H^{-1}_{\mu\nu}N^{-1\; \nu\beta}.
\eqno(4.105)
$$
Finally, one has to fix the operator $H$ (i.e., the parameters $\alpha$ and
$\beta$) and to determine the parameter of the metric of the 
configuration space,
$\kappa$, (4.45), (4.46).

In the paper [45] some conditions on the metric $E_{ik}$ were formulated, that
make it possible
to fix the parameter $\kappa$. First, the metric $E_{ik}$ must be
contained in the term with highest derivatives in the action $S(\varphi)$.
Second, the operator $N_{\mu\nu}$, (4.28), (4.48), must be 
non--degenerate within
the perturbation theory. To find the metric $E_{ik}$, (4.45), i.e., the parameter
$\kappa$, (4.46), one should consider the second variation of the action on the
physical quantum fields, $h_\perp=\Pi_\perp h$, that satisfy the De~Witt gauge
conditions, $R_{i\mu}h^i_\perp=0$, and identify the metric with 
the matrix $E$ in the highest
derivatives
\begin{eqnarray}
\setcounter{equation}{106}
h^i_\perp(-S_{,ik})h^k_\perp&=&{1\over 2f^2}\int d^4 x\;
h^\perp_{\mu\nu}g^{1/2}E^{\mu\nu,\alpha\beta}(\kappa)\Square^2
h^\perp_{\alpha\beta}
\nonumber\\[10pt]
& &
+ \ {\rm terms \ with \ the \ curvature.}
\end{eqnarray}

\noindent
This condition leads to a quadratic equation for $\kappa$ that has two
solutions
$$
\kappa_1=3{\omega\over \omega+1}, \qquad \kappa_2=3.
\eqno(4.107)
$$
As we already noted above, the value $\kappa=\kappa_2=3$ is unacceptable, since
the operator $N$, (4.48), in this case is degenerate on the flat background.
Therefore, we find finally
$$
\kappa=\tilde\kappa\equiv 3{\omega\over \omega+1}.
\eqno(4.108)
$$

Let us note that this value of $\kappa$ coincides with the minimal one,
$\tilde\kappa=\kappa_0$, (4.61). 
Thus, if we choose the matrix $H$ in the same form,
(4.58), with the minimal parameters, $\alpha=\alpha_0$ and $\beta=\beta_0$,
(4.61), then the operator $\tilde\Delta_{\rm loc}$ becomes a minimal operator
of the form (4.62)
$$
\tilde\Delta^{\rm loc}_{ik}=\Delta_{ik}
+\left\{ {}^j_{\ ik}\right\}\varepsilon_j,
\eqno(4.109)
$$
where $\Delta_{ik}$ is given by the expressions (4.62) and (4.63).

The divergences of the determinant of the operator $\tilde\Delta_{\rm loc}$,
(4.109), can be calculated either by direct application of the algorithm (4.65)
or by means of the expansion in the extremal
$$
\log\det \tilde\Delta_{\rm loc}\Big\vert^{\rm div}=
\log\det \Delta_{\rm loc}\Big\vert^{\rm div}
+U_4^{\rm div},
\eqno(4.110)
$$
where
$$
U_4=\Delta^{-1\; mn}\left\{ {}^i_{\ mn}\right\}\varepsilon_i.
$$
Using the formulas (4.99), (4.105), (4.110) and (4.52) we obtain the
divergences of the unique effective action
$$
\tilde\Gamma_{(1)}^{\rm div}=
\Gamma_{(1)}^{\rm div}(\kappa_0,\alpha_0,\beta_0)
+{1\over 2i}(U_3^{\rm div}-U_4^{\rm div}),
\eqno(4.111)
$$
where $\Gamma_{(1)}^{\rm div}(\kappa_0,\alpha_0,\beta_0)$ are the divergences
of the effective action in the minimal gauge, (4.71)--(4.75). The quantities
$U_3^{\rm div}$ and $U_4^{\rm div}$ are calculated by using the free
propagators, (4.86)-(4.88), in the minimal gauge, (4.61),
\begin{eqnarray*}
U_3^{\rm div}&=&U_2^{\rm div}(\kappa_0,\alpha_0,\beta_0)
-4f^2\int d^4 x\;\varepsilon^{\mu\nu}
\left\{ {}^{\alpha\beta,\rho\sigma}_{\mu\nu}\right\}
\nonumber\\[10pt]
& &
\times\nabla_\alpha\nabla_\rho\Biggl(-g_{\beta\sigma}\Square
+{1\over 3}(1-2\omega)
\nabla_\beta\nabla_\sigma\Biggl)\Square^{-4}g^{-1/2}
\delta(x,y)\Bigg\vert^{\rm div}_{y=x},
\end{eqnarray*}
$$
U_4^{\rm div}=\int d^4 x\;\varepsilon^{\mu\nu}
\left\{ {}^{\alpha\beta,\gamma\delta}_{\mu\nu}\right\}
2f^2E^{-1}_{\alpha\beta,\gamma\delta}\Square^{-2}g^{-1/2}
\delta(x,y)\Bigg\vert^{\rm div}_{y=x},
\eqno(4.112)
$$
where $U_2^{\rm div}(\kappa_0,\alpha_0,\beta_0)$ is given by the formula
(4.95) in the minimal gauge (4.61). Using the divergences of the coincidence
limits of the derivatives of the Green functions, (4.94), and the Christoffel
connection, (4.47), and substituting the minimal values of the parameters
$\kappa$, $\alpha$ and $\beta$, (4.61), we obtain
$$
U_3^{\rm div}=
-i{1\over (n-4)(4\pi)^2}{4\over 3}(2\nu^2+f^2)
\int d^4 x\;g^{1/2}{1\over k^2}
(R-4\Lambda),
$$
$$
U_4^{\rm div}=
-i{1\over (n-4)(4\pi)^2}6(\nu^2-f^2)\int d^4 x\;g^{1/2}{1\over k^2}
(R-4\Lambda).
\eqno(4.113)
$$

Thus, the off--shell divergences of the one--loop effective action,
$\tilde\Gamma_{(1)}$, have the standard form (4.71), where the
$\beta$--coefficients are determined by the same expressions (4.72)--(4.74) and
the $\gamma$--coefficient, (4.75), has an extra contribution due to the quantities
$U_3^{\rm div}$ and $U_4^{\rm div}$, (4.113), in (4.111). It has the form
$$
\tilde\gamma=\gamma(\kappa_0,\alpha_0,\beta_0)
+{1\over 3}(11f^2-5\nu^2)
={5\over 3}{f^4\over \nu^2}+{3\over 2}f^2-{13\over 6}\nu^2,
\eqno(4.114)
$$
where $\gamma(\kappa_0,\alpha_0,\beta_0)$ is given by (4.75).

%
%
%
%
\section{Renormalization group equations and the ul\-tra\-vio\-let
asym\-pto\-tics of the coupling con\-stants}

The structure of the divergences of the effective action, (4.71), indicates that
the higher--derivative quantum gravity is renormalizable off mass shell. Thus
one can apply the renormalization group methods to study the high--energy
behavior of the effective (running) coupling constants [48, 49, 163, 109]. The
dimensionless constants $\epsilon$, $f^2$, $\nu^2$ and $\lambda$ are the
`essential' coupling constants [109] but the Einstein dimensionful constant
$k^2$ is `non--essential' because its variation reduces to a reparametrization
of the quantum field, i.e., up to total derivatives
$$
{\partial S\over \partial k^2}\Bigg\vert_{\rm on-shell}=0.
\eqno(4.115)
$$

Using the one--loop divergences of the effective action, (4.71),  we obtain in the
standard way the renormalization group equations for the coupling constants of
the renormalized effective action [48, 109]
$$
{d\over d t} \epsilon=\beta_1,
\qquad
{d\over d t} f^2=-2\beta_2 f^4,
\eqno(4.116) 
$$
$$
{d\over d t} \nu^2=6\beta_3\nu^4={5\over 3}f^4+5f^2\nu^2+{5\over 6}\nu^4,
\eqno(4.117)
$$
$$
{d\over d t} \lambda={1\over 2}\beta_4={1\over 4}(\nu^4+5f^4)
+{1\over 3}\lambda\left(
10{f^4\over \nu^2}+15f^2-\nu^2\right),
\eqno(4.118)
$$
$$
{d\over d t} k^2=\gamma k^2,
\eqno(4.119)
$$
where $t=(4\pi)^{-2}\log(\mu/\mu_0)$, $\mu$ is the renormalization parameter
and $\mu_0$ is a fixed energy scale.

The ultraviolet behavior of the essential coupling constants $\epsilon(t)$,
$f^2(t)$, $\nu^2(t)$ and $\lambda(t)$ as $t\to \infty$ is determined by the
coefficients (4.72)--(4.74). They play the role of generalized Gell--Mann--Low
$\beta$--functions, (1.47), and do not depend neither on the gauge condition nor
on the parametrization of the quantum field. The non--essential coupling
constant $k^2(t)$ is, in fact, simply a field renormalization constant. Thus
the $\gamma$--coefficient, (4.75), (4.97) and (4.114), play in (4.119) the role of
the
anomalous dimension (1.48). Correspondingly, the ultraviolet behavior of the
constant $k^2(t)$ depends essentially both on the gauge and the parametrization
of the quantum field. It is obvious that one can choose the gauge condition in
such a way that the coefficient $\gamma$, (4.97), is equal to zero, $\gamma=0$.
In this case the Einstein coupling constant is not renormalized at all, i.e.,
$k^2(t)=k^2(0)={\rm const}$.

The solution of the eqs. (4.116) is trivial and reads
$$
\epsilon(t)=\epsilon(0)+\beta_1 t,
$$
$$
f^2(t)={f^2(0)\over 1+2\beta_2f^2(0) t}.
\eqno(4.120)
$$

Noting that $\beta_1<0$ and $\beta_2>0$, (4.72), we find the following. First,
the topological coupling constant $\epsilon(t)$  becomes negative in the
ultraviolet limit ($t\to \infty$) and its absolute value grows logarithmically
regardless of the initial value $\epsilon(0)$. Second, the Weyl coupling
constant $f^2(t)$ is either asymptotically free (at $f^2>0$) or has a
`zero--charge' singularity (at $f^2<0$). We limit ourselves to the first case,
$f^2>0$, since, on the one hand, this condition ensures the stability of the
flat background under the spin--2 tensor excitations, and, on the other hand, it
leads to a positive contribution of the Weyl term to the Euclidean action
(4.49).

The solution of the equation (4.117) can be written in the form
$$
\nu^2(t)={c_1f^{2p}(t)-c_2f_*^{2p}\over f^{2p}(t)-f_*^{2p}}f^2(t),
\eqno(4.121)
$$
where
\begin{equation}\setcounter{equation}{122}
c_{1,2}={1\over 50}(-549\pm\sqrt{296401})
\approx\left\{\begin{array}{ll}
-0.091\\
-21.87
\end{array}
\right.,
\end{equation}
$$
p={\sqrt{296401}\over 399}\approx 1.36,
\qquad
f_*^{2p}\equiv {\nu^2(0)-c_1f^2(0)\over \nu^2(0)-c_2f^2(0)}f^{2p}(0).
\eqno(4.123)
$$
There are also two special solutions
$$
\nu_{1,2}^2(t)=c_{1,2}f^2(t),
\eqno(4.124)
$$
that correspond to the values $f_*^{2p}=0, \ \infty$ in (4.121). These
solutions are asymptotically free but only $\nu_2^2(t)$ is stable in the
ultraviolet limit.

The behavior of the conformal coupling constant $\nu^2(t)$ depends essentially
on its initial value $\nu^2(0)$. In the case $\nu^2(0)>c_1f^2(0)$ we have
$f^{2p}(0)>f_*^{2p}>0$ and, therefore, the function $\nu^2(t)$, (4.121), has a
typical `zero--charge' singularity at a finite scale $t=t_*$ determined from
$f^{2p}(t_*)=f_*^{2p}$:
$$
\nu^2(t)\Bigg\vert_{t\to t_*}=c_3{f_*^{2(p+1)}\over f^{2p}(t)-f_*^{2p}}
+O(1),
\eqno(4.125)
$$
where
$$
c_3=c_1-c_2={\sqrt{296401}\over 25}\approx 21.78.
\eqno(4.126)
$$

In the opposite case, $\nu^2(0)<c_1f^2(0)$, the function $\nu^2(t)$, (4.121),
does not have any singularities and is asymptotically free,
$$
\nu^2(t)\Big\vert_{t\to
\infty}=c_2f^2(t)-c_3f_*^{-2p}f^{2(p+1)}(t)+O\left(f^{2(1+2p)}\right).
\eqno(4.127)
$$

Thus, to the contrary to the conclusions of the papers [152--155], we find that in the region $\nu^2>0$ there are no asymptotically
free solutions.
The asymptotic freedom for the conformal coupling constant $\nu^2(t)$ can be
achieved only in the negative region $\nu(0)<0$, (4.127).

The exact solution of the equation for the dimensionless cosmological constant,
(4.118), has the form
$$
\lambda(t)=\Phi(t)\left\{\Phi^{-1}(0)\lambda(0)+\int\limits_0^t d\tau
A(\tau)\Phi^{-1}(\tau)\right\},
\eqno(4.128)
$$
where
$$
A(\tau)={1\over 4}\left(5f^4(t)+\nu^4(t)\right),
$$
$$
\Phi(t)=\left\vert c_1f^{2p}(t)-c_2f_*^{2p}\right\vert^2
\left\vert f^{2p}(t)-f_*^{2p}\right\vert^{2/5}
\left\vert f^{2}(t)\right\vert^{-q},
\eqno(4.129)
$$
$$
q={2\over 665}(-241+\sqrt{296401})\approx 0.913.
$$

The ultraviolet behavior of the cosmological constant $\lambda(t)$, (4.128),
crucially depends on the initial values of both the conformal coupling constant,
$\nu^2(0)$, and the cosmological constant itself, $\lambda(0)$.
In the region $\nu^2(0)>c_1f^2(0)$ the solution (4.128) has a
`zero--charge' pole at the same scale $t_*$, similarly to the conformal coupling constant $\nu^2(t)$, (4.125),
$$
\lambda(t)\Big\vert_{t\to t_*}={3\over 14}c_3{f_*^{2(1+2p)}\over
f^{2p}(t)-f_*^{2p}}
+O(1).
\eqno(4.130)
$$
In the opposite case, $\nu^2(0)<c_1f^2(0)$, the function $\lambda(t)$, (4.128),
grows in the ultraviolet limit
$$
\lambda(t)\Big\vert_{t\to\infty}=c_4f^{-2q}(t)+O(f^2).
\eqno(4.131)
$$
The sign of the constant $c_4$ in (4.131) depends on the initial value
$\lambda(0)$, i.e., $c_4>0$ for $\lambda(0)>\lambda_2(0)$ and $c_4<0$ for
$\lambda(0)>\lambda_2(0)$, where
$$
\lambda_2(0)=-\Phi(0)\int\limits_0^\infty d\tau A(\tau)\Phi^{-1}(\tau).
\eqno(4.132)
$$
In the special case $\lambda(0)=\lambda_2(0)$ the constant $c_4$ is equal to
zero ($c_4=0$) and the solution (4.128) takes the form
$$
\lambda(t)=\lambda_2(t)=-\Phi(t)\int\limits_t^\infty d\tau
A(\tau)\Phi^{-1}(\tau).
\eqno(4.133)
$$
The special solution (4.133) is asymptotically free in the ultraviolet limit
$$
\lambda_2(t)\Big\vert_{t\to\infty}=c_5f^2(t)+O\left(f^{2(1+p)}\right),
\eqno(4.134)
$$
where
$$
c_5=-{5\over 266}\cdot{5+c_2^2\over q+1}\approx -4.75.
\eqno(4.135)
$$
However, the special solution (4.133) is unstable because of the presence of
growing mode (4.131). Besides, it exists only in the negative region 
$\lambda<0$. In
the positive region $\lambda>0$ the cosmological constant is not asymptotically
free, (4.131).

Our conclusions about the asymptotic behavior of the cosmological constant
$\lambda(t)$ also differ essentially from the results of the papers [152--155]
where the asymptotic freedom for the cosmological constant in the region
$\lambda>0$ and $\nu^2>0$ for any initial values of $\lambda(0)$ was
established.

Let us discuss the influence of arbitrary low--spin matter (except for 
spin--$3/2$ fields) interacting with the 
quadratic gravity (4.49) on the ultraviolet
behavior of the theory. The system of renormalization group equations in
presence of matter involves the equations (4.116)--(4.119) with the total
$\beta$--functions
$$
\beta_{i,{\rm tot}}=\beta_i+\beta_{i,{\rm mat}},
\eqno(4.136)
$$
where $\beta_{i,{\rm mat}}$ is the contribution of matter fields in the
gravitational divergences of the effective action, (4.71), and the equations for
the masses and the matter coupling constants. The values of the first three
coefficients at the terms quadratic in the curvature have the form [54--56]
$$
\beta_{1,{\rm mat}}=-{1\over 360}\left(62N_1^{(0)}+63N_1+11N_{1/2}+N_0\right),
$$
$$
\beta_{2,{\rm mat}}={1\over 120}\left(12N_1^{(0)}+13N_1+6N_{1/2}+N_0\right),
\eqno(4.137)
$$
$$
\beta_{3,{\rm mat}}={1\over 72}\left(N_1+(1-6\xi)^2N_0\right),
$$
where $N_s$ is the number of the fields with spin $s$, $N_1^{(0)}$ is the
number of massless vector fields, $\xi$ is the coupling constant of scalar
fields with the gravitational field. In the formula (4.137) the spinor fields
are taken to be two--component. The coefficients (4.137) possess important
general properties
$$
\beta_{1, \rm mat}<0, \qquad \beta_{2, \rm mat}> 0, \qquad
\beta_{3, \rm mat}>0.
\eqno(4.138)
$$
The gravitational $\beta$--functions (4.72)--(4.74) obtained in previous sections
have analogous properties for $f^2>0$ and $\nu^2>0$. Therefore, the total
$\beta$--functions, (4.136), also satisfy the conditions (4.138) for $f^2>0$ and
$\nu^2>0$. The properties (4.138) are most important for the study of the
ultraviolet asymptotics of the topological coupling constant $\epsilon(t)$, the
Weyl coupling constant $f^2(t)$ and the conformal  one $\nu^2(t)$.

The solution of the renormalization group equations for the topological and
Weyl coupling constants in the presence of matter have the same form (4.120)
with the substitution $\beta\to \beta_{\rm tot}$. Thus the presence of matter
does not change qualitatively the ultraviolet asymptotics of these constants:
the coupling $\epsilon(t)$ becomes negative and grows logarithmically and the
Weyl coupling constant is asymptotically free at $f^2>0$.

The renormalization group equation for the conformal coupling constant
$\nu^2(t)$ in the presence of the matter takes the form
$$
{d\over dt}\nu^2={5\over 3}f^4+5f^2\nu^2+{1\over
12}\left(10+N_1+(1-6\xi)^2N_0\right)\nu^4.
\eqno(4.139)
$$
Therefrom one can show that at $\nu^2>0$ the coupling constant $\nu^2(t)$ has a
`zero--charge' singularity at a finite scale.

The other properties of the theory (in particular, the behavior of the constant
$\nu^2(t)$ at $\nu^2<0$) depend essentially on the particular form of the
matter model. However, the strong conformal coupling, $\nu^2\gg 1$, at
$\nu^2>0$ leads to singularities in the cosmological constant as well as in all
coupling constants of matter fields.

Thus, we conclude that the higher--\-derivative quantum gravity interacting with
any low--spin matter necessarily goes out of the limits of weak conformal
coupling at high energies in the case $\nu^2>0$. This conclusion is also
opposite to the results of the papers [152--155] where the asymptotic freedom of
the higher--\-derivative quantum gravity in the region $\nu^2>0$ in the presence
of rather arbitrary matter was established.

Let us also find the ultraviolet behavior of the non--essential Einstein
coupling constant $k^2(t)$. The solution of the equation (4.119) has the form
$$
k^2(t)=k^2(0)\exp\left\{\int\limits_0^t d\tau \gamma(\tau)\right\}.
\eqno(4.140)
$$
The explicit expression depends on the form of the function $\gamma$ and,
hence, on the gauge condition and the parametrization of the quantum field,
(4.97).
We cite the result for two cases: for the standard effective action in the
minimal gauge and the 
standard parametrization (4.75) and for the unique effective
action (4.114). In both cases the solution (4.140) has the form
$$
k^2(t)=k^2(0){\Psi(t)\over \Psi(0)},
\eqno(4.141)
$$
where
$$
\Psi(t)=\left\vert c_1f^{2p}(t)-c_2f_*^{2p}\right\vert^2
\left\vert f^{2p}(t)-f_*^{2p}\right\vert^{s}
\left\vert f^{2}(t)\right\vert^{-r},
$$
\vskip10pt
\begin{equation}
\setcounter{equation}{142}
s=\left\{\begin{array}{ll}
{13\over 5}\\
{3\over 5}
\end{array}
\right.,
\qquad
r=\left\{\begin{array}{ll}
{3\over 665}(269+\sqrt{296401})\approx 3.67\\
{2\over 1995}(-437+2\sqrt{296401})\approx 0.653
\end{array}
\right..
\end{equation}
Here and below the upper values correspond to the unique
effective action  and the lower values correspond to the standard effective
action  in the minimal gauge and the standard parametrization.

Therefrom it is immediately seen that the Einstein coupling constant $k^2(t)$
grows in the ultraviolet limit ($t\to \infty$)
$$
k^2(t)\Big\vert_{t\to \infty}=c_6f^{-2r}(t)+O\left(f^{2(p-r)}\right).
\eqno(4.143)
$$

Let us note that the ultraviolet behavior of the dimensionful cosmological
con\-stant, $\Lambda(t)=\lambda(t)/k^2(t)$, is essentially different in the case of
the unique effective action and in the standard case
$$
\Lambda(t)\Big\vert_{t\to \infty}=c_7f^{2\alpha}+O\left(f^{2(\alpha+p)}\right),
\eqno(4.144)
$$
where
$$
\alpha=r-q\approx \left\{\begin{array}{ll}
2.76\\
-0.26
\end{array}
\right..
$$
In the first case $\Lambda(t)$ rapidly approaches zero $\sim t^{-2.76}$ and in
the second case it grows as $t^{0.26}$.

It is well known that the functional formulation of the quantum field theory
assumes the Euclidean action to be positive definite [49, 50, 64]. Otherwise,
(what happens, for example, in the conformal sector of the Einstein gravity),
one must resort to  the complexification of the configuration space to achieve
the divergence of the functional integral [64, 115, 116].

The Euclidean action of the higher--\-derivative theory of gravity differs only
by sign from the action  (4.49) we are considering. It is positive definite in
the case
$$
\epsilon>0,
\eqno(4.145)
$$
$$
\nu^2<0,
\eqno(4.146)
$$
$$
f^2>0, \qquad
\lambda>-{3\over 4}\nu^2.
\eqno(4.147)
$$
It is not necessary to impose the condition (4.145) if one restricts oneself to
a fixed topology. However, when including in the functional integral of quantum
gravity the topologically non--trivial metrics with large Euler characteristic,
the violation of the condition (4.145) leads to the exponential growth of their
weight and, therefore, to a foam--like structure of the spacetime at
micro--scales [64]. It is this situation that occurs in the ultraviolet limit,
when $\epsilon(t)\to -\infty$, (4.120).

The condition (4.146) is usually held to be ``non--physical'' [140--157]. The
point is, the conformal coupling constant $\nu^2$ plays the role of the
dimensionless square of the mass  of the conformal mode on the flat background.
In the case $\nu^2<0$ the conformal mode becomes tachyonic and leads to the
instability of the flat space (i.e., oscillations of the static potential,
unstable solutions etc. [144]).

As we showed above, the higher--\-derivative quantum gravity in the region
$\nu^2>0$ has unsatisfactory `zero--charge' behavior  in the conformal sector,
(4.125), (4.130). In the region of strong conformal coupling ($\nu^2\gg 1$) one
cannot make anything definitive conclusions 
on the basis of the perturbative calculations.
However, on the qualitative level it seems likely that the singularity in the
coupling constants $\nu^2(t)$ and $\lambda(t)$ can be interpreted as a
reconstruction of the ground state of the theory, i.e.,  the conformal mode
`freezes' and a condensate is formed.

We find the arguments against the `non--physical' condition (4.146) to be not
strong enough. First, the higher--\-derivative quantum gravity, 
strictly speaking, cannot be treated as a physical
theory within the limits of perturbation theory
because of the presence of the ghost states in the tensor sector that
violate the unitarity of the theory [143--155, 157]. This is not surprising in an
asymptotically free theory (that always takes place in the tensor sector),
since, generally speaking, the true physical asymptotic states have nothing to
do with the excitations in the perturbation theory [158]. Second, the
correspondence with the macroscopic gravitation is a rather fine problem that
needs a special investigation of the low--energy limit of the high derivative
quantum gravity. Third, the cosmological constant is always not asymptotically
free.
This means, presumably, that the expansion around the flat space in the high
energy limit is not valid anymore. Hence, the solution of the unitarity problem
based on this expansion by summing the radiation corrections and analyzing the
position of the poles of the propagator in momentum representation is not valid
too. In this case the flat background cannot present the ground state of the theory
any longer.

{}From this standpoint, in high energy region the higher--\-derivative quantum
gravity with positive definite Euclidean action, i.e., with an extra condition
$$
\nu^2<c_1f^2\approx-0.091 f^2,
\eqno(4.148)
$$
seems to be more intriguing. Such theory has an unique stable ground state that
minimizes the functional of the classical Euclidean action. It is
asymptotically free both in the tensor and conformal sectors. Besides, instead 
of the contradictory `zero--charge' behavior the cosmological constant
just grows logarithmically at high energies.

Let us stress once more the main conclusion of the present section. \
Not\-with\-standing the fact that the higher--\-derivative quantum gravity is
asymptotically free in the tensor sector of the theory with the natural condition
$f^2>0$, that ensures the stability of the flat space under the tensor
perturbations, the condition of the conformal stability of the flat background,
$\nu^2>0$, is incompatible with the asymptotic freedom in the conformal sector.
Thus, the flat background cannot present the ground state of the theory in the
ultraviolet limit. The problem with the conformal mode does not appear in the
conformally invariant models [97, 98]. Therefore, they are asymptotically free
[152--156]; however, the appearance of the $R^2$--divergences at higher loops
leads to their non--renormalizability [161].

%
%
%

\section{Effective potential of higher--derivative
quantum gravity}

Up to now the background field (i.e., the spacetime metric) was held to be
arbitrary. To construct the $S$--matrix one needs the background fields to be
the solutions of the classical equations of motion, (4.5), i.e., to lie on mass
shell. It is obvious that the flat space does not solve the equations of motion
(4.5) and (4.50) for non--vanishing cosmological constant. The most simple and
maximal symmetric solution of the equations of motion (4.5) and (4.50) is the De
Sitter space
$$
R^\mu_{\ \nu\alpha\beta}
={1\over 12}(\delta^\mu_\alpha g_{\beta\nu}-\delta^\mu_\beta g_{\alpha\nu})R,
\eqno(4.149)
$$
$$
R_{\mu\nu}={1\over 4}g_{\mu\nu}R, \qquad R={\rm const},
$$
with the condition
$$
R=4\Lambda.
\eqno(4.150)
$$

On the other hand, in quantum gravity De Sitter background, (4.149), plays the
role of covariantly constant field strength in gauge theories,
 $\nabla_\mu R_{\alpha\beta\gamma\delta}$ $=\nobreak0$. 
Therefore, the effective action on De Sitter
background determines, in fact, the effective potential of the
higher--derivative quantum gravity. Since in this case
the background field is characterized only by one constant $R$, the effective
potential is an usual function of one variable. The simplicity of De Sitter
background makes it possible to calculate the one--loop effective potential exactly.

In the particular case of De Sitter background one can also check our
result for the $R^2$--divergence of the one--loop effective action in general
case, (4.71), i.e., the coefficient $\beta_3$, (4.73), that differs from the
results of [152--155] and radically changes the ultraviolet behavior of the
theory in the conformal sector (see Sect. 4.4).

For the practical calculation of the effective potential we go to the Euclidean
sector of the spacetime. Let the spacetime be a compact four--dimensional sphere
$S^4$ with the volume
$$
V=\int d^4 x g^{1/2}=24\left({4\pi\over R}\right)^2, \qquad (R>0),
\eqno(4.151)
$$
and the Euler characteristic
$$
\chi={1\over 32\pi^2}\int d^4 x g^{1/2}R^*R^*=2.
\eqno(4.152)
$$
In present section we will always use the Euclidean action that differs only by
sign from the pseudo--Euclidean one, (4.49). All the formulas of the previous
sections remain valid by changing the sign of the action $S$, the extremal
$\varepsilon_i=S_{,i}$ and the effective action $\Gamma$.

On De Sitter background (4.149) the classical Euclidean action takes the form
$$
S(R)=24(4\pi)^2\left\{{1\over 6}\left(\epsilon-{1\over \nu^2}\right)
-{1\over x}+2\lambda{1\over x^2}\right\},
\eqno(4.153)
$$
where $x\equiv Rk^2$ and $\lambda=\Lambda k^2$. For  $\lambda>0$ it has a minimum on
the mass shell, $R=4\Lambda$, ($x=4\lambda$), (4.150), that reads
$$
S_{\rm on-shell}
=(4\pi)^2\left\{4\left(\epsilon-{1\over \nu^2}\right)
-{3\over \lambda}\right\}.
\eqno(4.154)
$$

Our aim is to obtain the effective value of the De Sitter curvature
$R$ from the full effective equations
$$
{\partial \Gamma(R)\over \partial R}=0.
\eqno(4.155)
$$
Several problems appear on this way: the dependence of the effective action
and, therefore, the effective equations on the gauge condition and the
parametrization of the quantum field, validity of the one--loop approximation
etc. [61].

First of all, we make a change of field variables $h_{\mu\nu}$
$$
h_{\mu\nu}=\bar h_{\mu\nu}^\perp+{1\over
4}g_{\mu\nu}\varphi+2\nabla_{(\mu}\varepsilon_{\nu)},
\eqno(4.156)
$$
$$
\varphi=h-\Square\sigma, \qquad h=g^{\mu\nu}h_{\mu\nu},
\eqno(4.157)
$$
$$
\varepsilon_\mu=\varepsilon_\mu^\perp+{1\over 2}\nabla_\mu\sigma,
\eqno(4.158)
$$
where the new variables, $\bar h_{\mu\nu}^\perp$ and $\varepsilon_\mu^\perp$,
satisfy the differential constraints
$$
\nabla^\mu \bar h_{\mu\nu}^\perp=0,
\qquad \bar h^\perp_{\mu\nu}g^{\mu\nu}=0,
\eqno(4.159)
$$
$$
\nabla^\mu\varepsilon^\perp_\mu=0.
\eqno(4.160)
$$
In the following we will call the initial field variables $h_{\mu\nu}$,
without any restrictions imposed on, the `unconstrained' fields and the fields
$\bar h_{\mu\nu}^\perp$ and $\varepsilon^\perp_\mu$, which satisfy the
differential conditions (4.159) and (4.160), the `constrained' ones.

When the unconstrained field is transformed under the gauge transformations with parameters $\xi_\mu$, (4.43), (4.44), 
$$
\delta h_{\mu\nu}=2\nabla_{(\mu}\xi_{\nu)},
\qquad \xi_\mu=\xi_\mu^\perp+\nabla_\mu\xi,
\eqno(4.161)
$$
the constrained fields transform in the following way
$$
\delta \bar h^\perp_{\mu\nu}=0, \qquad
\delta\varphi=0,
\eqno(4.162)
$$
$$
\delta\varepsilon_\mu^\perp=\xi_\mu^\perp, \qquad
\delta\sigma=2\xi.
\eqno(4.163)
$$
Therefore, the transverse traceless tensor field $\bar h^\perp_{\mu\nu}$ and
the conformal field $\varphi$ are the physical gauge--invariant components of
the field $h_{\mu\nu}$, whereas $\varepsilon_\mu^\perp$ and $\sigma$ are pure
gauge non--physical ones.

Let us write down the second variation of the Euclidean action on De Sitter
background
\begin{eqnarray}
\setcounter{equation}{164}
\lefteqn{
S_2(g+h)\equiv{1\over 2}h^iS_{,ik}h^k
}\nonumber\\[10pt]
& &
=\int d^4 x\;g^{1/2}\Biggl\{{1\over 4f^2}
\bar h^\perp\Biggl[\Delta_2\left(m_2^2+{f^2+\nu^2\over 3\nu^2}R\right)
\Delta_2\left({R\over 6}\right)
+{1\over 2}m_2^2(R-4\Lambda)\Biggr]\bar h^\perp
\nonumber\\[10pt]
& &
-{3\over 32\nu^2}\Biggl[\varphi\left(\Delta_0(m_0^2)\Delta_0\left(-{R\over
3}\right)
+{1\over 3}m_0^2(R-4\Lambda)\right)\varphi
\nonumber\\[10pt]
& &
-{2\over 3}m_0^2(R-4\Lambda)\varphi\Delta_0(0)\sigma
-{2\over 3}m_0^2(R-4\Lambda)\sigma\Delta_0(0)\Delta_0\left(-{R\over
2}\right)\sigma\Biggr]
\nonumber\\[10pt]
& &
+{1\over 4k^2}(R-4\Lambda)\varepsilon^\perp\Delta_1\left(-{R\over
4}\right)\varepsilon^\perp\Biggr\},
\end{eqnarray}
where
$$
m_2^2={f^2\over k^2}, \qquad
m_0^2={\nu^2\over k^2},
$$
and
$$
\Delta_s(X)=-\Square+X
\eqno(4.165)
$$
are the constrained differential operators acting on the constrained fields of spin
$s=0,1,2$.

When going to the mass shell, (4.150), the dependence of $S_2$, (4.164), on the
non--physical fields $\varepsilon^\perp$ and $\sigma$ disappears and (4.164)
takes the form
\begin{eqnarray}
\setcounter{equation}{166}
S_2(g+h)\Big\vert_{\rm on-shell}
\hskip-3pt&=&\hskip-4pt\int d^4 x g^{1/2}\Biggl\{{1\over 4f^2}
\bar h^\perp \Delta_2\left(m_2^2+{4\over 3}{(f^2+\nu^2)\over
\nu^2}\Lambda\right)
\Delta_2\left({2\over 3}\Lambda\right)\bar h^\perp
\nonumber\\[10pt]
& &
-{3\over 32\nu^2}\varphi\Delta_0(m_0^2)
\Delta_0\left(-{4\over 3}\Lambda\right)\varphi
\Biggr\}.
\end{eqnarray}
Herefrom it follows, in particular, the gauge invariance of the second
variation, (4.164), on the mass shell, (4.150),
$$
\delta S_2(g+h)\Big\vert_{\rm on-shell}=0.
\eqno(4.167)
$$

The eigenvalues $\lambda_n$ of the constrained d'Alambert operator
$\Delta_s(0)=-\Square$ and their multiplicities $d_n$ are [61]
$$
\lambda_n=\rho^2\bar\lambda_n,
\qquad \bar\lambda_n=n^2+3n-s, \qquad \rho^2={R\over 12},
$$
$$
d_n={1\over 6}(2s+1)(n+1-s)(n+2+s)(2n+3), \qquad n=s, s+1, \dots.
\eqno(4.168)
$$

Thus the condition of stability of the De Sitter background (4.149),
$$
S_2(g+h)\Big\vert_{\rm on-shell}>0,
\eqno(4.169)
$$
imposes the following restrictions (for $\lambda>0$): in the tensor sector
$$
f^2>0, \qquad -{1\over f^2}-{1\over 3\nu^2}<{1\over 4\lambda},
\eqno(4.170)
$$
and in the conformal sector
$$
\nu^2<0, \qquad {1\over 4\lambda}<-{1\over 3\nu^2}.
\eqno(4.171)
$$
However, even in the case when these conditions are fulfilled there are still
left five zero modes of the operator $\Delta_0\left(-{4\over 3}\Lambda\right)$
in the conformal sector, (4.166). Along these conformal directions, $\varphi_1$,
in the configuration space the Euclidean action does not grow
$$
S_2(g+\varphi_1)\Big\vert_{\rm on-shell}=0.
\eqno(4.172)
$$
This means that, in fact, the De Sitter background (4.149) does not give the
absolute minimum of the positive definite Euclidean action. This can be
verified by calculating the next terms in the expansion of $S(g+\varphi_1)$.
However, in the one--loop approximation these terms do not matter.

To calculate the effective action one has to find the Jacobian of the change of
variables (4.156)--(4.158). Using the simple equations
\begin{eqnarray}
\setcounter{equation}{173}
\int d^4 x g^{1/2} h^2_{\mu\nu}
&=&\int d^4 x g^{1/2}\Biggl\{\bar h^{\perp 2}
+2\varepsilon^\perp\Delta_1\left(-{R\over 4}\right)\varepsilon^\perp
\nonumber\\[10pt]
& &
+{3\over 4}\sigma\Delta_0(0)\Delta_0\left(-{R\over 3}\right)\sigma
+{1\over 4}h^2\Biggr\},
\end{eqnarray}
$$
\int d^4 x g^{1/2} \varepsilon^2_{\mu}
=\int d^4 x g^{1/2}\Biggl\{\varepsilon^{\perp 2}
+{1\over 4}\sigma\Delta_0(0)\sigma\Biggr\},
\eqno(4.173)
$$
we obtain
$$
d h_{\mu\nu}=d\bar h^\perp\, d\varepsilon^\perp\, d\sigma\, d\varphi\,
\left(\det J_2\right)^{1/2},
$$
$$
d\varepsilon_{\mu}=d\varepsilon^\perp\,d\sigma\,
\left(\det J_1\right)^{1/2},
\eqno(4.174)
$$
where
$$
J_2=\Delta_1\left(-{R\over 4}\right)\otimes 
\Delta_0\left(-{R\over 3}\right)\otimes\Delta_0(0),
$$
$$
J_1=\Delta_0(0).
$$

Let us calculate the effective action (4.12) in De~Witt gauge (4.57). The ghost
operator $F$, (4.16), in this gauge equals the operator $N$, (4.48). On De Sitter
background (4.149) it has the form
$$
F_{\mu\nu}=N_{\mu\nu}=2g^{1/2}
\left\{g_{\mu\nu}
\left(-\Square-{R\over 4}\right)
+{1\over 2}(\kappa-1)\nabla_\mu\nabla_\nu\right\}\delta(x,y).
\eqno(4.175)
$$
The operator of `averaging over the gauges' $H$, (4.58), and the gauge fixing
term have the form
$$
H^{\mu\nu}={1\over 4\alpha^2}g^{-1/2}
\left\{g^{\mu\nu}
\left(-\Square+{R+P\over 4}\right)
+\beta\nabla^\mu\nabla^\nu\right\}\delta(x,y),
\eqno(4.176)
$$
\begin{eqnarray}
\setcounter{equation}{177}
S_{\rm gauge}&\equiv &{1\over 2}\chi_\mu H^{\mu\nu}\chi_\nu
=\int d^4 x g^{1/2}{1\over 2\alpha^2}\Biggl\{
\varepsilon^\perp\Delta_1\left({R+P\over 4}\right)
\Delta_1^2\left(-{R\over 4}\right)\varepsilon^\perp
\nonumber\\[10pt]
& &
+{1\over 16}(1-\beta)\Biggl[\kappa^2\varphi\Delta_0(P')\Delta_0(0)\varphi
\nonumber\\[10pt]
& &
-2\kappa(\kappa-3)\varphi\Delta_0(P')
\Delta_0\left({R\over \kappa-3}\right)\Delta_0(0)\sigma
\nonumber\\[10pt]
& &
+(\kappa-3)^2\sigma\Delta_0(P')
\Delta_0^2\left({R\over \kappa-3}\right)\Delta_0(0)\sigma\Biggr]\Biggr\},
\end{eqnarray}
where
$$
P=g_{\mu\nu}P^{\mu\nu}=(p_1+4p_2)R+4p_3{1\over k^2}, \qquad P'\equiv {P\over
1-\beta}.
$$
Using the equation
\begin{eqnarray}
\setcounter{equation}{178}
& &
\int d^4 x g^{1/2}\varepsilon_\mu\left\{g^{\mu\nu}
\left(-\Square+X\right)
+\gamma\nabla^\mu\nabla^\nu\right\}\varepsilon_\nu=
\nonumber\\[10pt]
& &
\int d^4 x g^{1/2}\left\{\varepsilon^\perp\Delta_1(X)\varepsilon^\perp
+{1-\gamma\over 4}\sigma
\Delta_0\left({X-{R\over 4}\over 1-\gamma}\right)
\Delta_0(0)\sigma\right\},
\end{eqnarray}
we find the determinants of the ghost operators (4.175) and (4.176)
$$
\det F=\det \Delta_1\left(-{R\over 4}\right)
\det\Delta_0\left({R\over \kappa-3}\right),
$$
$$
\det H=\det \Delta_1\left({R+P\over 4}\right)
\det\Delta_0\left(P'\right).
\eqno(4.179)
$$
For the operators $F$ and $H$ to be positive definite we assume $\beta<1$
and $\kappa<3$.

Thus using the determinants of the ghost operators, (4.179), and the Jacobian of
the change of variables, (4.174), and integrating $\exp(-S_2-S_{\rm gauge})$ we
obtain the one--loop effective action off mass shell in De~Witt gauge (4.57)
with arbitrary gauge parameters $\kappa$, $\alpha$, $\beta$ and $P$
$$
\Gamma_{(1)}=I_{(2)}+I_{(1)}+I_{(0)},
\eqno(4.180)
$$
where
$$
I_{(2)}={1\over 2}\log\det\left[\Delta_2\left({R\over 6}\right)
\Delta_2\left(m_2^2+{f^2+\nu^2\over 3\nu^2}R\right)
+{1\over 2}m_2^2(R-4\Lambda)\right],
\eqno(4.181)
$$
$$
I_{(1)}={1\over 2}\log{
\det\left[\Delta_1\left(-{R\over 4}\right)
\Delta_1\left({R+P\over 4}\right)
+{1\over 2}{\alpha^2\over k^2}(R-4\Lambda)\right]
\over
\left(\det\Delta_1\left(-{R\over 4}\right)\right)^2
\det\Delta_1\left({R+P\over 4}\right)},
\eqno(4.182)
$$
$$
I_{(0)}={1\over 2}\log{
\det\left[\Delta_0^2\left({R\over \kappa-3}\right)
\Delta_0\left(P'\right)\Delta_0\left(m_0^2\right)
+{\cal D}\left(\Lambda-{R\over 4}\right)
+{\cal C}\left(\Lambda-{R\over 4}\right)^2\right]
\over
\left(\det\Delta_0\left({R\over \kappa-3}\right)\right)^2
\det\Delta_0\left(P'\right)},
\eqno(4.183)
$$
\begin{eqnarray}
\setcounter{equation}{184}
{\cal D}&=&4m_0^2{\kappa^2-3\over (\kappa-3)^2}
\Delta_0\left({R\over \kappa^2-3}\right)
\Delta_0\left(P'\right)
\nonumber\\[10pt]
& &
-{8\alpha^2\over k^2(1-\beta)(\kappa-3)^2}\Delta_0\left(m_0^2\right)
\Delta_0\left(-{R\over 2}\right),
\end{eqnarray}
$$
{\cal C}=16{\alpha^2\nu^2\over k^4(1-\beta)(\kappa-3)^2}.
$$

The quantity $I_{(2)}$ describes the contribution of two tensor fields,
$I_{(1)}$ gives the contribution of the vector ghost and $I_{(0)}$ is the
contribution of the scalar conformal field off mass shell. The contribution of
the tensor fields $I_{(2)}$, (4.181), does not depend on the gauge, the
contribution of the vector ghost $I_{(1)}$, (4.182), depends on the parameters
$\alpha$ and $P$ and the contribution of the scalar field $I_{(0)}$, (4.183),
(4.184), depends on all gauge parameters $\kappa$, $\alpha$, $\beta$ and $P$.

The expressions for $I_{(1)}$ and $I_{(0)}$, (4.182), (4.183), are simplified
in some particular gauges
$$
I_{(1)}\Big\vert_{\alpha=0}=
-{1\over 2}\log\det\Delta_1\left(-{R\over 4}\right),
\eqno(4.185)
$$
$$
I_{(0)}\Bigg\vert_{\alpha=0 \atop \kappa=0\hfill}
=I_{(0)}\Bigg\vert_{\beta=-\infty \atop \kappa=0}
={1\over 2}\log{
\det\left[\Delta_0\left(-{R\over 3}\right)
\Delta_0\left(m_0^2\right)
+{1\over 3}m_0^2(R-4\Lambda)\right]
\over
\det\Delta_0\left(-{R\over 3}\right)},
\eqno(4.186)
$$
$$
I_{(0)}\Bigg\vert_{\alpha=0 \atop \kappa=1}
=I_{(0)}\Bigg\vert_{\beta=-\infty \atop \kappa=1\hfill}
={1\over 2}\log{
\det\left[\Delta_0\left(-{R\over 2}\right)
\Delta_0\left(m_0^2\right)
+{1\over 2}m_0^2(R-4\Lambda)\right]
\over
\det\Delta_0\left(-{R\over 2}\right)},
\eqno(4.187)
$$
$$
I_{(0)}\Big\vert_{\kappa=-\infty}
={1\over 2}\log{
\det\left[\Delta_0(0)\Delta_0\left(m_0^2\right)
-m_0^2(R-4\Lambda)\right]
\over \det\Delta_0(0)},
\eqno(4.188)
$$

\ Let us also calculate the gauge--independent and 
reparametriza\-tion\---in\-va\-ri\-ant unique effective action, (4.21),
on De Sitter background
in the orthogonal gauge, (4.36),  (4.38). In the one--loop approximation, (4.40), it
differs from the standard effective action in De~Witt gauge, (4.14), (4.56),
(4.58), with $\alpha=0$ only by an extra term in the operator $\tilde\Delta$,
(4.41), due to the Christoffel connection of the configuration space, (4.27), (4.47).
Therefore, the unique effective action, (4.38), (4.40), can be obtained from the
standard one, (4.12), (4.14), in De~Witt gauge, (4.56), (4.58), for $\alpha=0$ by
substituting the operator ${\cal D}_i{\cal D}_kS$, (4.41), for the operator
$S_{,ik}$, i.e., by the replacing the quadratic part of the action $S_2(g+h)$,
(4.164), by
$$
\tilde S_2(g+h)=S_2(g+h)-{1\over 2}h^i\left\{{}^j_{ik}\right\}\varepsilon_j
h^k,
$$
\begin{eqnarray}
\setcounter{equation}{189}
h^i\left\{{}^j_{ik}\right\}\varepsilon_j h^k
&=&-{1\over 4}(\kappa^{-1}-1){1\over k^2}(R-4\Lambda)\int d^4
xg^{1/2}\left(h_{\mu\nu}^2-{1\over 4}h^4\right)
\nonumber\\[10pt]
&=&
-{1\over 4}(\kappa^{-1}-1){1\over k^2}(R-4\Lambda)\int d^4 xg^{1/2}
\Biggl\{\bar h^{\perp\ 2}
\nonumber\\[10pt]
& &
+2\varepsilon^\perp\Delta_1\left(-{R\over 4}\right)\varepsilon^\perp
+{3\over 4}\sigma\Delta_0(0)\Delta_0\left(-{R\over 3}\right)\sigma\Biggr\},
\end{eqnarray}
where $\kappa$ is the parameter of the configuration space metric that is given,
according to the paper [45], by the formula (4.108). Let us note
that in the Einstein gravity one obtains for the parameter $\kappa$ the value
$\kappa=1$ [45, 46]. Therefore, the additional contribution of the connection
(4.189) vanishes and the unique effective action on De Sitter background
coincides with the standard one computed in De~Witt gauge, (4.56), (4.58), with
$\alpha=0$ and $\kappa=1$ (i.e., in the harmonic De Donder--Fock--Landau gauge
(4.60)).

Taking into account the Jacobian of the change of variables (4.174) and the
ghosts determinant (4.179) the functional measure in the constrained
variables takes the form
\begin{eqnarray}
\setcounter{equation}{190}
d h_{\mu\nu}\delta(R_{i\mu}h^i)\det N
&=&d\bar h^\perp\, d\varepsilon^\perp\, d\varphi\,d\sigma\,
\delta(\varepsilon^\perp)
\nonumber\\[10pt]
& &
\times \delta\left[\kappa\varphi
-(\kappa-3)\Delta_0\left({R\over \kappa-3}\right)\sigma\right]
\det\Delta_0\left({R\over \kappa-3}\right)
\nonumber\\[10pt]
& &
\times\left[\det\Delta_1\left(-{R\over 4}\right)
\det\Delta_0\left(-{R\over 3}\right)\right]^{1/2}.
\end{eqnarray}

\noindent
Integrating $\exp(-\tilde S_2)$ we obtain the one--loop unique effective action
$$
\tilde\Gamma_{(1)}=\tilde I_{(2)}+\tilde I_{(1)}+\tilde I_{(0)},
\eqno(4.191)
$$
where
$$
\tilde I_{(2)}={1\over 2}\log\det\left[\Delta_2\left({R\over 6}\right)
\Delta_2\left(m_2^2+{f^2+\nu^2\over 3\nu^2}R\right)
+{1\over 2\kappa }m_2^2(R-4\Lambda)\right],
\eqno(4.192)
$$
$$
\tilde I_{(1)}=-{1\over 2}\log\det\Delta_1\left(-{R\over 4}\right),
\eqno(4.193)
$$
$$
\tilde I_{(0)}={1\over 2}\log{
\det\left[\Delta_0\left({R\over \kappa-3}\right)
\Delta_0\left(m_0^2\right)
-{1\over \kappa-3}m_0^2(R-4\Lambda)\right]
\over
\det\Delta_0\left({R\over \kappa-3}\right)}.
\eqno(4.194)
$$
The eqs. (4.191)--(4.194) for $\kappa=1$ do coincide indeed with the
standard effective action in the gauge $\alpha=0 $, $\kappa=1$, (4.180),
(4.181), (4.185), (4.187). However, to obtain the unique effective action in
our case one has to put in (4.191)--(4.194) $\kappa=3f^2/(f^2+2\nu^2)$, (4.108).

On the mass shell (4.150) the dependence on the gauge disappears and we have
$$
I_{(2)}^{\rm on-shell}=
{1\over 2}\log\det\Delta_2\left({2\over 3}\Lambda\right)
+{1\over 2}\log\det
\Delta_2\left(m_2^2+{4\over 3}\cdot{(f^2+\nu^2)\over \nu^2}\Lambda\right),
\eqno(4.195)
$$
$$
I^{\rm on-shell}_{(1)}=-{1\over 2}\log\det\Delta_1\left(-\Lambda\right),
\eqno(4.196)
$$
$$
I^{\rm on-shell}_{(0)}={1\over 2}\log\det\Delta_0(m_0^2).
\eqno(4.197)
$$

Herefrom one sees immediately the spectrum of the physical excitations
of the theory: one massive tensor field of spin $2$ ($5$ degrees of freedom),
one massive scalar field ($1$ degree of freedom) and the Einstein graviton,
i.e., the massless tensor field of spin $2$ ($5-3=2$ degrees of freedom).
Altogether the higher-\-derivative quantum gravity (4.49) has $5+2+1=8$ degrees
of freedom.

To calculate the functional determinants of the differential operators we will
use the technique of the generalized $\zeta$--function [55, 61, 64, 79, 80, 115,
116]. Let us define the $\zeta$--function by the functional trace of the complex
power of the differential operator of order $2k$,  $\Delta^{(k)}$,
$$
\zeta_s\left(p;\Delta^{(k)}/\mu^{2k}\right)\equiv
{\rm tr} \left(\Delta^{(k)}/\mu^{2k}\right)^{-p},
\eqno(4.198)
$$
where
$$
\Delta^{(k)}=P_k(-\Square),
\eqno(4.199)
$$
 $P_k(x)$ is a polynomial of order $k$, $\mu$ is a dimensionful mass parameter
and $s$ denotes the spin of the field, the operator $\Delta^{(k)}$
is acting on.

For ${\rm Re} \, p>2/ k$ \ the $\zeta$--function is determined by the convergent
series over the eigenvalues (4.168)
$$
\zeta_s\left(p;\Delta^{(k)}/\mu^{2k}\right)
=\sum\limits_n d_n\left(P_k(\lambda_n)/\mu^{2k}\right)^{-p},
\eqno(4.200)
$$
where the summation runs over all modes of the d'Alambert operator, (4.168),
with positive multiplicities, $d_n>0$, including the negative and zero modes of
the operator $\Delta^{(k)}$. The zero modes  give an infinite constant that
should be simply subtracted, whereas the negative modes lead to an imaginary
part indicating to the instability [61]. For ${\rm Re} \, p\le 2/ k$ the
analytical continuation of (4.200) defines a meromorphic function with poles on
the real axis. It is important, that the $\zeta$--function is analytic at the
point $p=0$. Therefore, one can define the finite values of the total number of
modes of the operator $\Delta^{(k)}$ (taking each mode $k$ times) and its
functional determinant
$$
k\;{\rm tr}\;1=B(\Delta^{(k)}),
\eqno(4.201)
$$
$$
\log\det\left(\Delta^{(k)}/\mu^{2k}\right)=-\zeta'_s(0),
\eqno(4.202)
$$
where
$$
B(\Delta^{(k)})=k\zeta_s(0),
$$
$$
\zeta'_s(p)={d\over dp}\zeta_s(p).
\eqno(4.203)
$$

Under the change of the scale parameter $\mu$ the functional determinant
behaves as follows
$$
\zeta'_s\left(0;\Delta^{(k)}/\mu^{2k}\right)
=-B(\Delta^{(k)})\log{\rho^2\over \mu^2}
+\zeta'_s\left(0;\Delta^{(k)}/\rho^{2k}\right).
\eqno(4.204)
$$

Using the spectrum of the d'Alambert operator (4.168) we rewrite (4.200) in
the form
$$
\zeta_s\left(p;\Delta^{(k)}/\rho^{2k}\right)
={2s+1\over 3}\sum\limits_{\nu\ge s+{3\over 2}, \ \Delta\nu=1} \nu(\nu^2-j^2)
\left[P_k\left(\rho^2\left(\nu^2-{9\over 4}-s\right)\right)\right]^{-p},
\eqno(4.205)
$$
where
$$
\rho^2={R\over 12}, \qquad j=s+{1\over 2}.
$$

The sum (4.205) can be calculated for ${\rm Re}\, p>{2\over k}$ by means of the
Abel--Plan summation formula [168]
\begin{eqnarray}
\setcounter{equation}{206}
\sum\limits_{\nu\ge {1\over 2}}f(\nu)
&=&\sum\limits_{{1\over 2}\le\nu\le k-{1\over 2}}f(\nu)
+\int\limits_{k+\varepsilon}^\infty d t\, f(t)
\nonumber\\[10pt]
& &
+\int\limits_0^\infty {dt\over e^{2\pi t}+1}
\left[if(k+\varepsilon-it)-if(k+\varepsilon+it)\right],
\end{eqnarray}
where the integer $k$ should be chosen in such a way that for ${\rm Re}\, \nu
>k$ the function $f(\nu)$ is analytic, i.e., it does not have any poles. The
infinitesimal parameter $\varepsilon>0$ shows the way how to get around the
poles (if any) at ${\rm Re}\, \nu =k$. The formula (4.206) is valid for the
functions $f(\nu)$ that fall off sufficiently rapidly at the infinity:
$$
f(\nu)\Big\vert_{|\nu|\to\infty}\sim|\nu|^{-q}, \qquad {\rm Re} \, q >1.
\eqno(4.207)
$$
When applying the formula (4.206) to (4.205) the second integral in (4.206)
gives an analytic function of the variable $p$. All the poles of the
$\zeta$--function are contained in the first integral. By using the analytical
continuation and integrating by parts one can calculate both $\zeta(0)$ and
$\zeta'(0)$.

As a result we obtain for the operator of second order,
$$
\Delta_s(X)=-\Square+X,
\eqno(4.208)
$$
and for the operator of forth order,
$$
\Delta_s^{(2)}(X,Y)=\Square^2-2X\Square+Y,
\eqno(4.209)
$$
the finite values of the total number of modes (4.201) and the
determinant  (4.202)
$$
B(\Delta_s(X))={2s+1\over 12}\left\{(b^2+j^2)^2
-{2\over 3}j^2+{1\over 30}\right\},
\eqno(4.210)
$$
$$
B\left(\Delta_s^{(2)}(X,Y)\right)=2\cdot{2s+1\over 12}\left\{(b^2+j^2)^2
-{2\over 3}j^2+{1\over 30}-a^2\right\},
\eqno(4.211)
$$
$$
\zeta_s'(0;\Delta_s(X)/\rho^2)={2s+1\over 3}F_s^{(0)}(\bar X),
\eqno(4.212)
$$
$$
\zeta_s'(0;\Delta^{(2)}_s(X,Y)/\rho^4)={2s+1\over 3}F_s^{(2)}(\bar X, \bar Y),
\eqno(4.213)
$$
where
$$
b^2=\bar X-s-{9\over 4}, \qquad \bar X={X\over \rho^2},
$$
$$
a^2=\bar Y-\bar X^2, \qquad \bar Y={Y\over \rho^4},
$$
and
\begin{eqnarray}
\setcounter{equation}{214}
F_s^{(0)}(\bar X)&=&-{1\over 4}b^2(b^2+2j^2)\log b^2+{1\over 2}j^2b^2
+{3\over 8}b^4
\nonumber\\[10pt]
& &
+2\int\limits_0^\infty {dt\;t\over e^{2\pi t}+1}(t^2+j^2)\log|b^2-t^2|
\nonumber\\[10pt]
& &
+\sum\limits_{{1\over 2}\le \nu\le s-{1\over 2}}
\nu(\nu^2-j^2)\log(\nu^2+b^2),
\end{eqnarray}
\begin{eqnarray}
\setcounter{equation}{215}
F_s^{(2)}(\bar X, \bar Y)&=&{1\over 4}(a^2-b^4-2j^2b^2)\log(b^4+a^2)
\nonumber\\[10pt]
& &
-a(b^2+j^2)\left[\arctan\left({b^2\over a}\right)-{\pi\over 2}\right]
-{1\over 4}a^2+{3\over 4}b^4+j^2b^2
\nonumber\\[10pt]
& &
+2\int\limits_0^\infty {dt\;t\over e^{2\pi t}+1}(t^2+j^2)
\log\left[(b^2-t^2)^2+a^2\right]
\nonumber\\[10pt]
& &
+\sum\limits_{{1\over 2}\le \nu\le s-{1\over 2}}
\nu(\nu^2-j^2)\log\left[(\nu^2+b^2)^2+a^2\right].
\end{eqnarray}
The introduced functions, (4.214) and (4.215), are related by the equation
$$
F_s^{(0)}(\bar X)+F_s^{(0)}(\bar Y)=
F_s^{(2)}\left({\bar X+\bar Y\over 2}; \bar X\bar Y\right).
\eqno(4.216)
$$

In complete analogy one can obtain the functional determinants of the operators
of higher orders and even non--local, i.e., integro--differential, operators.

Using the technique of the generalized $\zeta$--function and
separating the dependence on the
renormalization parameter $\mu$
we get the one--loop
effective action, (4.180)--(4.184),
$$
\Gamma_{(1)}={1\over 2}B_{\rm tot}\log{R\over 12\mu^2}+\Gamma_{(1)\ \rm ren},
\eqno(4.217)
$$
where
$$
\Gamma_{(1)\ \rm ren}=\Gamma_{(1)}\Big\vert_{\mu^2=\rho^2={R\over 12}}.
\eqno(4.218)
$$

For the study of the effective action one should calculate, first of all, the
coefficient $B_{\rm tot}$. To calculate the contributions of the tensor, (4.181),
and vector, (4.182), fields in $B_{\rm tot}$ it suffices to use the formulas
(4.210) and (4.211) for the operators of the second and the forth order.
Although the contribution of the scalar field in arbitrary gauge, (4.183),
(4.184), contains an operator of eighth order, it is not needed to calculate the
coefficient $B$ for the operator of the eighth order. Noting that on mass shell,
(4.150), the contribution of the scalar field (4.197) contains only a second
order operator, one can expand the contribution of the scalar field off mass
shell in the extremal, i.e., in $(R-4\Lambda)$, limiting oneself only to linear
terms.

One should note, that the differential change of the variables (4.156)--(4.158)
brings some new zero modes that were not present in the non--constrained
operators. Therefore, when calculating the total number of modes (i.e., the
coefficient $B_{\rm tot}$) one should subtract the number of zero modes of the
Jacobian of the change of variables (4.174):
$$
B_{\rm tot}=\sum\limits_i B(\Delta_i)-{\cal N}(J).
\eqno(4.219)
$$
Using the number of zero modes of the operators entering the Jacobians, (4.174),
$$
{\cal N}(\Delta_0(0))=1, \qquad
{\cal N}\left(\Delta_0\left(-{R\over 3}\right)\right)=5, \qquad
{\cal N}\left(\Delta_1\left(-{R\over 4}\right)\right)=10,
\eqno(4.220)
$$
we obtain
$$
{\cal N}(J)=2\times 15-1=29.
\eqno(4.221)
$$
Thus we obtain the coefficient $B_{\rm tot}$
\begin{eqnarray}
\setcounter{equation}{222}
B_{\rm tot}&=&{20\over 3}{f^4\over \nu^4}+20{f^2\over \nu^2}-{634\over
45}+24\gamma{1\over x}
\nonumber\\[10pt]
& &
+16\left[{3\over 4}(\nu^4+5f^4)
+\lambda\left(10{f^4\over \nu^2}+15f^2-\nu^2
-6\gamma\right)\right]{1\over x^2},
\end{eqnarray}
where  $x=Rk^2$ and the coefficient $\gamma$ is given by the formula (4.97).

The unique effective action (4.191) has the same form, (4.217), with the
coefficient $\tilde B_{\rm tot}$ of the form (4.222) but with the change
$\gamma\to \tilde\gamma$, where $\tilde\gamma$ is given by the formula (4.114).

On the other hand, the coefficient $B_{\rm tot}$ can be obtained from the
general expression for the divergences of the effective action (4.71) on De
Sitter background (4.149),
$$
B_{\rm tot}=4\beta_1+24\beta_3+24\gamma{1\over x}
+16\left({3\over 2}\beta_4-6\gamma\lambda\right){1\over x^2},
\eqno(4.223)
$$
where the coefficients $\beta_1$, $\beta_3$ and $\beta_4$ are given by the
formulas (4.72)--(4.74), and the coefficient $\gamma$ is given by the formula
(4.97) for standard effective action in arbitrary gauge and by the expression
(4.114) for the unique effective action.

Comparing the expressions (4.222) and (4.223) we convince ourselves that our
result for the coefficient $\beta_3$, (4.73), that differs from the results of
other authors [152--155] (see Sect. 4.2), and our results for the divergences of
the effective action in arbitrary gauge, (4.97), and for the  divergences of the
unique effective action, (4.114), are correct.

On mass shell (4.150) the coefficient (4.222) does not depend on the gauge and
we have a single--valued expression
\begin{eqnarray}
\setcounter{equation}{224}
B_{\rm tot}^{\rm on-shell}
&=&{20\over 3}{f^4\over \nu^4}+20{f^2\over \nu^2}-{634\over 45}
\nonumber\\[10pt]
& &
+\left(10{f^4\over \nu^2}+15f^2-\nu^2\right){1\over \lambda}
+{3\over 4}(\nu^4+5f^4){1\over \lambda^2}.
\end{eqnarray}

Let us also calculate the finite part of the effective action (4.218). Since it
depends essentially on the gauge, (4.180)--(4.184), we limit ourselves to the
case of the unique effective action, (4.191)--(4.194). Using the results (4.202)
and (4.212)--(4.215) we obtain
\begin{eqnarray}
\setcounter{equation}{225}
\tilde\Gamma_{(1)\rm ren}&=&-{1\over 6}\Biggl\{5F_2^{(2)}(Z_1,Z_2)
-3F_1^{(0)}(-3)
\nonumber\\[10pt]
& &
+F_0^{(2)}(Z_3,Z_4)
-F_0^{(0)}\left(-2{f^2\over \nu^2}-4\right)\Biggr\},
\end{eqnarray}
where
$$
Z_1=6f^2{1\over x}+2{f^2\over \nu^2}+3,
$$
$$
Z_2=-96(f^2+2\nu^2)\lambda{1\over x^2}
+48(f^2+\nu^2){1\over x}
+8{f^2\over \nu^2}+8,
\eqno(4.226)
$$
$$
Z_3=6\nu^2{1\over x}-{f^2\over \nu^2}-2,
$$
$$
Z_4=-96(f^2+2\nu^2)\lambda{1\over x^2},
$$
 $F_s^{(k)}$ are the functions introduced above, (4.214), (4.215), and $x=Rk^2$.

 On mass shell (4.150) the effective action does not depend on the gauge and
has the form, (4.13), (4.154), (4.195)-(4.197), (4.217),
\begin{eqnarray}
\setcounter{equation}{227}
\Gamma_{\rm on-shell}&=&(4\pi)^2
\left\{4\left(\epsilon-{1\over \nu^2}\right)-{3\over \lambda}\right\}
\nonumber\\[10pt]
& &
+\hbar\left\{{1\over 2}B_{\rm tot}^{\rm on-shell}\log{\lambda\over 3\mu^2 k^2}
+\Gamma^{\rm on-shell}_{(1) \rm ren}\right\}+O(\hbar^2),
\end{eqnarray}
where
\begin{eqnarray}
\setcounter{equation}{228}
\Gamma^{\rm on-shell}_{(1) \rm ren}
&=&-{1\over 6}\Biggl\{
5F_2^{(0)}\left(3{f^2\over \lambda}+4{f^2\over \nu^2}+4\right)
\nonumber\\[10pt]
& &
+5F_2^{(0)}(2)-3F_1^{(0)}(-3)
+F_0^{(0)}\left(3{\nu^2\over \lambda}\right)\Biggr\}.
\end{eqnarray}

The expression (4.227) gives the vacuum action on De Sitter background with
quantum corrections. It is real, since the operators in (4.195)--(4.197) do not
have any negative modes provided the conditions (4.170) and 
 (4.171) are fulfilled.
Although the operator $\Delta_0(m_0^2)$ has one negative mode, $\varphi={\rm
const}$, subject to the condition (4.171), it is non--physical, since it is just
the zero mode of the Jacobian of the change of variables (4.174). All other
modes of the operator  $\Delta_0(m_0^2)$ are positive subject to the condition
(4.171) in spite of the fact that $m_0^2<0$. Depending on the value of De
Sitter curvature $R$
off mass shell there can appear negative modes leading to an imaginary part of
the effective action.

Differentiating the effective action, (4.13), (4.153), (4.217), (4.222), (4.225),
we obtain the effective equation for the background field, i.e., the curvature
of De Sitter space,
\begin{eqnarray}
\setcounter{equation}{229}
{1\over \hbar k^2}{\partial \Gamma\over \partial R}
&=&{1\over \hbar}24(4\pi)^2\cdot{x-4\lambda\over x^3}
+{1\over 2x}B_{\rm tot}(x)
\nonumber\\[10pt]
& &
+{1\over 2}B'_{\rm tot}(x)\log{x\over 12\mu^2 k^2}
+\Gamma'_{(1)\rm ren}(x)
+O(\hbar)=0,
\end{eqnarray}
where
\begin{eqnarray}
B'_{\rm tot}(x)={\partial B_{\rm tot}\over \partial x}
&=&
-{32\over x^3}\Biggl[{3\over 4}(\nu^4+5f^4)
\nonumber\\[10pt]
& &
+\lambda\left(10{f^4\over \nu^2}+15f^2-\nu^2-6\gamma\right)\Biggr]
-24\gamma{1\over x^2},
\end{eqnarray}
$$
\Gamma'_{(1)\rm ren}(x)={\partial \Gamma_{(1)\rm ren}\over \partial x},
\qquad x=Rk^2.
\eqno(4.231)
$$
The perturbative solution of the effective equation (4.229) has the form
\begin{eqnarray}
\setcounter{equation}{232}
Rk^2&=&4\lambda
-\hbar{\lambda^2\over 3(4\pi)^2}\Biggl\{B_{\rm tot}^{\rm on-shell}
+4\lambda B'_{\rm tot}(4\lambda)\log{\lambda\over 3\mu^2k^2}
\nonumber\\[10pt]
& &
+8\lambda\Gamma'_{(1)\rm ren}(4\lambda)\Biggl\}
+O(\hbar^2).
\end{eqnarray}
It gives the corrected value of the curvature of De Sitter space with regard to
the quantum effects.

Perturbation theory near this solution is applicable for $\lambda\ne 0$ in the
region $f^2\sim\nu^2\sim\lambda\ll 1$. For $\lambda\sim 1$, i.e., when $\Lambda$ is of Planck mass order ${1/k^2}$, the contributions of higher loops are essential
and the perturbation theory is not adequate anymore.

Apart from the perturbative solution (4.232) the equation (4.229) can also have
non-perturbative ones. In the special case $\lambda=0$ non-perturbative
solution $R\ne 0$ means the spontaneous creation of De Sitter space from the
flat space due to quantum gravitational fluctuations. Therefore, it seems quite
possible that De Sitter space, needed in the inflational cosmological scenarios
of the evolution of the Universe [170], has quantum--gravitational origin [169].
However, almost any non--perturbative solution has the order $Rk^2\sim 1$ and,
therefore, is inapplicable in the one--loop approximation.

%
%
%
%
%
%
%
%
%
%
%
%
%
%
%
%
%
%

\chapter{Conclusion}
\markboth{\sc Chapter 5. Conclusion}{\sc Chapter 5. Conclusion}

The following main results are obtained in the present dissertation.
\begin{itemize}
\item[1.] The methods for the covariant expansions of arbitrary fields in a curved
space with arbitrary connection in generalized Taylor series and the Fourier
integral  in most general form are formulated.

\item[2.] A manifestly covariant technique for the calculation of De~Witt
coefficients on the basis of the method of covariant expansions is elaborated.
The corresponding diagrammatic formulation of this technique is given.

\item[3.] The De~Witt coefficients $a_3$ and $a_4$ at coinciding points are calculated.

\item[4.] The renormalized one--loop effective action for the 
massive scalar, spinor
and vector fields in an external gravitational field up to the 
terms of order $1/ m^4$ is calculated.

\item[5.] Covariant methods for studying the non--local structure of the
effective action are developed.

\item[6.] The terms of first order in background fields in De~Witt coefficients
are calculated. The summation of these terms is carried out and a non--local
covariant expression for the Green function at coinciding points up to terms of
second order in background fields is obtained. It is shown that in the
conformally invariant case the Green function at coinciding points is finite in the
first order in the background fields.

\item[7.] The terms of second order in background fields in De~Witt coefficients
are calculated. The summation of these terms is carried out and a manifestly
covariant non--local expression for the one--loop effective action up to the 
terms of
third order in background fields is obtained. All formfactors, their asymptotics
and imaginary parts (for standard definition of the asymptotic regions, ground
states and causal boundary conditions) are calculated. A finite effective
action in the conformally invariant case of massless scalar field in
two--dimensional space is obtained.

\item[8.] The covariantly constant terms in De~Witt coefficients for the case
of scalar field are calculated. It is shown that the corresponding 
Schwinger--De~Witt series diverges. The Borel summation of the covariantly 
constant terms is
carried out and an explicit non--analytic expression in the background fields for
the one--loop effective action up to the terms with covariant derivatives of the
background fields is obtained.

\item[9.] The off--shell one--loop divergences of the effective action in
arbitrary covariant gauge as well as those of the unique
effective action in higher--derivative quantum gravity are calculated.

\item[10.] The ultraviolet asymptotics of the coupling constants of the
higher--derivative quantum gravity are found. It is shown that in the
`physical' region of the coupling constants, that is characterized by the absence
of the tachyons on the flat background, the conformal sector has `zero--charge'
behavior. Therefore, the higher--derivative quantum gravity at higher energies
goes beyond the limits of weak conformal coupling. This conclusion does not
depend on the presence of the matter fields of low spins. In other words, the
condition of the 
conformal stability of the flat background, which is held usually as
`physical', is incompatible with the asymptotic freedom in the conformal
sector. Therefore, the flat background cannot present the ground state of the
theory in the ultraviolet region.

\item[11.] It is shown, that the theory of gravity with a quadratic in the
curvature and positive definite Euclidean action possesses a stable non--flat
ground state and is asymptotically free both in the tensor sector and the
conformal one. A
physical interpretation of the nontrivial ground state as a condensate of
conformal excitations, that is formed as a result of a phase transition, is
proposed.

\item[12.] The effective potential, i.e., the off--shell one--loop effective
action  in arbitrary covariant gauge, and the unique effective action in the
higher\---derivative quantum gravity on De Sitter background, is calculated. 
The determinants of the operators of second and forth orders are 
obtained by means of the generalized $\zeta$--function.

\item[13.] The gauge-- and parametrization--independent `unique' effective
equations for the background field, i.e., for the curvature of De Sitter space,
are obtained. The perturbative solution of the effective equations, that gives
the corrected value of the curvature of De Sitter background space due to
quantum effects, is found.

In conclusion, I would like to thank my scientific supervisor, Prof. Dr. V. R.
Khalilov, for the general guidance, all--round help and many pieces of good
advice on all stages of my work. I am also much obliged to A. O. Barvinsky for
many valuable discussions and constant interest in the work, to all members of
the Department of Theoretical Physics, Physics Faculty, of Moscow State
University for the attention to my work and friendly support, and to the chiefs
of the scientific seminars, where the results obtained in this dissertation
were presented, Associate Member of the USSR Academy of Sciences E. S. Fradkin,
Academician M. A. Markov
and Prof. Dr. V. N. Ponomarev, as well as to the participants of these seminars
for fruitful discussions.

\end{itemize}

%
%
%
%
%
%
%
%
%
%
%
%
%
%
%


\end{document}